\documentclass[10pt]{article}

\usepackage{amsmath}
\usepackage{amssymb}
\usepackage[nosort]{cite}
\usepackage{color}
\usepackage{epsfig}
\usepackage{float}
\usepackage{graphicx}
\usepackage[latin1]{inputenc}
\usepackage{times}
\usepackage{url}

\begin{document}

\baselineskip = 14pt
\parskip 1.0mm
\renewcommand{\baselinestretch}{1.0}


\def\det{\mathop{\rm det}\nolimits}
\def\tr{\mathop{\rm tr}\nolimits}

\newcommand{\Ga}{{\Gamma}}
\newcommand{\De}{{\Delta}}
\newcommand{\Lm}{{\Lambda}}

\newcommand{\al}{{\alpha}}
\newcommand{\bt}{{\beta}}
\newcommand{\ga}{{\gamma}}
\newcommand{\de}{{\delta}}
\newcommand{\ep}{{\epsilon}}
\newcommand{\vep}{{\varepsilon}}
\newcommand{\zt}{{\zeta}}
\newcommand{\te}{{\theta}}
\newcommand{\ka}{{\kappa}}
\newcommand{\lm}{{\lambda}}
\newcommand{\sig}{{\sigma}}
\newcommand{\vphi}{{\varphi}}
\newcommand{\om}{{\omega}}

\newcommand{\alt}{{\rm alt}}
\newcommand{\bdy}{{\rm bdy}}
\newcommand{\blan}{{\Big\langle}}
\newcommand{\bran}{{\Big\rangle}}
\newcommand{\bsa}{{\boldsymbol{a}}}
\newcommand{\bsb}{{\boldsymbol{b}}}
\newcommand{\bsD}{{\boldsymbol{D}}}
\newcommand{\bsk}{{\boldsymbol{k}}}
\newcommand{\bsM}{{\boldsymbol{M}}}
\newcommand{\bulk}{{\rm bulk}}
\newcommand{\cN}{{\mathcal{N}}}
\newcommand{\cont}{{\rm cont.}}
\newcommand{\ct}{{\rm ct}}
\newcommand{\cdN}{{\mathcal{N}_d}}
\newcommand{\I}{{\rm I}}
\newcommand{\IR}{{\rm \, IR}}
\newcommand{\lan}{{\langle}}
\newcommand{\ol}{\overline}
\newcommand{\pd}{{\partial}}
\newcommand{\pr}{{\prime}}
\newcommand{\R}{{\rm R}}
\newcommand{\ra}{{\rm a}}
\newcommand{\rad}{{\rm rad}}
\newcommand{\ran}{{\rangle}}
\newcommand{\rs}{{\rm s}}
\newcommand{\sh}{{\sharp}}
\newcommand{\srel}[2]{{\stackrel{\scriptstyle #1}{\scriptstyle #2}}}
\newcommand{\std}{{\rm std}}
\newcommand{\U}{{\rm U}} 
\newcommand{\UV}{{\rm UV}}
\newcommand{\wg}{{\wedge}}
\newcommand{\wh}{\widehat}
\newcommand{\wt}{\widetilde}

\newcommand{\hchi}{{\wh{\chi}}}
\newcommand{\hochi}{{\wh{\ol{\chi}}}}
\newcommand{\shchi}{{\wh{\chi}^{\sh}}}
\newcommand{\shochi}{{\wh{\ol{\chi}}^{\sh}}}


\newcommand{\EQ}[1]{\begin{equation} #1 \end{equation}}
\newcommand{\AL}[1]{\begin{subequations} \begin{align} #1
\end{align}\end{subequations}}
\newcommand{\ALlabel}[2]{\begin{subequations} \label{#2} \begin{align} #1
\end{align}\end{subequations}}
\newcommand{\SP}[1]{\begin{equation}\begin{split} #1
\end{split}\end{equation}}

\def\mPhi{{\mathbf \Phi}}

\def\gap#1{\vspace{#1 ex}}
\def\be{\begin{equation}}
\def\ee{\end{equation}}
\def\bal{\begin{array}{l}}
\def\ba#1{\begin{array}{#1}}  
\def\ea{\end{array}}
\def\bea{\begin{eqnarray}}
\def\eea{\end{eqnarray}}
\def\beas{\begin{eqnarray*}}
\def\eeas{\end{eqnarray*}}
\def\del{\partial}
\def\eq#1{(\ref{#1})}
\def\fig#1{Fig \ref{#1}} 
\def\re#1{{\bf #1}}
\def\bull{$\bullet$}
\def\nn{\\\nonumber}
\def\ub{\underbar}
\def\nl{\hfill\break}
\def\ni{\noindent}
\def\bibi{\bibitem}
\def\ket#1{|#1 \rangle}
\def\bra#1{\langle #1 |}
\def\bbra{{\langle\kern-2.5pt\langle}}
\def\kket{{\rangle\kern-2.5pt\rangle}}
\def\Bbra{{\Big\langle\kern-3.5pt\Big\langle}}
\def\Kket{{\Big\rangle\kern-3.5pt\Big\rangle}}

\def\vev#1{\langle #1 \rangle} 
\def\lsim{\stackrel{<}{\sim}}
\def\gsim{\stackrel{>}{\sim}}
\def\mattwo#1#2#3#4{\left(
\begin{array}{cc}#1&#2\\#3&#4\end{array}\right)} 
\def\tgen#1{T^{#1}}
\def\half{\frac12}
\def\floor#1{{\lfloor #1 \rfloor}}
\def\ceil#1{{\lceil #1 \rceil}}

\def\mysec#1{\gap1\ni{\bf #1}\gap1}
\def\mycap#1{\begin{quote}{\footnotesize #1}\end{quote}}

\def\bit{\begin{item}}
\def\eit{\end{item}}
\def\benu{\begin{enumerate}}
\def\eenu{\end{enumerate}}
\def\a{\alpha}
\def\as{\asymp}
\def\ap{\approx}
\def\b{\beta}
\def\bp{\bar{\partial}}
\def\cA{{\cal{A}}}
\def\cD{{\cal{D}}}
\def\cL{{\cal{L}}}
\def\cP{{\cal{P}}}
\def\cR{{\cal{R}}}
\def\da{\dagger}
\def\de{\delta}
\def\tD{\tilde D}
\def\e{\eta}
\def\ep{\epsilon}
\def\eqv{\equiv}
\def\f{\frac}
\def\g{\gamma}
\def\G{\Gamma}
\def\h{\hat}
\def\hs{\hspace}
\def\i{\iota}
\def\k{\kappa}
\def\lf{\left}
\def\l{\lambda}
\def\la{\leftarrow}
\def\La{\Leftarrow}
\def\Lla{\Longleftarrow}
\def\Lra{\Longrightarrow}
\def\L{\Lambda}
\def\m{\mu}
\def\na{\nabla}
\def\nn{\nonumber\\}
\def\mm{&&\kern-18pt}  
\def\om{\omega}
\def\O{\Omega}
\def\P{\phi}
\def\pa{\partial}
\def\pr{\prime}
\def\r{\rho}
\def\ra{\rightarrow}
\def\Ra{\Rightarrow}
\def\ri{\right}
\def\s{\sigma}
\def\sq{\sqrt}
\def\S{\Sigma}
\def\si{\simeq}
\def\st{\star}
\def\t{\theta}
\def\ta{\tau}
\def\ti{\tilde}
\def\tm{\times}
\def\Tr{{\rm Tr}}
\def\T{\Theta}
\def\up{\upsilon}
\def\Up{\Upsilon}
\def\v{\varepsilon}
\def\vh{\varpi}
\def\vk{\vec{k}}
\def\vp{\varphi}
\def\vr{\varrho}
\def\vs{\varsigma}
\def\vt{\vartheta}
\def\w{\wedge}
\def\z{\zeta}

\thispagestyle{empty}
\addtocounter{page}{-1}
\vskip-0.35cm
\begin{flushright}
TIFR/TH/11-11\\
\end{flushright}
\vspace*{0.2cm} 
\begin{center}
{\LARGE \bf Holographic Wilsonian flows and emergent fermions 
\\
\vspace*{0.2cm} 
in extremal charged black holes}
\end{center}

\vspace*{0.1cm}

\vspace*{1.0cm} 
\centerline{\bf 
Daniel~Elander, Hiroshi~Isono, and Gautam~Mandal}
\vspace*{0.7cm}
\centerline{\rm Department of Theoretical Physics,}
\vspace*{0.2cm}
\centerline{\it Tata Institute of Fundamental Research,} 
\vspace*{0.2cm}
\centerline{\it Mumbai 400 005, \rm INDIA}
\vspace*{0.4cm}
\centerline{\tt email: daniel, isono, mandal@theory.tifr.res.in,
}

\vspace*{0.8cm}
\centerline{\bf Abstract}
\vspace*{0.3cm}
\vspace*{0.5cm} 

We study holographic Wilsonian RG in a general class of asymptotically
AdS backgrounds with a $U(1)$ gauge field. We consider free charged
Dirac fermions in such a background, and integrate them up to an
intermediate radial distance, yielding an equivalent low energy dual
field theory. The new ingredient, compared to scalars, involves a
`generalized' basis of coherent states which labels a particular half
of the fermion components as coordinates or momenta, depending on the
choice of quantization (standard or alternative).  We apply this
technology to explicitly compute RG flows of charged fermionic
operators and their composites (double trace operators) in field
theories dual to (a) pure AdS and (b) extremal charged black hole
geometries.  The flow diagrams and fixed points are determined
explicitly. In the case of the extremal black hole, the RG flows
connect two fixed points at the UV AdS boundary to two fixed points at
the IR AdS$_2$ region.  The double trace flow is shown, both
numerically and analytically, to develop a pole singularity in the
AdS$_2$ region at low frequency and near the Fermi momentum, which can
be traced to the appearance of massless fermion modes on the low
energy cut-off surface. The low energy field theory action we derive
exactly agrees with the semi-holographic action proposed by Faulkner
and Polchinski in \cite{Faulkner:2010tq}.  In terms of field theory,
the holographic version of Wilsonian RG leads to a quantum theory with
random sources. In the extremal black hole background the random
sources become `light' in the AdS$_2$ region near the Fermi surface
and emerge as new dynamical degrees of freedom.

\vspace{3ex}

\tableofcontents

\section{\label{sec:intro}Introduction and Summary}

Recently, a holographic description of field theories with a finite
cut-off has been proposed in \cite{Heemskerk:2010hk, Faulkner:2010jy,
Bredberg:2010ky,  Nickel:2010pr}. One of the imports of this proposal
is a holographic version of Wilsonian renormalization group (hWRG), in
which non-trivial RG flows of strongly interacting field theories are
described with consummate ease in terms of free fields in the
bulk. The basic idea of hWRG is to integrate out the bulk field from
the boundary up to some intermediate radial distance, yielding an
equivalent dual field theory at a lower cut-off. This procedure has
been exactly implemented for a free scalar field in a fixed background
geometry \cite{Heemskerk:2010hk,
  Faulkner:2010jy}\footnote{\cite{Heemskerk:2010hk, Faulkner:2010jy}
  also discuss gauge field fluctuations. See \cite{Brattan:2011my} for
  a related discussion on metric fluctuations, although from a
  somewhat different viewpoint.}; see \cite{Sin:2011yh, Harlow:2011ke,
  Fan:2011wm, Radicevic:2011py, Evans:2011zd} for further related
developments, as well as, {\em e.g.}, \cite{Alvarez:1998wr,
  Balasubramanian:1999jd,Freedman:1999gp,deBoer:1999xf,Bianchi:2001kw,Skenderis:2002wp} for earlier
treatments of holographic renormalization group (the relationship
between the earlier treatments and hWRG has been discussed in
\cite{Heemskerk:2010hk}).  Although it is not easy to identify the
specific dual field theories which arise in this fashion,\footnote{See
  Section~\ref{sec:FT} for a brief qualitative comparison with a
  matrix field theory with a Wilson-Fisher fixed point; details will
  appear in \cite{EIM2}.}  it is of interest to explore the general
structure of RG flows and fixed points in these theories. Indeed, this
program, with the inclusion of metric fluctuations and interactions,
can provide an RG landscape of cut-off field theories and their
universality classes holographically: something which is of obvious
importance in applied AdS/CFT.
 
The main purpose of this note is to extend this method to fermions
(see also \cite{Laia:2011wf}). A technical motivation is that, because
the Dirac action is of first order, the fermion components contain
both coordinates and momenta; thus, it is not {\em a priori} clear how
to define the `Dirichlet boundary condition' for fermions (see Section
\ref{sec:scalar} for comparison with scalars). As we will see below,
the definition needs the construction of some suitably `generalized'
coherent states.  A more physical motivation for considering the
fermion problem is to understand the origin of the semi-holographic
theory of \cite{Faulkner:2010tq} which gives a simple understanding of
the models of \cite{Liu:2009dm, Faulkner:2009wj} near the Fermi
surface\footnote{See \cite{Faulkner:2010jy} for an elucidation of the
  semi-holographic bosonic model of \cite{Nickel:2010pr}.}. In
\cite{Faulkner:2010tq}, it is argued that the UV theory flows to an IR
theory which, near the Fermi surface, is described by a dual AdS$_2$
geometry plus a `domain wall' fermion. We will discover both the
relevance of the Fermi surface and the origin of this additional
fermion in hWRG: (a) the Fermi surface can be precisely characterized
by a singular behaviour of the double trace coupling as the hWRG flow
is continued to the near horizon AdS$_2$ region, (b) the result of
hWRG can be interpreted as a field theory with a lower cut-off with a
random source: when the cut-off surface is in the AdS$_2$ region the
random source becomes massless and becomes an emergent degree of
freedom in the low energy theory, which is to be identified with the
`domain wall' fermion of \cite{Faulkner:2010tq}.

We describe below the plan of the paper and the basic results:

Section~\ref{sec:scalar} gives a review of hWRG for scalars, including
some new material which include explicit RG flow diagrams in pure AdS
and in extremal charged black hole (ECBH) backgrounds. The flow
diagrams show, respectively, two and four fixed points in these two
backgrounds (see Figure \ref{fig-scalar-flow}). The field theory
interpretation of these findings is described in Section~\ref{sec:FT}.

In Section~\ref{sec:beta} we consider free fermions in a general class
of asymptotically AdS backgrounds with a $U(1)$ gauge field, which are
translationally invariant in the `boundary' directions. As in the case
of the bosonic hWRG, the idea is to integrate out the bulk fermions
from the boundary down to an intermediate radial distance, yielding an
equivalent low energy dual field theory. A crucial difference from the
bosonic hWRG is the need to use a basis of states in the bulk Hilbert
space which are eigenstates of one half of the fermion components (the
`coordinates'); we construct such a basis by using `generalized'
coherent states which distinguish the `appropriate half' in
question.\footnote{Similar states were proposed and constructed in
  \cite{Floreanini:1987gr, Mansfield:1999cc}.} The hWRG gives rise to
flows involving products of pairs of single-trace fermionic operators;
in contrast with the bosonic example, there are different double trace
operators in the fermionic case. We write down the general flow
equations for these operators in the general background and a set of
simplified flow equations in rotationally symmetric backgrounds.

In Section~\ref{sec:AdS}, we provide an exact solution of the hWRG
equations in pure AdS backgrounds (with a constant electrostatic
potential). As in the scalar case, double trace flows interpolate
between two fixed points (with 0 and 1 relevant directions), which,
inside the Klebanov-Witten window \cite{Klebanov:1999tb}, correspond
to two different choices of quantization: standard and alternative
(see Figure \ref{fig:fermionicflowf0f1}). More details of the field
theory interpretation are provided in Section~\ref{sec:FT}.

In Section~\ref{sec:BH}, we compute RG flows in an ECBH background.  
The ECBH background has two AdS
regions: AdS$_{d+1}$ at the boundary and AdS$_2 \times \mathbb
R^{d-1}$ in the near-horizon region.  Developing further on the pure
AdS results of the previous section, we first compute RG flows in the
two AdS regions.  We then complete these flows in the full background
numerically. The RG flows feature four fixed points, two each in the
two AdS regions. The two fixed points in the asymptotically
AdS$_{d+1}$ region both become unstable in the ECBH background
(acquiring, respectively, 2 and 1 relevant directions) and flow down
to the two fixed points in the interior AdS$_2$ region (which continue
to have 0 and 1 relevant directions). We parametrize the flow down
from AdS$_{d+1}$ to AdS$_2$ by the flow of a geometric quantity,
namely the red-shift factor (as in Section~\ref{sec:scalar}); near the
AdS$_{d+1}$ boundary, this can be regarded in the dual field theory as
the flow of the stress tensor $T_{tt}$, as described further in
Section~\ref{sec:FT}. There are parameter ranges 
in which the two fixed points in the AdS$_2$ region merge and
annihilate, which is reminiscent of \cite{Kaplan:2009kr}.

In Section~\ref{sec:FP}, we use the formalism developed above to give
a first-principles derivation of the semi-holographic action of
Faulkner and Polchinski \cite{Faulkner:2010tq} for the extremal black
hole. We show, numerically and analytically, that the double trace
coupling, in the process of hWRG flow, develops a pole singularity at
zero frequency and Fermi momentum on a cut-off surface in the AdS$_2$
region, leading to a non-local dual field theory.  We trace the origin
of the non-locality to integration over a fermion present in the
theory, which becomes massless on the cut-off surface at the Fermi
momentum. The low energy theory, therefore, can be written in a local
form, by combining this emergent fermion with the operators of the
dual field theory, precisely as in \cite{Faulkner:2010tq}. Remarks to
this effect were briefly mentioned earlier in \cite{Faulkner:2010jy}
(see also \cite{Sachdev:2010uz}). Effects of double trace deformations
to AdS/CMT have been considered in \cite{Faulkner:2010gj,Iqbal:2011aj}.

In Section~\ref{sec:FT} we briefly summarize the various results of
the previous sections in field theoretic terms. In particular, we
briefly mention a comparison of the bosonic hWRG with a matrix field
theory in $4-\ep$ dimensions with a Wilson-Fisher fixed point. We also
briefly mention an interpretation of hWRG in terms of a random source,
and interpret the emergence of a massless fermionic degree of freedom
in terms of a fluctuating fermionic source. We include a comparison of
the random source fields with mesons in the Gross-Neveu model
\cite{Gross:1974jv} and in the Nambu$-$Jona-Lasinio approach to QCD
\cite{Dhar:1983fr,Dhar:1985gh}.

In Section~\ref{sec:future} we conclude with a few open questions. The
appendices describe various technical details, including a summary of
the construction of the `generalized' coherent states.

Work described in this paper has been presented at various stages of
development in various meetings \cite{confs}. Fermionic Wilsonian
flows in pure AdS have been discussed in \cite{Laia:2011wf} using
somewhat different methods. 

\section{\label{sec:scalar}Holographic Wilsonian RG: scalars}

This section will contain a review of some of the results in
\cite{Heemskerk:2010hk,Faulkner:2010jy} and a brief account of some
new computations (details will appear in \cite{EIM2}). We will
consider \cite{Heemskerk:2010hk,Faulkner:2010jy} a free bulk scalar
$\phi(z,x^\mu)$ \footnote{\label{ftnt:scalar action} Described by
    an action $S(\phi)= -\frac12 \int~d^dx~dz~ \sqrt g~ [\del_M \phi
      \del^M \phi + m^2 \phi^2]$} in a fixed asymptotically AdS
geometry
\begin{align}
&{}
ds^2
= g_{MN}dx^Mdx^N
= g_{zz}(z) dz^2+ g_{\mu\nu}(z) dx^\mu dx^\nu \,.
\label{gen_met-s}
\end{align}
$\phi$ is dual to a single-trace operator $O(x^\mu)$. Wilsonian RG is
implemented by (a) considering the gravity partition function
(effectively a partition function over $\phi$) as a path integral with
a radial Hamiltonian, from a UV cut-off $z=\ep_0$ to an IR cut-off 
$z=\ep_{\IR}$, 
and (b) performing the integral over a slice $z \in [\ep_0,  \ep]$:
\begin{align}
Z_{\rm grav}(\ep_0) := \langle \IR | \hat U(\ep_{\IR}, \ep_0) | \UV \rangle
= \int\!\!D\tilde\phi 
\langle \IR | \hat U(\ep_{\IR}, \ep) |\tilde \phi \rangle
\langle \tilde \phi| \hat U(\ep, \ep_0) | \UV \rangle 
= \int \!\!D\tilde\phi \,Z_{\rm grav}(\ep, \tilde \phi) 
e^{S_{\rm uv}(\tilde \phi, \ep, \ep_0)} \,.
\label{slicing}
\end{align}
Here $ \hat U(\ep_1, \ep_2):= {\rm P} e^{- \int_{\ep_1}^{\ep_2} \wh{H}} $ (${\rm P}$
denotes radial time ordering) and we will use the parametrization
\begin{align}
e^{S_{\rm uv}} \equiv \langle \tilde \phi|\hat U(\ep, \ep_0) |\UV \rangle 
:= \exp\left[ \int_k \left(- \frac12 
f(k,\ep)\tilde \phi(k) \tilde
\phi(-k) + J(k,\ep)\tilde \phi(-k) \right) + C(\ep)\right] \,,
\label{s-uv}
\end{align}
where $\int_k := \frac{1}{\ka^2}\int\!\!\frac{d^dk}{(2\pi)^d}$.  The wave-functions
$| \IR \rangle$ and $| \UV \rangle$ implement the IR and UV boundary
conditions. For example, if $| \UV \rangle = | \phi_0
\rangle$ we have a Dirichlet b.c.  at UV. In that case,
the holographic dual to \eq{slicing} is obtained by replacing $Z_{\rm
  grav}$s in \eq{slicing} with field theory partition functions with
sources (assuming that the usual GKP-Witten relation \cite{Gubser:1998bc, Witten:1998qj} holds at finite cut-offs)
\begin{align}
\Bbra \exp\int_k\!\phi_0 O \Kket^{\rm std}_{\ep_0}
&= 
\int\!\! D\tilde \phi \Bbra \exp\int_k \!\tilde \phi O \Kket_{\!\!\ep}^{\rm std}~
e^{S_{\rm uv}(\tilde \phi, \ep, \ep_0)} 
\nonumber\\
&=
\Bbra \exp\left(\int_k \frac12 g(k, \ep) O^2 + \int_k h(k,\ep) O + 
C'(\ep)\right) \Kket^{\rm std}_{\ep} \,.
 \label{slicing-ft}
\end{align}
In the last step, we have performed the $\tilde \phi$ integration,
leading to a double trace coupling $g$ and a single trace coupling
$h$, where $g(k)= 1/f(k),\, h(k)=J(k)/f(k).$ Equation
\eq{slicing-ft} relates the field theory at the UV cut-off $\ep_0$ to
that at the intermediate cut-off $\ep$ (for small $\ep_0,\ep$ we can
roughly imagine $\ep_0 \sim \L_0^{-1} \,, \, \ep\sim \L^{-1} \,, \, \L
<\L_0$
\cite{Susskind:1998dq,Peet:1998wn}).\footnote{\label{bare-multi-trace}
  Double trace deformations can also be {\em introduced in the bare
    theory}, by choosing the wave-function 
  $|\rm UV \rangle= \int D\phi_0 \Psi(\phi_0) \ket{\phi_0}$, with $\Psi(\phi_0)$ a Gaussian,
  like in the RHS of \eq{s-uv}. For other formulations of multi-trace deformations in the scalar case,
  see, for example, \cite{Witten:2001ua,Papadimitriou:2007sj,Vecchi:2010dd,Allais:2010qq,Faulkner:2010jy}.}  
The symbol $\bbra .. \kket$ denotes
     {\em unnormalized} correlators which we do not divide by the
     partition function; the subscript $_\ep$ or $_{\ep_0}$ denotes
     the field theory cut-off; the superscript $^{\rm std}$ denotes
     the fact that a Dirichlet b.c. in the bulk corresponds to {\em
       standard} quantization in the field theory (a Neumann boundary
     condition, in an appropriate window, corresponds to the so-called
     {\em alternative} quantization; see below \eq{fn-flow} for more
     details; see footnote \ref{bare-multi-trace} for more general
     choices of boundary conditions).

Clearly $e^{S_{\rm uv}}$ satisfies a Schr\"{o}dinger equation, which, in
the semi-classical limit, becomes the Hamilton-Jacobi equation 
$\del_\ep S =- H(\del S/\del\tilde \phi, \tilde \phi)$. 
In terms of the parametrization
\eq{s-uv}, the Schr\"{o}dinger (or Hamilton-Jacobi) equation becomes
\begin{align}
& \frac{\sqrt{g^{zz}}}{\sqrt{\g}} \del_\ep\left(\sqrt{\g} f(k, \ep)
\right)
= -  f^2(k, \ep)  + k^\mu k_\mu + m^2 \,,
\label{f-flow}
\\
&  \frac{\sqrt{g^{zz}}}{\sqrt{\g}} \del_\ep\left(\sqrt{\g} J(k, \ep)
\right)
= - J(k, \ep) f(k, \ep), ~~\frac{\sqrt{g^{zz}}}{\sqrt{\g}} \del_\ep C(\ep)
= \frac12 \int_k J(k, \ep) J(-k, \ep) \,,
\label{scalar-flow}
\end{align}
where $\gamma$ is the determinant of the induced metric at the cut-off surface.
In view of the comments below \eq{slicing-ft} the above equations are
analogous to beta-function equations; we will elaborate more on this
in Section~\ref{sec:FT}.

\paragraph{AdS$_{d+1}$}

In this case, \eq{f-flow} becomes (using the notation $\dot{(~)}:=
\ep\del_\ep (~)$)
\begin{align}
\dot f=  -f^2 + d\,f + \ep^2 k^2 + m^2 \,.
\label{f-ads-flow}
\end{align}
Now since $\ep$ is a cut-off scale (see comments below
\eq{slicing-ft}), {the explicit appearance of a cut-off in
  the above equation} appears to spoil an immediate interpretation as
a beta-function. To remedy this, we define dimensionless couplings
$\bar f$ by $f(\ep,k) := \bar f(\ep, \bar k)$ where $\bar k := \ep k$
is the dimensionless momentum. Making a formal Taylor expansion $\bar
f= f_0(\ep) + f_1(\ep) (\bar k)^2 + ...$ we get the following coupled
flow equations
\begin{align}
\dot f_0  =  -f_0^2 + d~f_0 + m^2 & = -(f_0 - \Delta_+)(f_0 - \Delta_-) \,, \quad
\dot f_1 = (d-2 - 2 f_0) f_1 + 1, \quad \ldots
\nn 
\Delta_\pm & \equiv \frac{d}2 \pm \nu, \quad \nu =  \sqrt{
\left(\frac d 2\right)^2 +  m^2} \,.
\label{fn-flow}
\end{align} 
There are clearly two fixed points given by $f_0 = \Delta_\pm$ (all
higher modes $f_n$ are uniquely determined in terms of $f_0$). Clearly
$f_0= \Delta_+$ is an attractive, hence IR, fixed point of the double
trace flow (consider $\delta f:= f- \Delta_+$), while $f_0= \Delta_-$
is a repulsive, hence UV, fixed point. These correspond, respectively,
to the standard and alternative quantizations; the field theory
interpretations of these statements are detailed in
Section~\ref{sec:FT}, and in Table \ref{table-anomalous}.

\begin{table}[h]
\begin{tabular}{l|l|l|l|l}
\hline
& dimension of $O(x)$ & dimension of $O(x)^2$ & nature of $O(x)^2$
& nature of fixed point
\\
\hline
`standard' CFT & $\Delta_+$ & $2\Delta_+ = d + 2\nu > d$ & irrelevant
&  IR (attractive)
\\
`alternative' CFT & $\Delta_-$ & $2\Delta_- = d - 2\nu <  d$ &
relevant & UV (repulsive)
\\
\hline
\end{tabular}
\caption{Dimensions of operators at the two fixed points. The
  dimensions can be read off either by a computation of two-point
  functions, or by using the fact that for an interaction $\int g
  \mathbb{O}$ with beta-function $\b := -\dot g$, the dimension of
  the operator $\mathbb{O}$ satisfies $\g = \del\b/\del g$ (see
Section \ref{sec:FT} for more details).}
\label{table-anomalous}
\end{table}

The RG flow in the $f_0$-$f_1$ plane (for $d=3, m^2=-
2$) \footnote{For $d=3$, unitarity requires $m^2 \ge -9/4$
(the Breitenlohner-Freedman (BF) bound)
whereas the Klebanov-Witten window is $-5/4 \ge m^2 \ge
  -9/4$.} is shown in the left panel of Figure \ref{fig-scalar-flow}.

\begin{figure}[t]
\begin{center}
\begin{picture}(500,135)
\put(10,0){\includegraphics[height=4.5cm]{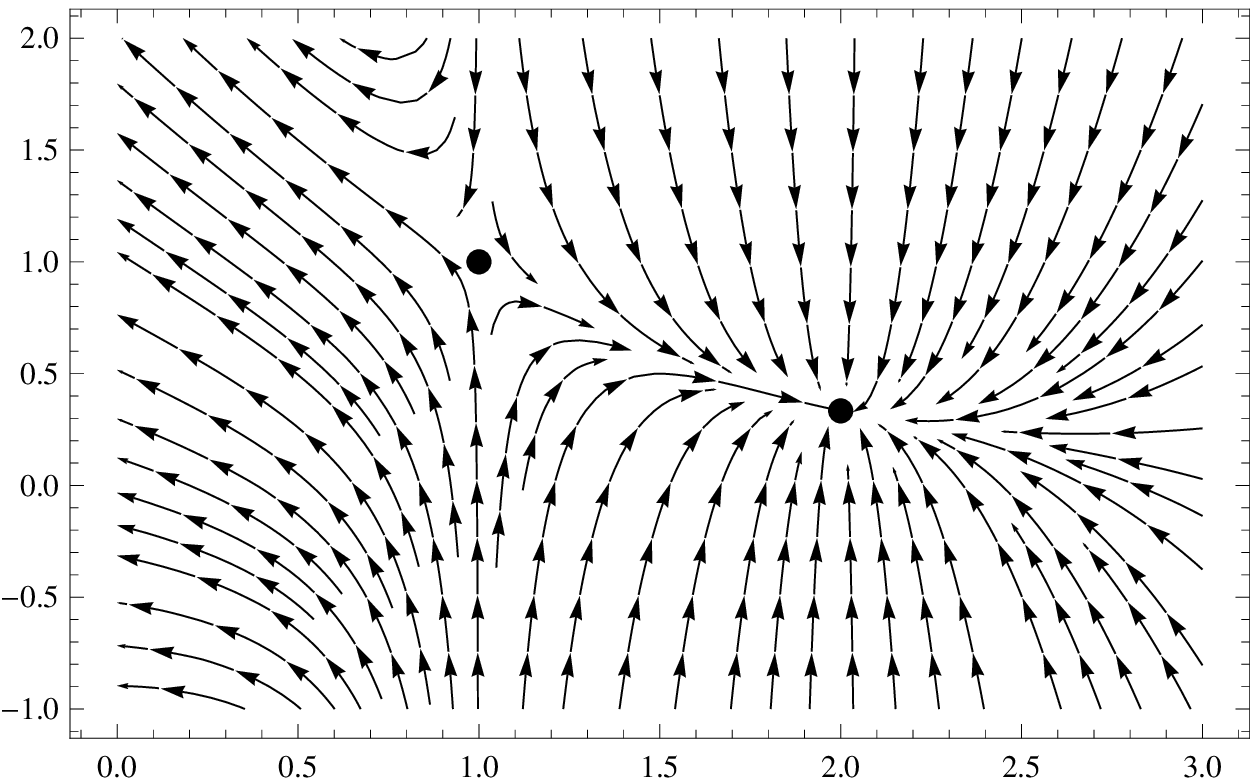}}
\put(245,0){\includegraphics[height=4.5cm]{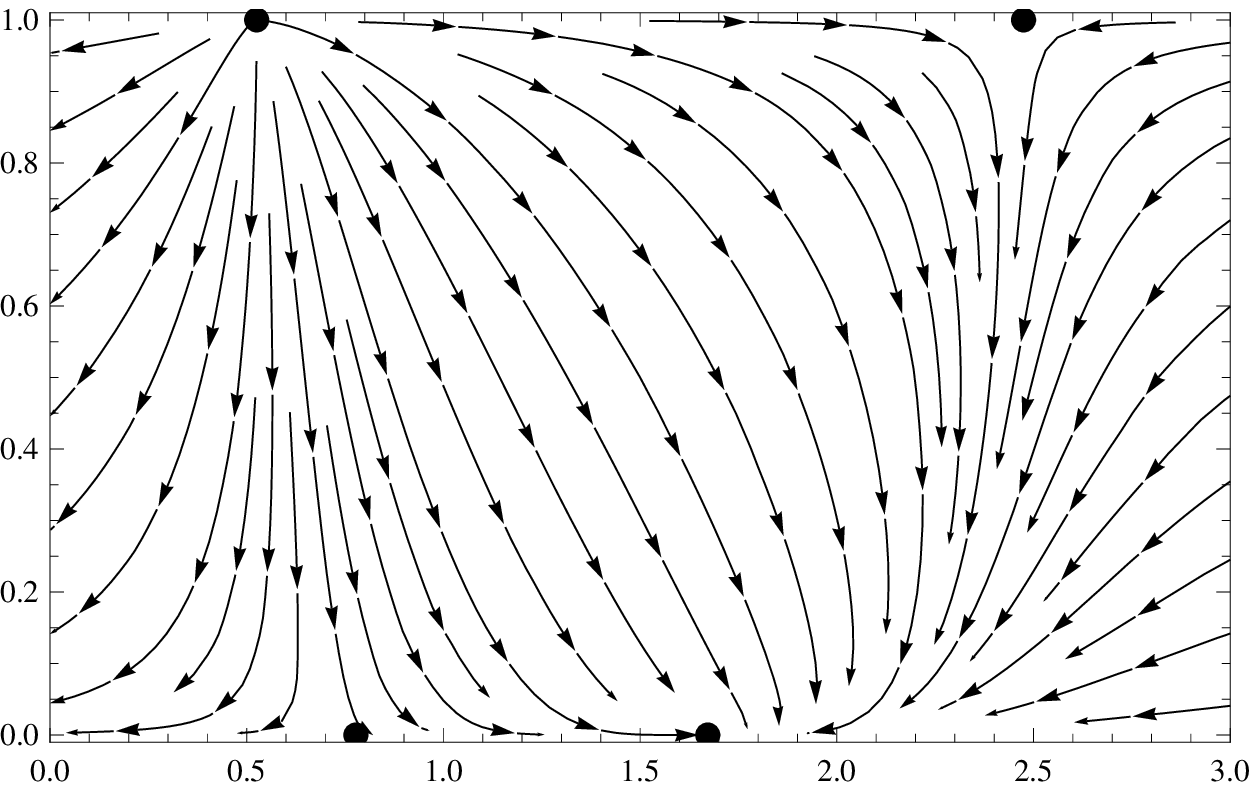}}
\put(220,6){$f_0$}
\put(449,6){$f_0$}
\put(18,135){$f_1$}
\put(250,135){$H$}
\end{picture}
\caption{RG flow for double trace couplings: (a) the left panel shows
  the case of a scalar field in pure AdS background, with double trace
  flows \eq{fn-flow} connecting two fixed points; (b) the right panel shows the
  case of the extremal BH background, where double trace flows at the
  top (AdS$_4$) boundary and at the bottom (AdS$_2$) boundary are
  connected by a geometric flow, that of the red-shift factor $H$.}
\label{fig-scalar-flow}
\end{center}
\end{figure}

\paragraph{Extremal Charged BH background}
In this case, the metric and the gauge potential are given by  (using units where the
AdS$_4$ radius is equal to 1)
\begin{align}
ds^2&= \frac{1}{z^2}(H(z) dt^2 + dx_idx_i)+\frac{dz^2}{z^2H(z)} \,, \quad
A = \mu \left( 1 - \frac{z}{z_{\ast}} \right)dt, \quad
\nn
H(z)&= 
1 + \frac{d}{d-2}\left(\frac{z}{z_{\ast}}\right)^4 
- \frac{2(d-1)}{d-2}\left(\frac{z}{z_{\ast}}\right)^3 \,,
\label{bh-metric}
\end{align}
where $z_*$ is the black hole horizon and $i=1,\cdots,d-1$.  The flow
equation \eq{f-flow} for $f$ becomes (writing $k_\mu=(\om, k_i)$)
\begin{align}
\ep\del_\ep (f \sqrt H) = d~f \sqrt H - f^2 + \ep^2 ( k_i k_i +
\om^2/H )+ m^2 \,.
\label{r-rgf-bh}
\end{align} 
There are now two explicit sources of cut-off dependence on the RHS.
We have learned above how to handle the $\ep^2$-term, by introducing a
dimensionless coupling $\bar f$. {\em How about the explicit
  dependence of $H$ on $\ep$?} The answer is that $H(\ep)$ simply
reflects the presence of a new Wilsonian-`type' \footnote{The `quotes'
  reflect the fact that $\delta H(z) \sim \delta g_{tt}$ is actually a
  normalizable deformation, so $\b_H$ is to be identified with a flow
  of the VEV of a stress tensor, rather than with the running of a
  coupling; see Section~\ref{sec:FT} for details.} coupling in the
black hole background, whose `{beta-function}' is independent of the
double trace flow.\footnote{$\b_H := \ep\del_\ep H$, which is a
  function of $\ep$, can be recast as a function of $H$ by inverting
  $H(\ep)$ as $\ep(H)$; $\b_H(H)$ satisfies the implicit equation
\begin{align}
(1- H + \beta_H/3)^3 = (1- H + \beta_H/4)^4 \,.
\label{implicit-betaH}
\end{align}}
Using these facts, and considering the translationally invariant mode
of $f(k,\ep)$, which is $f_0(\ep)$, we get an RG flow in the $f_{0}$-$H$ plane as
in the right panel of Figure \ref{fig-scalar-flow}.  The plot
(computed for $d=3, m^2=-1.3$) clearly shows four fixed points: two of them
at the AdS$_4$ boundary ($H=1$), with $f_0 = \Delta_\pm$; and
the remaining two at the AdS$_2$ boundary ($H=0$), with $f_0 = \delta_\pm =
( 1/2 \pm \nu_2)/R_2$, $\nu_2= \sqrt{1/4 + m^2 R_2^2}, \, R_2^2= 1/6$;
where $R_2$ is the value of the effective radius of curvature of the
emergent AdS$_2$ in the limit $z \to z_*$. See Section~\ref{sec:FT} for a field
theory interpretation of this flow diagram.

\section{\label{sec:beta}
Holographic Wilsonian flow equations for Dirac fermions}


\paragraph{Summary of this section}

In this section, we will consider a free charged Dirac fermion $\psi$
propagating in an asymptotically AdS$_{d+1}$ spacetime \eq{gen_met}.
The holographic dual of this system will be understood to be a large
$N$ field theory with a global $U(1)$ current, with a certain charged
fermionic single-trace operator $O$ which couples to the bulk fermion
$\psi$. The precise nature of the coupling is described in equations
\eq{gkp-dual} and \eq{standard-coupling},
\eq{alternative-coupling}. It will be assumed that the geometry is not
affected by the bulk fermion. We will describe how to integrate out
the bulk fermion up to an intermediate radial cut-off $z=\ep$. As
described in the Introduction and in Section~\ref{sec:scalar}, this
procedure implements a Wilsonian RG in the dual field theory, and
induces a flow of double trace operators (see
\eq{double-trace-fermion} and \eq{double-trace-fermion-alt}) whose
beta-function is computed in \eq{fermionic-flow} and
\eq{eq:floweqsnicegammas}.

\paragraph{Dirac action}

Consider a free Dirac fermion, of charge $q$, in an asymptotically AdS
background, with a metric and a $U(1)$ gauge field of the form
\begin{align}
&{}
ds^2
= g_{MN}dx^Mdx^N
= g_{zz}(z) dz^2+ g_{\mu\nu}(z) dx^\mu dx^\nu, \quad
A_z=0, \quad A_{\mu}= A_\mu(z). 
\label{gen_met}
\end{align}
The Dirac action is given by\footnote{
See Appendix \ref{notation} for definitions of the covariant derivatives and so on.}
\begin{align}
S = \int\!dzd^dx\sqrt{g}\,\left(\ol{\psi}\Ga^M\cD_M\psi-m\ol{\psi}\psi\right) \,.
\end{align}
For simplicity, we work with Euclidean signature.\footnote{ For a
  general form of the Dirac action which is applicable to both
  Lorentzian and Euclidean metrics, see Appendix \ref{gen_act}.}
The kinetic term in the
$z$-direction of the Dirac action can be made into a canonical one,
\begin{align}
S =
\int \! dzd^dx 
\left(
\ol{\Phi}\Ga^{\wh{z}}\pd_z\Phi
+ \sqrt{g_{zz}}
\left[ \ol{\Phi}(\Ga^{\mu}\pd_{\mu} - iq \Ga^\mu A_\mu)\Phi - m\ol{\Phi}\Phi \right]
\right) \,,
\label{action-phi}
\end{align}
after field rescalings 
$\Phi := (gg^{zz})^{1/4}\psi \,, ~ \ol{\Phi} := (gg^{zz})^{1/4}\ol{\psi}$.

In order to implement the strategy indicated in \eq{slicing}, we need
to compute the radial Hamiltonian and a complete set of states.  A
basic difference from the scalar case is that the components of
$\psi$ contain both `coordinates' and `momenta'. Thus, we cannot
have states analogous to field-eigenstates $| \phi \rangle$; rather,
it turns out that we need to introduce some kind of coherent states. 
We will first describe the radial Hamiltonian and the coherent states.

\paragraph{Radial Hamiltonian}

We adopt the following representations of $(d+1)$-dimensional Dirac matrices,
\begin{align}
\quad d=~\mbox{even} &{} \quad\quad
\Ga^{\wh{z}} := \ga^{d+1}  \,, \quad
\Ga^{\wh{\mu}} := \ga^{\wh{\mu}} \,, \quad
\chi_{\pm} := \frac{1 \pm \Ga^{\wh{z}}}{2}\Phi \,, \quad
\ol{\chi}_{\pm} := \ol{\Phi}\frac{1 \pm \Ga^{\wh{z}}}{2} \,, \label{deven}\\
&{} \quad\quad
\mbox{where } \ga^{d+1} \mbox{ is the chirality matrix in } d\mbox{-dimension.} \nonumber\\
\quad d=~\mbox{odd} &{} \quad\quad
\Ga^{\wh{z}} := \begin{pmatrix} 1 & 0 \\ 0 & -1 \end{pmatrix} \,, \quad
\Ga^{\wh{\mu}} := \begin{pmatrix} 0 & \ga^{\wh{\mu}} \\ \ga^{\wh{\mu}} & 0 \end{pmatrix} \,, \quad
\Phi := \begin{pmatrix} \chi_+ \\ \chi_- \end{pmatrix} \,, \quad
\ol{\Phi} := (\ol{\chi}_+ , \ol{\chi}_- ) \,, \label{dodd}
\end{align}
In these conventions, the anticommutation relations and the
Hamiltonian are given by 
\begin{align}
&{} 
\wh{H} = \int\!\frac{d^dk}{(2\pi)^d}\sqrt{g_{zz}}
\left[ 
i\wh{\ol{\chi}}_+\ga^{\mu}K_{\mu}\wh{\chi}_- + i\wh{\ol{\chi}}_-\ga^{\mu}K_{\mu}\wh{\chi}_+
+m\wh{\chi}_+\wh{\ol{\chi}}_+ - m\wh{\ol{\chi}}_-\wh{\chi}_-
\right] \,, \label{rad_H_ads} \\
&{} 
\{ \wh{\chi}_{\pm}(z,k_{\mu}), \, \pm\wh{\ol{\chi}}_{\pm}(z,k_{\mu}^{\pr}) \} 
= (2\pi)^d\ka^2\de^d(k-k^{\pr}) \,, \label{a-com_ads}
\end{align}
where we have moved into the momentum space and $K_{\mu}:=k_{\mu}-qA_{\mu}$.  
We can prove that the above radial Hamiltonian and anticommutation relations {\em indeed}
reproduce the classical equations of motion through $\ka^2\pd_z\wh{O}(z) =
[\wh{H}, \wh{O}(z)] \,$, as well as the path integral
\eq{path-integral}, thus constituting a proof of the equations
\eq{rad_H_ads} and \eq{a-com_ads} (see Appendix \ref{app_bdy_act}).

\paragraph{Boundary conditions and generalized coherent states}

Let us review boundary conditions imposed on Dirac fields at the
boundary $z\sim 0$. Since the Dirac action is of first order,
the wavefunction is a function of only half of the
components of $\psi$ and $\ol{\psi}$. Accordingly, boundary conditions
are imposed on only half of the components.  In the case $m>1/2$, the
components $\psi_- $ and $\ol{\psi}_+ $ become non-normalizable near
the boundary (they behave as $\sim z^{d/2-m}$, see \eq{dirac-sols} and
\eq{bessel-small-z}), and thus these components are set to be the
boundary sources. The states at the boundary,
analogous to $\ket{\UV}$ in \eq{slicing}, are chosen to be eigenstates
of these components, defined by\footnote{ A notation rule is that the
  first (second) column in the state vector corresponds to the $+(-)$
  chirality.}
\begin{align}
\wh{\ol{\chi}}_+|\ol{\chi}_+, \chi_-\ran = 
\ol{\chi}_+|\ol{\chi}_+, \chi_-\ran \,, \quad 
\wh{\chi}_-|\ol{\chi}_+, \chi_-\ran = \chi_-|\ol{\chi}_+, \chi_-\ran \,.
\label{standard}
\end{align}
We will call such states generalized coherent states (see Appendix
\ref{g_coh}). The above choice of boundary states turns out to
correspond to standard quantization, as shown shortly.

For $0< m< 1/2$,\footnote{This range of masses defines the
Klebanov-Witten window for fermions, where both standard
and alternative quantizations are allowed.} it is possible to choose the
other components (namely, $\chi_+, \ol{\chi}_-$) as the boundary
sources; the corresponding boundary states are the complementary
set
\begin{align}
\wh{\chi}_+|\chi_+, \ol{\chi}_-\ran 
= \chi_+|\chi_+, \ol{\chi}_-\ran \,, \quad 
\wh{\ol{\chi}}_-|\chi_+, \ol{\chi}_-\ran = 
\ol{\chi}_-|\chi_+, \ol{\chi}_-\ran \,.
\label{alternative}
\end{align}
This choice turns out to correspond to the alternative quantization.

\paragraph{Path integral}
Based on the generalized coherent states, we can show that 
\begin{align}
&{}
\lan \chi_+^1, \ol{\chi}_-^1 |\wh{U}(\ep_1,\ep_2)|\ol{\chi}_+^2, \chi_-^2 \ran 
=
\int [d^4\chi] e^{-\ka^{-2}S[\chi,\ol{\chi}] + S_{\bdy}} \,, \quad
S_{\bdy}
:=
-\int_k  \left( \ol{\chi}_-^1\chi_-^1 +  
\ol{\chi}_+^2\chi_+^2 \, \right) \,.
\label{path-integral}
\end{align}
where $\chi_-^1:=\chi_-(z=\ep_1)$ and $\chi_+^2:=\chi_+(z=\ep_2)$, and
$S[\chi,\ol{\chi}]$ is given by \eq{action-phi}.  The proof is given
in Appendix \ref{app_bdy_act}.  Since the derivation is independent of
details of the radial Hamiltonian, this result applies both to pure
AdS$_{d+1}$ case and to a general background case.  Thus, we can see
that the above choice of the boundary state \eqref{standard} correctly
gives the boundary terms for the standard quantization, which was
proposed and justified in a semi-classical manner in
\cite{Henningson:1998cd,Arutyunov:1998ve,Henneaux:1998ch}.

\paragraph{The AdS/CFT dictionary}

As mentioned in the beginning of this section, the bulk fermion $\psi$
is coupled to a certain single trace fermionic operator $O$ in the
field theory, which is charged under a global $U(1)$ current.  We can
now state the precise couplings of the various spinorial components
using the $\chi$-notation for the bulk fermions.  These are as
follows:
\begin{align}
\chi_- \longleftrightarrow \ol{O}_- \,, ~~ 
\ol{\chi}_+ \longleftrightarrow O_+ \,; ~~
\chi_+ \longleftrightarrow \ol{O}_+ \,, ~~
\ol{\chi}_- \longleftrightarrow O_- \,.
\label{gkp-dual}
\end{align}
In the standard quantization, AdS/CFT duality reads as \cite{Henningson:1998cd}
\begin{align}
\Bbra e^{\int_k \! (\ol{\chi}_+O_+ + \ol{O}_-\chi_-)} \Kket^{\std}_\ep
= \lan \IR | \wh{U}(\ep_{\IR}, \ep) | \ol{\chi}_+, \chi_- \ran \,,
\label{standard-coupling}
\end{align}
where, as in \eq{slicing-ft}, we have used $\bbra .. \kket$ to denote
an unnormalized partition function.  The subscript $_\ep$ denotes the
cut-off at which the field theory is defined (here we are identifying
the radial cut-off with the field theory cut-off, see comments below
\eq{slicing-ft}).  Note that the initial state is chosen as
eigenstates of $\chi_-, \ol{\chi}_+$, in keeping with the boundary
condition appropriate for the standard quantization, as indicated by
the superscript $^{\rm std}$. \\ In the case of the alternative
quantization we must choose an initial state which is the eigenstate
of the remaining half of the fermion components, giving rise to
\begin{align}
\Bbra e^{\int_k \! (\ol{O}_+\chi_+ + \ol{\chi}_-O_-)} \Kket^{\alt}_\ep
= \lan \IR | \wh{U}(\ep_{\IR}, \ep) | \chi_+, \ol{\chi}_- \ran \,.
\label{alternative-coupling}
\end{align}
This dictionary matches with the usual definition of the alternative
quantization as a Legendre transformation of the standard
quantization, which can be seen from \eqref{inpro} and \eqref{comp2}.
Both \eq{standard-coupling} and \eq{alternative-coupling} are `bare'
AdS/CFT relations; to define continuum field theories, we need to
consider counterterms, as discussed in Section~\ref{app:fermion-ct}.

\paragraph{Multi-trace deformations}
Following the scalar case (see footnote \ref{bare-multi-trace}), let
us consider a transition amplitude with a general initial state
$\lan \IR | \wh{U}(\ep_{\IR}, \ep)
| \UV \ran $ where $| \UV \ran$ is described by a wavefunction
\begin{align}
\lan \chi_+, \ol{\chi}_- | \UV \ran
=: \exp\left( W[\chi_+, \ol{\chi}_-]   \right) \,.
\label{general-w}
\end{align}
The AdS/CFT relation in this case, written in standard quantization,
can be shown to be\footnote{To derive this, we expand $|\UV \ran$ in
  the $|\ol{\chi}_+, \chi_- \ran$ basis, make a temporary replacement
  $W[\chi_+, \ol{\chi}_-]\!\!\longrightarrow$
  $W[\chi_+, \ol{\chi}_-] - \int
  \!(\ol{J}_+\chi_+ + \ol{\chi}_-J_-)$, and bring $W$ outside the
  $\int D^4\chi$ integral as $W(-\del/\del \ol{J}_+,\del/\del
  J_-)$. The $ \int D^4\chi$ integral now converts $|\ol{\chi}_+,
  \chi_- \ran \to |\ol{J}_+, J_- \ran $, so that using
  \eq{standard-coupling}, the $W$-term becomes $W[-O_+,
    -\ol{O}_-]$. At this stage, we can put $J=0$ and arrive at
  \eq{multr-f}.}
\begin{align}
\lan \IR | \wh{U}(\ep_{\IR}, \ep) | \UV \ran = \Bbra e^{W[-O_+,
    -\ol{O}_-] }\Kket_{\ep}^{\std} \,.
\label{multr-f}
\end{align}
To obtain a similar result in terms of alternative quantization, we
can expand the wavefunction in the complementary basis 
by inserting \eqref{comp1} or \eqref{comp2}:
\begin{align}
\lan \ol{\chi}_+, \chi_- | \UV \ran 
= \int D\eta_+D\ol{\eta}_-
\exp\left\{\int_k(\ol{\chi}_+ \eta_+ + \ol{\eta}_- \chi_-) 
+ W[\eta_+, \ol{\eta}_-] \right\}
=:  \exp\left( \widetilde{W}[\ol{\chi}_+,{\chi}_-]   \right),
\end{align}
and find the following representation for the transition amplitude
\begin{align}
\lan \IR | \wh{U}(\ep_{\IR}, \ep)
| \UV \ran 
= \Bbra e^{\widetilde{W}[\ol O_+, O_-]} \Kket_{\ep}^{\alt}
\end{align}
These results mean that the general initial state gives rise to
multi-trace deformations of the standard and the alternative
quantizations.\footnote{
For other formulations of multi-trace deformations in the fermion case,
see \cite{Allais:2010qq}.}  
Thus, for example, if we choose $W[\chi_+,
  \ol{\chi}_-]= \int_k  \ol{\chi}_-~a~ \chi_+$,
\begin{align}
\lan \IR | \wh{U}(\ep_{\IR}, \ep) | \UV \ran = \Bbra e^{\int_k
\ol{O}_- ~a~ O_+}\Kket_{\ep}^{\std} = \Bbra e^{-\int_k
\ol{O}_+ ~ a^{-1} ~ O_-}\Kket_{\ep}^{\rm alt} .
\label{multr-f-specific}
\end{align}

\subsection{Schr\"{o}dinger equation and RG flow equations}

Let us consider a field theory with a bare cut-off $\ep_0$, defined
holographically in the standard quantization, through equations such
as \eq{standard-coupling} or \eq{multr-f} with $\ep= \ep_0$.  As in
\eq{slicing} for the scalar case, we slice the bulk path
integral into two parts at $z=\ep$:
\begin{align}
\lan \IR | \wh{U}(\ep_{\IR}, \ep_0) | \UV \ran
&= 
\int \! D^4\chi \, \lan \IR | \wh{U}(\ep_{\IR}, \ep) | \ol{\chi}_+, \chi_- \ran
e^{\int_k (\ol{\chi}_+\chi_+ + \ol{\chi}_-\chi_-)}
\lan \chi_+, \ol{\chi}_- | \wh{U}(\ep, \ep_0) | \UV \ran \,,
\label{slicing-f}
\end{align}
where we used a completeness relation \eqref{comp1}. Using
\eq{standard-coupling}, the IR amplitudes
$\lan \IR | ... \ran$ on both sides of the above equation can be
replaced by field theory quantities at cut-offs $\ep_0$ and $\ep$
respectively. Making this replacement on the RHS explicit, we get
\begin{align}
\lan \IR | \wh{U}(\ep_{\IR}, \ep_0) | \UV \ran
= 
\int D^4\chi \, 
\Bbra e^{\int_k \! (\ol{\chi}_+O_+ + \ol{O}_-\chi_-)} \Kket_{\ep}^{\std}
e^{\int_k (\ol{\chi}_+\chi_+ + \ol{\chi}_-\chi_-)} 
\lan \chi_+, \ol{\chi}_- | \wh{U}(\ep, \ep_0) | \UV \ran \,.
\label{ep0-ep-f}
\end{align}
Thus the UV amplitude $\lan ... | \UV \ran$ on the RHS connects a
field theory at a lower cut-off $\ep$ with that at a higher cut-off
$\ep_0$.  Since we are considering here a quadratic Hamiltonian, the
UV amplitude can be written as an exponential of a quadratic
polynomial in $\chi_+, \ol{\chi}_-$,\footnote{In this and various
  other equations, the $k$-dependence of the $\chi$s and of the
  coefficients $F,J$ are suppressed.}
\begin{align}
\lan \chi_+, \ol{\chi}_- | \wh{U}(\ep, \ep_0) | \UV \ran
\equiv \Psi[\chi_+, \ol{\chi}_-] :=
\exp\left\{
-\int_k
\left[
\ol{\chi}_-F(\ep)\chi_+ + \ol{\chi}_-J_-(\ep) + \ol{J}_+(\ep)\chi_+ + C(\ep)
\right]
\right\} \,, 
\label{g_wf}
\end{align}
Then, applying the technique used in deriving \eqref{multr-f} to the
right hand side of \eqref{ep0-ep-f}, we get an important relation
\begin{align}
\lan \IR | \wh{U}(\ep_{\IR}, \ep_0) | \UV \ran
= 
\Bbra
\exp \int_k \left[
- \ol{O}_-F(\ep)O_+
+ \ol{O}_-J_-(\ep) + \ol{J}_+(\ep)O_+ - C(\ep)
\right]
\Kket_{\ep}^{\std} \,.
\label{double-trace-fermion}
\end{align}
Since the LHS is clearly independent of $\ep$, so must be the RHS.
This is the statement of RG invariance: the dependence on $\ep$ of the
coefficients $F(\ep), J(\ep)$ and $C(\ep)$ is such that the field
theory on the RHS of the above equation, for any choice of the cut-off
$\ep$, is independent of $\ep$, and is equivalent to the original
cut-off theory. In terms of alternative quantization, \eqref{double-trace-fermion} becomes
\begin{align}
\lan \IR | \wh{U}(\ep_{\IR}, \ep_0) | \UV \ran
= 
\Bbra
\exp \int_k \left[ \ol{O}_+ F(\ep)^{-1} O_-
- \ol{O}_+ F(\ep)^{-1} J_-(\ep) - \ol{J}_+ (\ep) F(\ep)^{-1} O_- - C(\ep)
\right]
\Kket_{\ep}^{\alt} \,.
\label{double-trace-fermion-alt}
\end{align}

To determine the $\ep$-dependence of $F,J, C$, note that the
wavefunction $\Psi$ in \eq{g_wf} satisfies the Schr\"{o}dinger
equation in the radial time $\ep$,
\begin{align}
-\ka^2\pd_{\ep}\lan \eta_+, \ol{\eta}_- | \wh{U}(\ep, \ep_0) | \UV \ran
= H\left[\eta_+,\ol{\eta}_-,\frac{\de}{\de\eta_+},\frac{\de}{\de\ol{\eta}_-}
\right]
\lan \eta_+, \ol{\eta}_- | \wh{U}(\ep, \ep_0) | \UV \ran \,, 
\label{schr-f}
\end{align}
where $ H\left[\eta_+,\ol{\eta}_-,\frac{\de}{\de\eta_+},
\frac{\de}{\de\ol{\eta}_-}\right]$ is defined by applying the following
replacements to the Hamiltonian $\wh{H}$ in
\eq{rad_H_ads},\footnote{ Note that  (neglecting momentum and spin indices) the
  replacements give a two-dimensional representation of the operator
  algebra \eqref{a-com_ads} while the Fock space basis gives a
  four-dimensional representation.  The former representation is
  related to a representation defined in \cite{Mansfield:1999cc},
  whereas the latter is similar to that defined in
  \cite{Floreanini:1987gr}.}
\begin{align}
\wh{\chi}_+ \longrightarrow \eta_+ \,, \quad 
\wh{\ol{\chi}}_+ \longrightarrow \ka_d^2\frac{\de}{\de\eta_+} \,, \quad 
\wh{\chi}_- \longrightarrow -\ka_d^2\frac{\de}{\de\ol{\eta}_-} \,, \quad 
\wh{\ol{\chi}}_- \longrightarrow \ol{\eta}_- \,, \quad
\ka_d^2:=\ka^2(2\pi)^d \,.
\label{replace}
\end{align}
For the derivation of the replacements, see the end of Appendix
\ref{g_coh}.  Furthermore, the same replacement laws can be
applied to a general Hamiltonian including interactions, as one can
see from the derivation.

It is easy to see that the Schr\"odinger equation \eq{schr-f} with the
Hamiltonian \eqref{rad_H_ads} is satisfied if and only if the
coefficients $(F,\ol J_+, J_-, C)$ in the UV amplitude \eqref{g_wf} satisfy the
following flow equations,
\SP{\label{fermionic-flow}
\sqrt{g^{zz}}\pd_\epsilon F &= F(i\ga^{\mu}K_{\mu})F+i\ga^{\mu}K_{\mu} - 2mF
\,, \\ 
\sqrt{g^{zz}}\pd_\epsilon J_- &= F(i\ga^{\mu}K_{\mu})J_- - mJ_- \,, \\ 
\sqrt{g^{zz}}\pd_\epsilon \ol{J}_+ &=
\ol{J}_+(i\ga^{\mu}K_{\mu})F-m\ol{J}_+ \,, \\ 
\sqrt{g^{zz}}\pd_\epsilon C &=
\ol{J}_+(i\ga^{\mu}K_{\mu})J_- +
\ka_d^2\tr(i\ga^{\mu}K_{\mu}F)\de^d(k=0) \,,
}
where the trace is taken over spinor indices.
Note that the first three equations are classical.
One can construct the solutions of the flow equations 
from classical solutions $\Phi_{\pm}, \ol{\Phi}_{\pm}$
of the equations of motion\footnote{Namely, $\Ga^{\wh{z}}\pd_z\Phi
+ \sqrt{g_{zz}}\left(
(\Ga^{\mu}\pd_{\mu} - iq \Ga^\mu A_\mu)\Phi - m\ol{\Phi}\Phi \right)
=0$.} as follows:
\begin{align}
F &= \Phi_-(\Phi_+)^{-1} = (\ol{\Phi}_-)^{-1}\ol{\Phi}_+ \,, \quad
J_- = (\ol{\Phi}_-)^{-1}j_- \,, \quad
\ol{J}_+ = \ol{j}_+(\Phi_+)^{-1} \,,
\label{flow-sol}
\end{align}
where $j_-, \ol{j}_+$ are spinors independent of $\ep$. 
\subsubsection{\label{sec:rot-sym}Rotationally symmetric case}
In the case where the background is rotationally symmetric in the
$d-1$ spatial boundary directions, the flow equations simplify
considerably after using a special choice of the Dirac matrices
\footnote{In this case we use Lorentzian signature.}. Details
are given in Appendix \ref{app:rot-sym}.\footnote{
Note that the special choice of the Dirac matrices in Appendix \ref{app:rot-sym}
has different structure from the representation \eqref{deven}, \eqref{dodd}.
The difference arises from the projection operators $\Pi_{\al}$ \eqref{proj}.} 
If the wave-functional $\Psi$
in \eq{g_wf} is parametrized by \eq{psi-ansatz}, then the
Schr\"{o}dinger equation \eq{app:schro} yields the following flow
equations 
\SP{\label{eq:floweqsnicegammas} \sqrt{g^{zz}} \pd_\epsilon
  F_{\al} &= -F_{\al}T_{\al}^-F_{\al} - 2m F_{\al} + T_{\al}^+ \,,
  \\ \sqrt{g^{zz}} \pd_\epsilon \ol{J}_{\al}^+ &=
  -\ol{J}_{\al}^+T_{\al}^-F_{\al} - m \ol{J}_{\al}^+ \,,
  \\ \sqrt{g^{zz}} \pd_\epsilon J_{\al}^- &=
  -F_{\al}T_{\al}^-J_{\al}^- - m J_{\al}^- \,, \\ \sqrt{g^{zz}}
  \pd_\epsilon C_{\al} &= -\ol{J}_{\al}^+T_{\al}^-J_{\al}^- -
  \ka_d^2T_{\al}^-F_{\al}\de^d(0)\, ,
}
where
\SP{
	T_{\al}^{\pm} := \pm\sqrt{-g^{tt}} \left(\omega + q A_t\right) - (-)^{\al}\sqrt{g^{11}}k.
}
Just like \eqref{flow-sol}, one can construct the
solutions to the flow equations from solutions of the
classical equations of motion \eqref{eq:Dirac}.  Let $\Phi_{\al}^{\pm},
\ol{\Phi}_{\al}^{\pm}$ be the classical solutions of \eqref{eq2ord}.
Then, the solutions of the flow equations are given by
\begin{align}
F_{\al} = \Phi_{\al}^-(\Phi_{\al}^+)^{-1} = (\ol{\Phi}_{\al}^-)^{-1}\ol{\Phi}_{\al}^+ \,, \quad
J_{\al}^- = (\ol{\Phi}_{\al}^-)^{-1}j_{\al}^- \,, \quad
\ol{J}_{\al}^+ = \ol{j}_{\al}^+(\Phi_{\al}^+)^{-1} \,. \label{flowsol}
\end{align}

\subsection{\label{sec:twisted}Flow equations in a twisted basis}
For later use, 
let us define the ``twisted'' operators as follows
\begin{align}
\shchi_+ &:= a\hchi_+ + b\hchi_- \,, \quad
\shchi_- := c\hchi_+ + d\hchi_- \,, \quad
\shochi_+ := d\hochi_+ + c\hochi_- \,, \quad
\shochi_- := b\hochi_+ + a\hochi_- \,, 
\label{twist}
\end{align}
with $ad-bc=1$.  One can easily show that these twisted operators also
satisfy the same anti-commutation relations $\{\shchi_{\pm},
\shochi_{\pm}\}=\pm 1$.  Thus, one can construct generalized coherent
states in terms of the {\it twisted} operators, which are related to 
an untwisted state \eq{standard} as follows:
\begin{align}
\lan \eta^{\sh}_+, \ol{\eta}^{\sh}_- | \ol \chi_+, \chi_- \ran 
= 
a\exp\left\{
\int_k\left(- \frac{1}{a}\ol{\chi}_+\eta^{\sh}_+ - \frac{1}{a}\ol{\eta}^{\sh}_-\chi_- 
+ \frac{b}{a}\ol{\chi}_+\chi_- - \frac{c}{a}\ol{\eta}^{\sh}_-\eta^{\sh}_+ \right)
\right\} \,.
\label{twist-standard}
\end{align} 
Note that since the twisting does not involve the annihilation and
the creation operators, the Fock vacuum $|0,0\kket$ (see Appendix
\ref{g_coh}) is invariant under the twist.  Let us parametrize the
wave-functional \eq{g_wf} in terms of the twisted basis, as follows
\begin{align}
\label{eq:wavefunctionaltheta}
\Psi[\chi^{\sh}_+, \ol{\chi}^{\sh}_-] 
:= \lan \chi^{\sh}_+, \ol{\chi}^{\sh}_- | \wh{U}(\ep, \ep_0) | \UV \ran
:=
\exp\left\{
-\int_k
\left[
\ol{\chi}^{\sh}_- F^{\sh} \chi^{\sh}_+ 
 + \ol{\chi}^{\sh}_- J^{\sh}_- + \ol{J}^{\sh}_+ \chi^{\sh}_+ + C^{\sh}
\right]
\right\} \,.
\end{align}
Using the inner product \eq{twist-standard}, the relation
\SP{
	\Psi[\chi^{\sh}_+, \ol{\chi}^{\sh}_-] 
= \int \mathcal D^4 \chi \lan \chi^{\sh}_+, \ol{\chi}^{\sh}_- | 
\ol \eta_+, \eta_- \ran e^{\int_k(\ol \eta_+ \eta_+ + \ol \eta_- \eta_-)} \Psi[\eta_+, \ol{\eta}_-],
}
and performing the integration over $\chi$, we obtain the following relations
\SP{
	F^{\sh} &= (c + d F)(a + bF)^{-1} \,, \quad
	\ol J^{\sh}_+ = \ol J_+ (a+bF)^{-1} \,, \quad
	J^{\sh}_- = (a+bF)^{-1}J_- \,.
	\label{twistfj}
}
The flow equation for $F^{\sh}$, by using \eq{schr-f}, takes the form
\begin{align}
\sqrt{g^{zz}} \partial_\epsilon F^{\sh} 
&= 
F^{\sh} \Big( i \gamma^\mu K_\mu (a^2+b^2) + 2mab \Big) F^{\sh}
-(ac+bd)(F^{\sh}i\ga^{\mu}K_{\mu} + i\ga^{\mu}K_{\mu}F^{\sh}) 
\nonumber\\
&{} \quad
- 2m (ad+bc)F^{\sh} + i \gamma^\mu K_\mu(c^2+d^2) + 2m cd \,.
\end{align}
A particularly convenient choice is given by
$a=b=-c=d=1/{\sqrt{2}}$\,, for which we obtain
\SP{\label{tong-flow} 
\sqrt{g^{zz}} \partial_\epsilon F^{\sh} =
F^{\sh} \Big( i \gamma^\mu K_\mu + m \Big) F^{\sh} + i \gamma^\mu
K_\mu - m \,.
}
In this basis, the standard and alternative fixed points of AdS$_{d+1}$, 
which correspond to $F=0$ and $F=\infty$, respectively
(see Section~\ref{sec:AdS}), are at $F^{\sh} = \pm 1$.
A special case of Equation \eq{tong-flow}, for pure AdS geometries and for the
homogeneous mode ($k_\mu=0$), is in agreement with the corresponding flow
equation  in \cite{Laia:2011wf}.

Application of the twisting to the rotationally symmetric case is 
straightforward.

\section{\label{sec:AdS}Fermionic flows in AdS$_{d+1}$}

\paragraph{Exact solution} In AdS$_{d+1}$ with a constant electric field $A_t = \mu$, the flow equation \eqref{fermionic-flow} for $F$ becomes
\SP{\label{eq:flowFAdS4}
	\epsilon \partial_\epsilon F = F \epsilon \left( i \gamma^{\wh \mu} K_\mu \right) F - 2mF + \epsilon \left( i \gamma^{\wh \mu} K_\mu \right),
}
an exact solution of which is given by
\SP{
\label{eq:FsolAdS4}
	F &= \frac{i \gamma^{\wh \mu} K_\mu }{K} \frac{I_{-\nu_+}(\epsilon K) + \chi(k_\mu) I_{\nu_+}(\epsilon K)}{I_{-\nu_-}(\epsilon K) + \chi(k_\mu) I_{\nu_-}(\epsilon K)}, \\
	K &:= \sqrt{k^2 - (\omega + q \mu)^2},
}
where $\chi(k_\mu)$ is an integration constant, and $\nu_\pm := m \pm \frac{1}{2}$.

\paragraph{Flows from the alternative to standard fixed point} In order to visualize the flow, let us now mimic the discussion in the bosonic case and make a derivative expansion of the double-trace coupling.\footnote{In essence, this focuses the attention to a particular class of flows which can be obtained from \eqref{eq:FsolAdS4} by choosing the integration constant $\chi(k_\mu)$ such that only integral powers of $\gamma^{\wh \mu} K_\mu$ are present in the expansion.} From \eqref{double-trace-fermion}, we see that in standard quantization $F$ appears in the effective action as $\ol O_- F O_+$. Viewing $F$ as a function of $\epsilon$ and $\bar K_\mu = \epsilon K_\mu$, the flow equation \eqref{eq:flowFAdS4} becomes
\SP{\label{eq:flowFAdS42}
	\epsilon \partial_\epsilon F = F \left( i \gamma^{\wh \mu} \bar K_\mu \right) F - \left( 2m + \bar K_\mu \frac{\partial}{\partial{\bar K_\mu}} \right) F + i \gamma^{\wh \mu} \bar K_\mu .
}
Expanding $F$ as \footnote{In \eq{rel} we consider Lorentz invariant
double trace deformations; deformations and
fixed points which are  Lorentz non-invariant are
considered at the end of this section (see \eq{non-rel}).}
\SP{\label{rel}
	F(\epsilon,\bar K_\mu) = f_0(\epsilon) + i \gamma^{\wh \mu} \bar K_\mu f_1(\epsilon) + \bar K^2 f_2(\epsilon) + \ldots,
}
leads to the flow equations
\SP{\label{eq:betaf}
	\epsilon \partial_\epsilon f_0 &= -2m f_0, \\
	\epsilon \partial_\epsilon f_1 &= f_0^2 - (1 + 2m) f_1 + 1, \\
	&\ldots
}
from which it is clear that there is a fixed point at $(f_0,f_1) = (0,1/(2m+1))$. From the eigenvalues of the flow, the fixed point is attractive in the IR; in Section~\ref{sec:FFT} we
show that the CFT defined by it indeed corresponds to standard quantization.

In alternative quantization, the double trace term
in the effective action is ({\em cf.} \eq{double-trace-fermion-alt}) is 
$\ol O_+ F^{-1} O_-$. Expanding the coupling $F^{-1}$ as 
\SP{
	F(\epsilon,\bar K_\mu)^{-1} = g_0(\epsilon) + i \gamma^{\wh \mu} \bar K_\mu g_1(\epsilon) + \bar K^2 g_2(\epsilon) + \ldots,
}
and following the above discussion, we then obtain the flow equations\footnote{Generically, $g_0 = f_0^{-1}$, $g_1 = - f_1 f_0^{-2}$, and so on.}
\SP{\label{eq:betag}
	\epsilon \partial_\epsilon g_0 &= 2m g_0, \\
	\epsilon \partial_\epsilon g_1 &= - g_0^2 - (1 - 2m) g_1 - 1, \\
	&\ldots
}
with a fixed point at $(g_0,g_1)=(0,1/(2m-1))$. From the eigenvalues
of the flow, the fixed point is clearly repulsive; in Section~\ref{sec:FFT}
we show that this UV fixed point indeed corresponds to alternative
quantization.

In order to be able to see flows from the alternative to the standard fixed point, it is practical to make the change of variables (corresponding to a specific choice \eq{tong-flow} of the twisted basis \eq{twist}) 
\SP{\label{special-basis}
	\tilde F = \frac{F-1}{F+1},
}
after which the flow equation becomes
\SP{
	\epsilon \partial_\epsilon \tilde F = \tilde F \left( i \gamma^{\wh \mu} \bar K_\mu + m \right) \tilde F - \bar K_\mu \frac{\partial \tilde F}{\partial {\bar K_\mu}} + i \gamma^{\wh \mu} \bar K_\mu - m.
}
Expanding $\tilde F$ as
\SP{
	\tilde F(\epsilon,\bar K_\mu) = \tilde f_0(\epsilon) + i \gamma^{\wh \mu} \bar K_\mu \tilde f_1(\epsilon) + \bar K^2 \tilde f_2(\epsilon) + \ldots,
}
we obtain the flow equations
\SP{
	\epsilon \partial_\epsilon \tilde f_0 &= m \left( \tilde f_0^2 - 1 \right), \\
	\epsilon \partial_\epsilon \tilde f_1 &= \tilde f_0^2 + \left( 2m \tilde f_0 - 1 \right) \tilde f_1 + 1, \\
	&\ldots,
}
whose fixed points are at $(\tilde f_0, \tilde f_1)=(\pm 1,2 / (1 \mp 2m))$ with minus and plus signs corresponding to standard and alternative quantizations, respectively. These equations can be solved exactly ($r := -\log z$):
\SP{\label{tilde-fn-exact}
	\tilde f_0 &= \tanh (m(r-r_0)), \\
	\tilde f_1 &= \frac{C e^{r-r_0}}{4\cosh^2\left(m \left(r-r_0\right)\right)} +\frac{\tanh ^2\left(m
   \left(r-r_0\right)\right)+4 m \tanh \left(m \left(r-r_0\right)\right)+1}{1-4 m^2},
}
where $r_0$ and $C$ are integration constants. In Figure~\ref{fig:fermionicflowf0f1r}, we show typical flows of $\tilde f_0$ and $\tilde f_1$, while Figure~\ref{fig:fermionicflowf0f1} shows the flow in the $\tilde f_0$-$\tilde f_1$ plane. As can be seen, the flows are from the alternative fixed point in the UV to the standard fixed point in the IR.

\begin{figure}[t]
\begin{center}
\begin{picture}(500,135)
\put(10,0){\includegraphics[height=4.5cm]{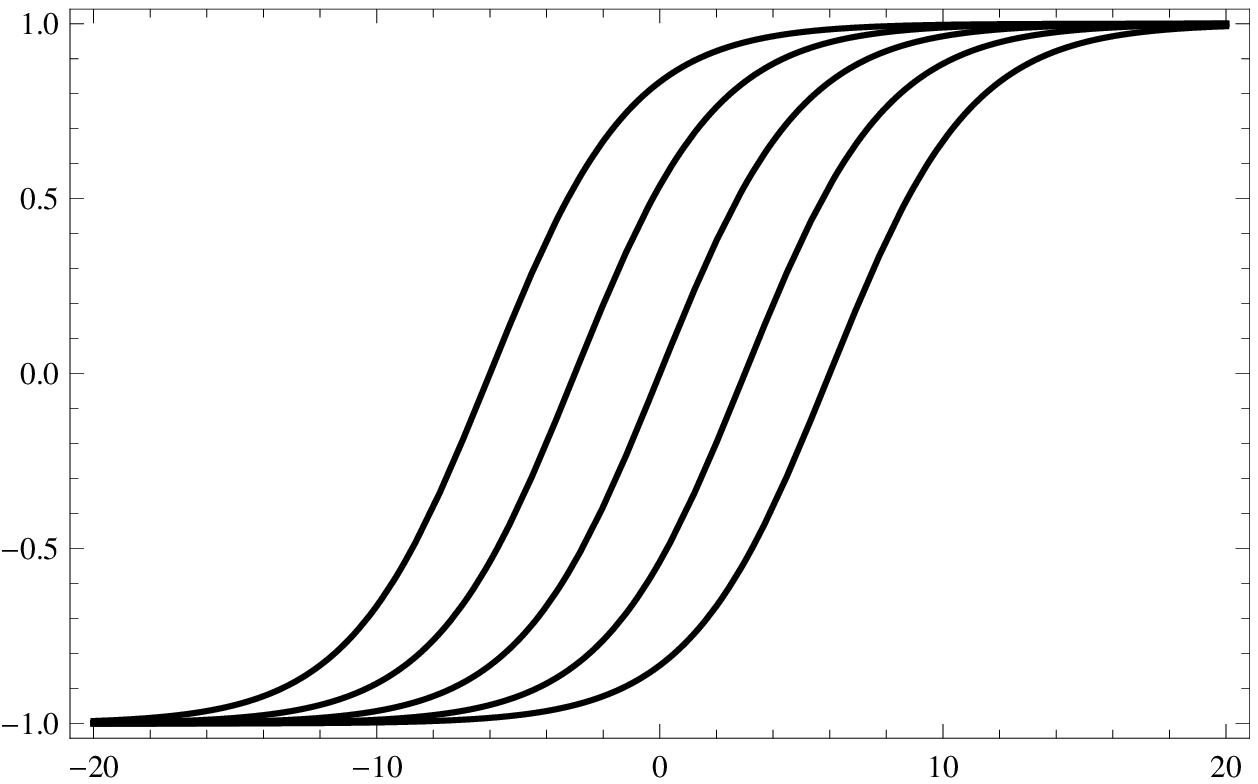}}
\put(245,0){\includegraphics[height=4.5cm]{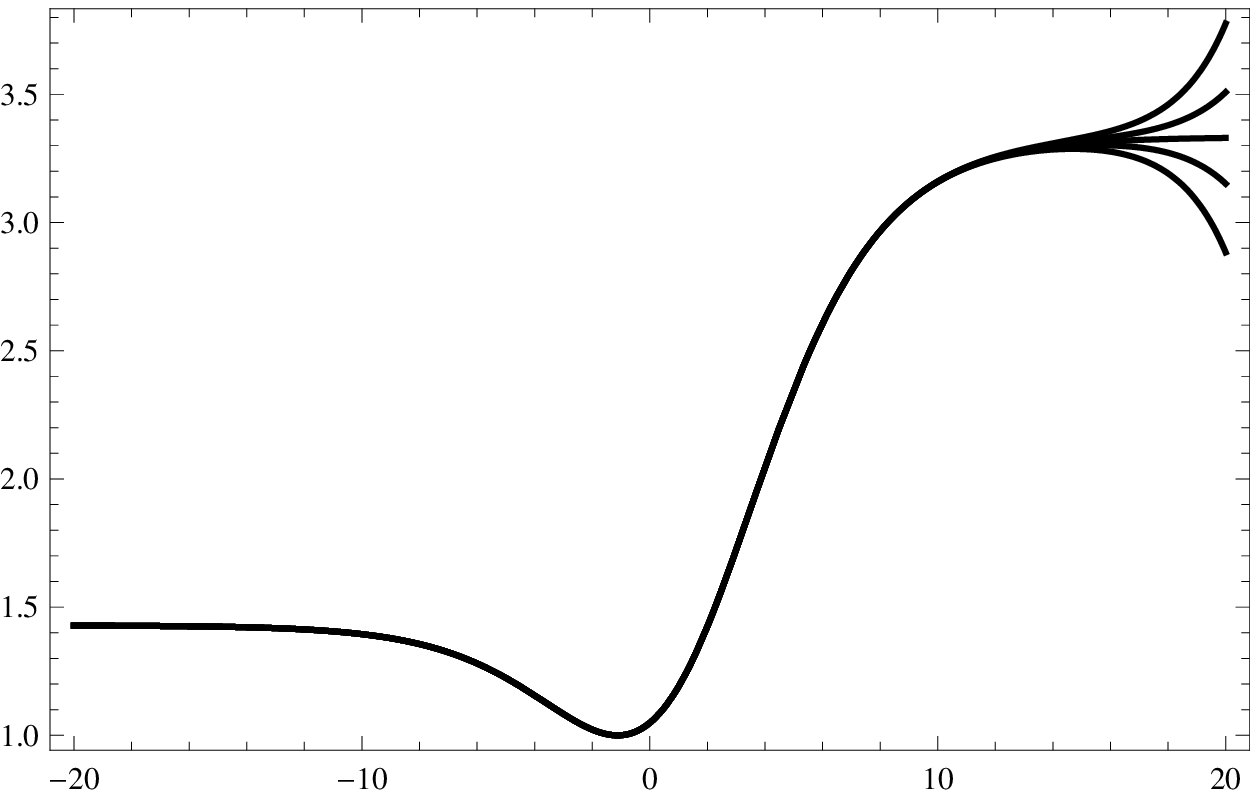}}
\put(218,6){$r$}
\put(453,6){$r$}
\put(18,135){$\tilde f_0$}
\put(250,135){$\tilde f_1$}
\end{picture}
\caption{Flows of $\tilde f_0$ and $\tilde f_1$ in AdS$_{d+1}$ as a
  function of $r = -\log z$ for $m = 0.2$. The UV is located at
  $r=\infty$ and the IR at $r=-\infty$. In the left panel, different
  curves correspond to different values of $r_0$ in
  \eq{tilde-fn-exact}. In the right panel, $r_0$ has been chosen to be
  1, and different curves correspond to different values of $C$.}
\label{fig:fermionicflowf0f1r}
\end{center}
\end{figure}

\begin{figure}[t]
\begin{center}
\begin{picture}(500,135)
\put(10,0){\includegraphics[height=4.5cm]{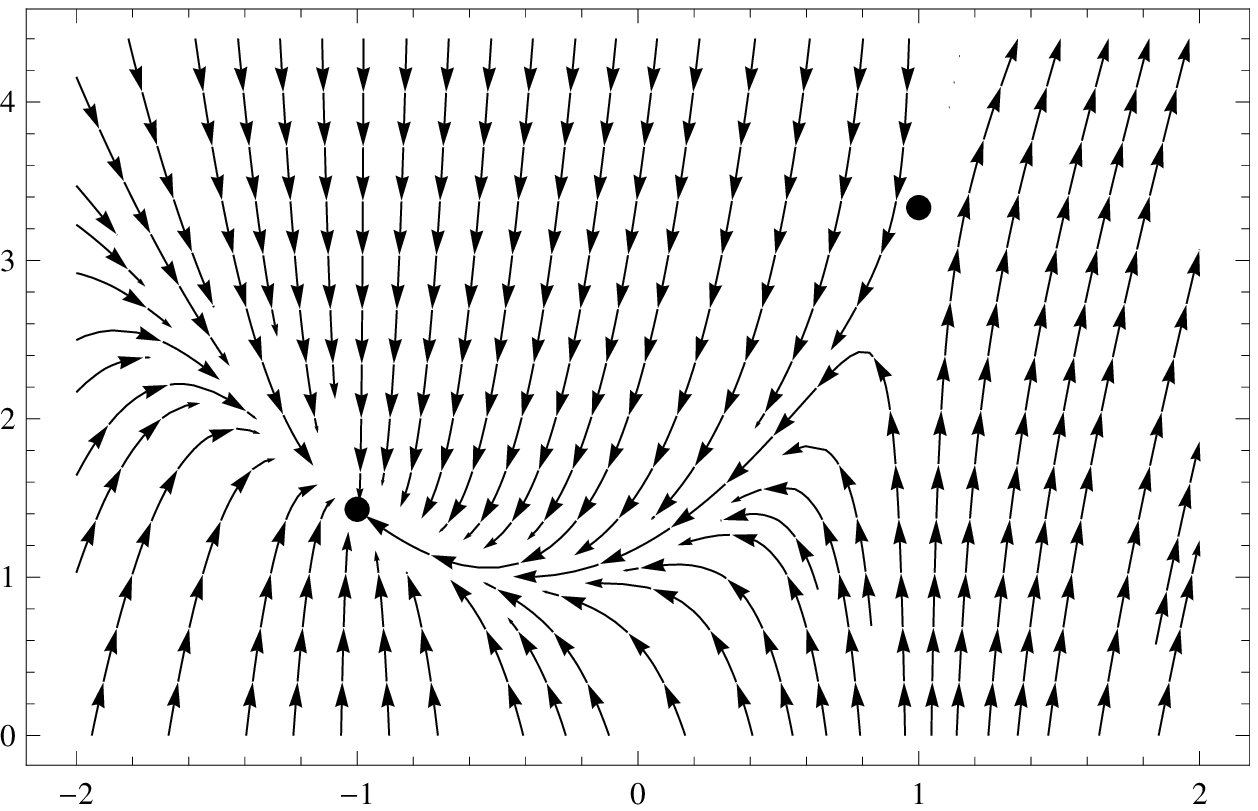}}
\put(245,0){\includegraphics[height=4.5cm]{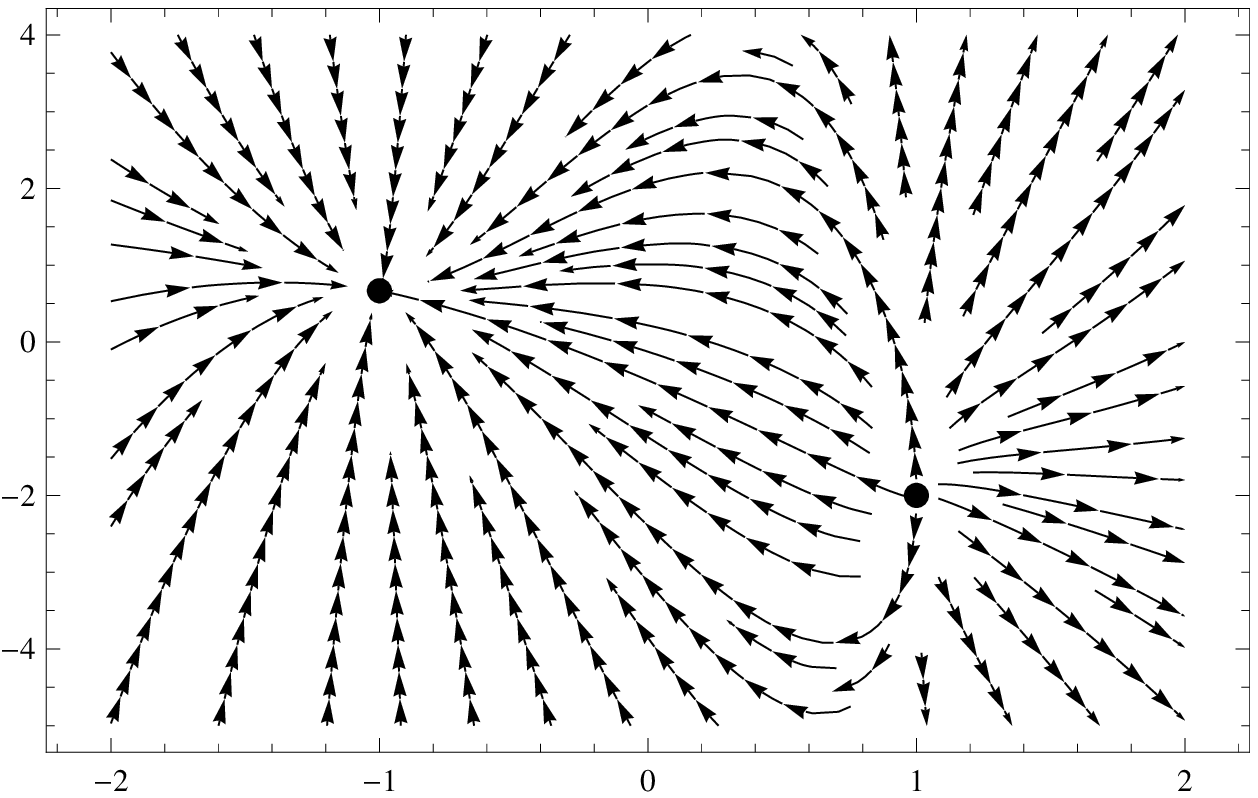}}
\put(213,6){$\tilde f_0$}
\put(451,6){$\tilde f_0$}
\put(10,135){$\tilde f_1$}
\put(250,135){$\tilde f_1$}
\end{picture}
\caption{RG flows in the $\tilde f_0$-$\tilde f_1$ plane for AdS$_{d+1}$ and $m=0.2$ (left panel) and $m=1$ (right panel). In the left panel, the
two fixed points, at $\tilde f_0=\mp 1$, correspond to standard
and alternative quantizations, respectively. In the right panel,
the value of $m$ is outside the Klebanov-Witten window and only
the standard fixed point defines a quantum theory.}
\label{fig:fermionicflowf0f1}
\end{center}
\end{figure}

\paragraph{$\bsk^{-1}$ mode} If one expands the exact solution \eqref{eq:FsolAdS4} for $F$ directly, the expansion generically starts at the order $\bsk^{-1}$, in the sense that it has the form
\SP{
	F(\epsilon,\bar K_\mu) = \left( i \gamma^{\wh \mu} \bar K_\mu \right)^{-1} f_{-1}(\epsilon) + f_0(\epsilon) + i \gamma^{\wh \mu} \bar K_\mu f_1(\epsilon) + \bar K^2 f_2(\epsilon) + \ldots
}
The introduction of a non-local operator like $\bsk^{-1} \ol O_- O_+$ near a fixed point is problematic. Fortunately, it is consistent to turn this operator off --- it would only start flowing if the theory was perturbed by it initially. This can be seen directly from the resulting flow equations
\SP{\label{eq:kminusonemodeflow}
	\epsilon \partial_\epsilon f_{-1} &= f_{-1}^2 + (1-2m) f_{-1}, \\
	\epsilon \partial_\epsilon f_0 &= 2 \left( f_{-1}  - m \right) f_0, \\
	&\ldots
}
Nevertheless, one might wonder whether these flows, which arise naturally from considering the evolution of the wavefunctional $\Psi$ in the bulk, have an interpretation in (a local) field theory. The double-trace deformation $\bsk^{-1} \ol O_- O_+$ can be viewed as arising from integrating out an emergent fermion (see Section~\ref{sec:FP}). The possibility remains to refrain from doing such an integration and study the `bigger' field theory, obtained by coupling the original one (corresponding to standard quantization) to the emergent fermion. This approach will be described in more detail in Section~\ref{sec:FP} where we will encounter the appearance of emergent fermions at the Fermi surface. After canonically normalizing the kinetic term of the emergent fermion $(\ol{\tilde \chi}_+, \tilde \chi_-)$ in \eqref{interim}, we obtain that the coupling of the emergent fermion to the sector corresponding to standard quantization is proportional to $\sqrt{f_{-1}}$. At the standard and alternative fixed points, $f_{-1}$ is equal to zero, so that the emergent fermion decouples from the system.

In addition to the standard and alternative fixed points, at $(f_{-1},f_0)=(0,0)$ and $(f_{-1},f_0)=(0,\infty)$, the flow equations \eqref{eq:kminusonemodeflow} have two other fixed points at $(f_{-1},f_0)=(2m-1,0)$ and $(f_{-1},f_0)=(2m-1,\infty)$. These two new fixed points are only there because we have coupled the original theory to an emergent fermion. We can visualize flows between these fixed points by making the change of variable $a_0 = (f_0 - 1)/(f_0 + 1)$. As can be seen from Figure~\ref{fig:fermionicflowkminus1}, outside the Klebanov-Witten window, {\it i.e.} for $m>1/2$, the two new fixed points are unstable, so that all flows end up at the standard fixed point. However, for $0 \leq m \leq 1/2$, one of the new fixed points becomes IR stable. At this IR fixed point, the coupling between the emergent fermion and the standard quantization sector is non-zero.

\begin{figure}[t]
\begin{center}
\begin{picture}(500,135)
\put(10,0){\includegraphics[height=4.5cm]{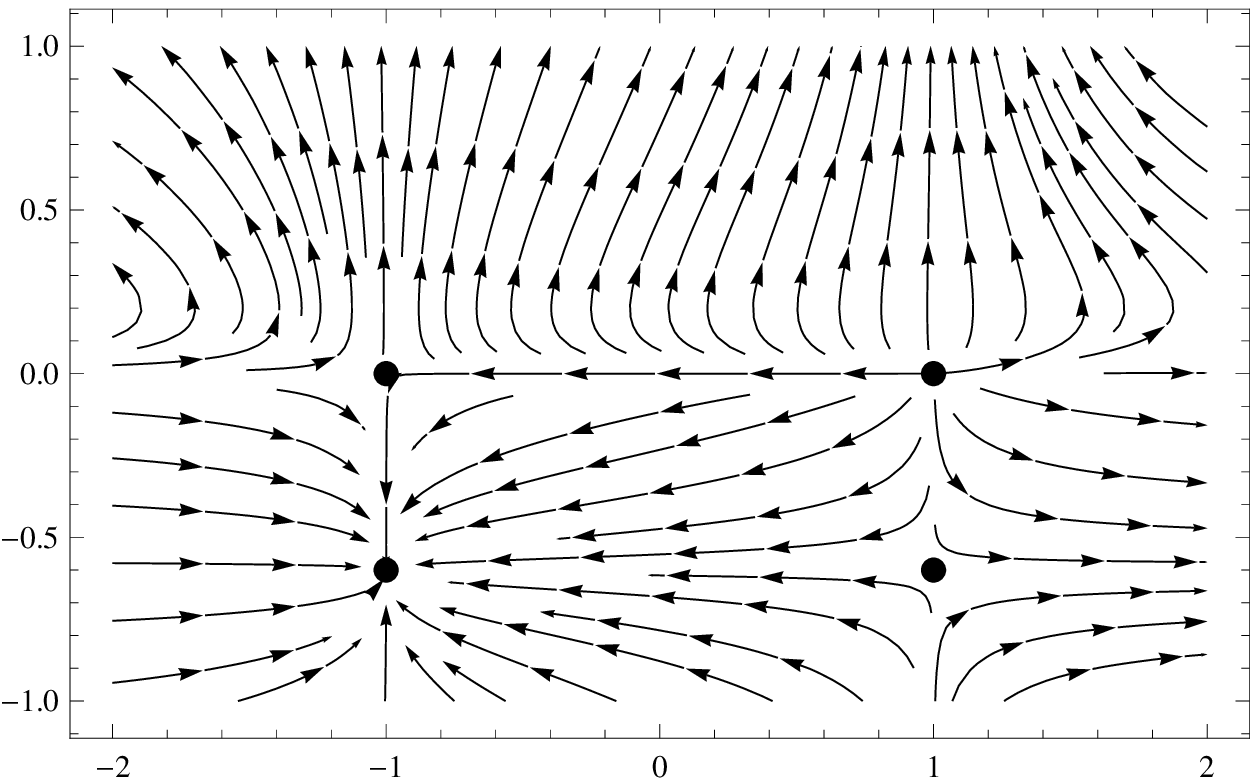}}
\put(245,0){\includegraphics[height=4.5cm]{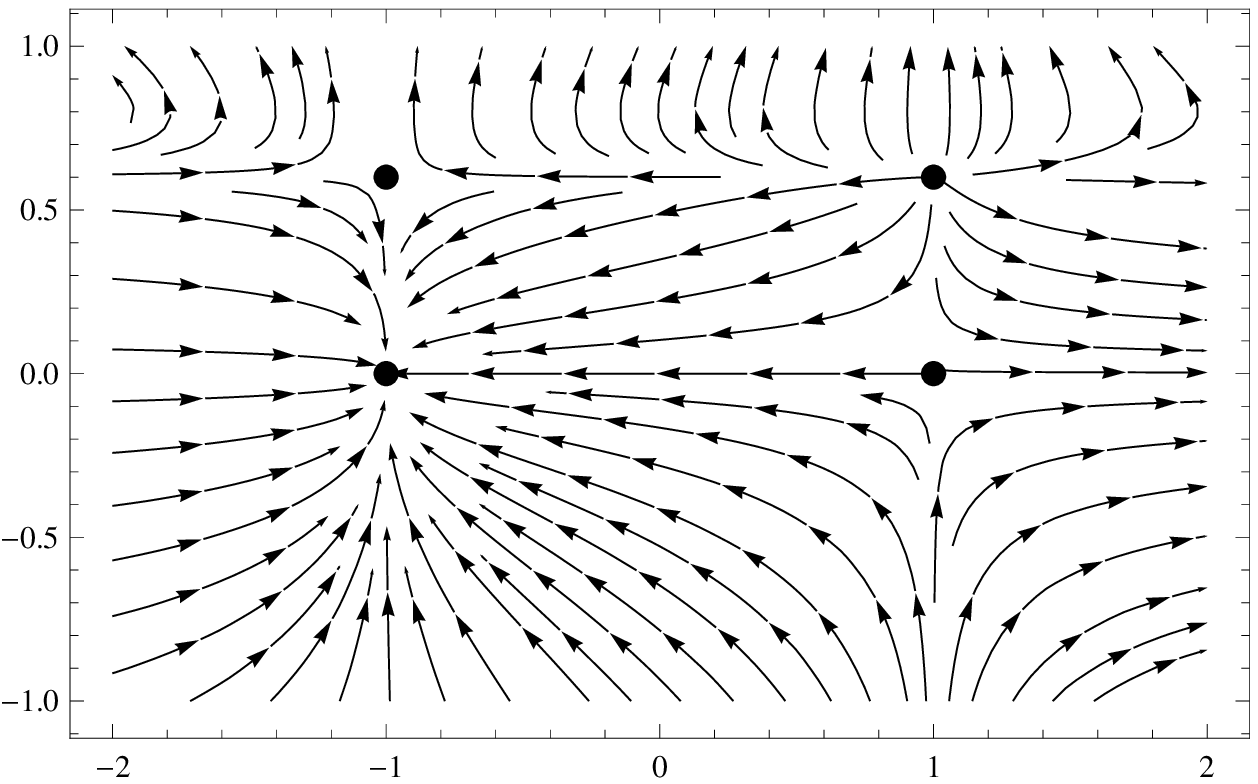}}
\put(220,6){$a_0$}
\put(455,6){$a_0$}
\put(15,135){$f_{-1}$}
\put(250,135){$f_{-1}$}
\end{picture}
\caption{RG flows in the $a_0$-$f_{-1}$ plane for AdS$_{d+1}$ and $m=0.2$ (left panel) and $m=0.8$ (right panel).}
\label{fig:fermionicflowkminus1}
\end{center}
\end{figure}

\paragraph{Non-relativistic deformations}

For simplicity, let us focus on $d=3$. Consider \eqref{eq:flowFAdS4} and make the ansatz $F = i \gamma^{\hat \mu} F_{\mu} + f$. Concentrating on the homogeneous modes ({\it i.e.} putting $k=\omega=0$), we find
\SP{\label{non-rel}
	\ep \del_\ep F_\mu = -2m F_\mu, \ \ \ \ep \del_\ep f = -2m f.
}
Non-zero $F_\mu$ violates Lorentz invariance. We see that there are new fixed points given by $F_\mu=0$ or $\infty$ for various $\mu$. Starting out with one of these components $F_\mu$ equal to infinity in the UV, {\it i.e.} at a Lorentz-violating fixed point, one however always ends up flowing to the usual IR fixed point given by standard quantization, where Lorentz symmetry is restored. Similar
flows from Lorentz non-invariant fixed points to Lorentz invariant fixed
points occur in discussions involving Lifshitz theories, {\em e.g.}, in \cite{Kachru:2008yh} in the
context of gravity, and in \cite{Dhar:2009dx, Dhar:2009am} in
the context of field theory.

\section{\label{sec:BH}Fermionic flows in the extremal charged black hole background}

In this section, we will study flows in the extremal charged black hole background given by \eq{bh-metric}. Since this background is rotationally symmetric, it is practical to use the form of the flow equation \eqref{eq:floweqsnicegammas} given in Section~\ref{sec:rot-sym}, which was obtained by making a convenient choice of Dirac matrices. Let us concentrate on the case $\alpha=1$,\footnote{The flow equation for $F_2$ is the same after making the switch $k \rightarrow -k$.} in the following omitting the $\alpha$-index of $F_\alpha$. We have that
\SP{\label{eq:floweqECBH}
	\sqrt{H} \epsilon \partial_\epsilon F = \left( \frac{\epsilon (\omega + q A_t)}{\sqrt{H}} - \epsilon k \right) F^2 - 2 m F + \frac{\epsilon (\omega + q A_t)}{\sqrt{H}} + \epsilon k.
}
In the far UV, the ECBH background is given by AdS$_{d+1}$ with a constant electric field, which is the case we studied in the last section. Let us start by analyzing the flow equation for $F$ in the near horizon limit.

\subsection{Near horizon limit}

\begin{figure}[t]
\begin{center}
\begin{picture}(500,135)
\put(13,0){\includegraphics[height=4.5cm]{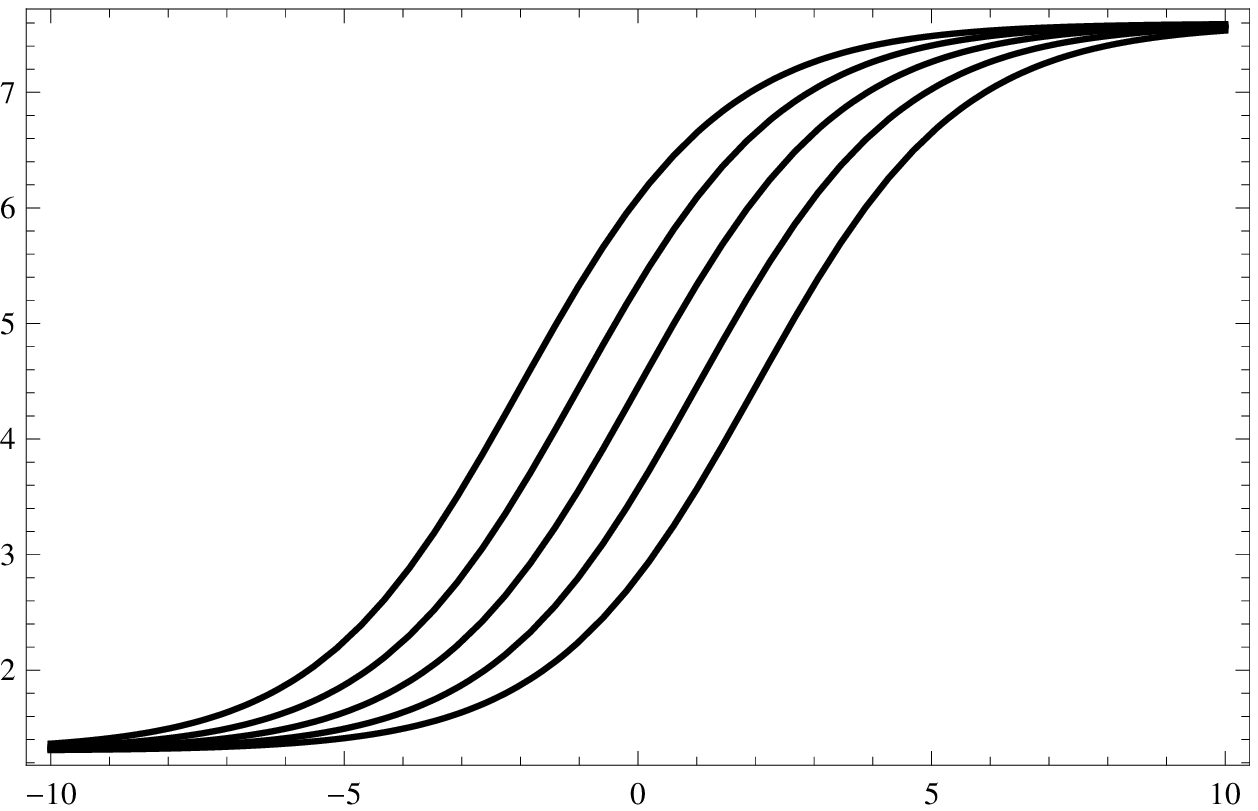}}
\put(243,0){\includegraphics[height=4.5cm]{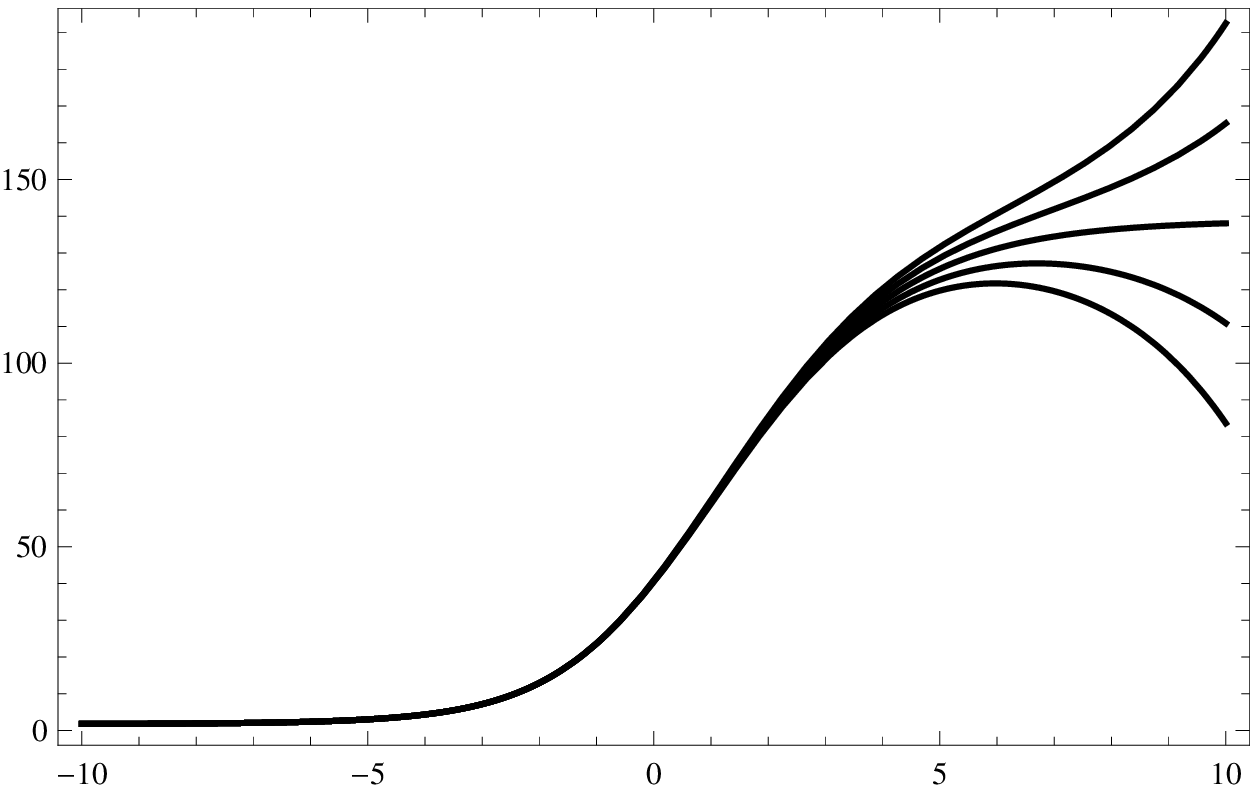}}
\put(218,6){$r$}
\put(453,6){$r$}
\put(14,135){$b_0$}
\put(250,135){$b_1$}
\end{picture}
\caption{Flows of $b_0$ and $b_1$ in the near horizon limit as a function of $r = - \log \zeta$ for $m = 1$, $k=1$, $q e_d = 0.5$, and $z_* = 1$, making $\nu_k = 0.3$. The UV of the AdS$_2$ region is at $r=\infty$, while its IR is at $r=-\infty$.}
\label{fig:fermionicflowb0rb1r}
\end{center}
\end{figure}

The background \eq{bh-metric} has an emergent AdS$_2 \times
\mathbb{R}^{d-1}$ geometry in the near horizon limit. To see
this \cite{Faulkner:2009wj}, define new coordinates $\zeta$ and $\tau$ through ($R_2 =
1/\sqrt{d(d-1)}$)
\SP{
	z - z_* = - \frac{\lambda R_2^2 z_*^2}{\zeta}, \ \ t = \lambda^{-1} \tau, \ \ \lambda \ll 1,
}
so that near the horizon
\SP{\label{near-horizon}
	ds^2 &= \frac{R_2^2}{\zeta^2} \left( d\zeta^2 - d\tau^2 \right) + \frac{1}{z_*^2} dx_2^2, \\
	A_\tau  &= \lambda^{-1} A_t = \frac{R_2^2 \mu z_*}{\zeta} =: \frac{e_d}{\zeta}.
}
Furthermore, in this limit we will hold $\Omega := \lambda^{-1} \omega$ fixed, thereby focusing on low frequencies $\omega$.

In this limit, the flow equation for $F$ becomes 
\SP{
	\zeta \partial_\zeta F = \left( \zeta \Omega + q e_d - R_2 z_* k \right) F^2 - 2 R_2 m F + \zeta \Omega + q e_d + R_2 z_* k.
}
Consider the case $\Omega = 0$. Then the solution is given by ($r = - \log \zeta$)
\SP{
	F &= \frac{R_2 m + \nu_k \tanh
   \left(\nu_k \left(r-r_0\right)\right)}{q e_d - R_2 z_* k}, \\
   \nu_k &:= \sqrt{ R_2^2 (m^2 + z_*^2 k^2) - q^2 e_d^2},
}
where, for real $\nu_k$, $r_0$ is an integration constant that sets the scale of where the kink is (see Figure \ref{fig:fermionicflowb0rb1r}.  This solution is a good approximation as long as $\zeta \Omega \ll 1$. This means that even for small $\Omega$, deviations will eventually be seen in the deep IR. For imaginary $\nu_k$, the solution becomes oscillatory, signalling that we are inside what is referred to as the oscillatory region \cite{Liu:2009dm}.

Viewing $F$ as a function of $\zeta$, $k$, and $\bar \Omega = \zeta \Omega$, we have that 
\SP{
	\zeta \partial_\zeta F = \left( \bar \Omega + q e_d - R_2 z_* k \right) F^2 - \left( 2 R_2 m + \bar \Omega \frac{\partial}{\partial{\bar \Omega}} \right) F + \bar \Omega + q e_d + R_2 z_* k.
}
Expanding
\SP{
	F(\zeta, \bar \Omega, k) = \sum_{n=0}^\infty b_n(\zeta, k) \bar \Omega^n,
}
we obtain
\SP{
	\zeta \partial_\zeta b_0 &= \left( q e_d - R_2 z_* k \right) b_0^2 - 2 R_2 m b_0 + q e_d + R_2 z_* k, \\
	\zeta \partial_\zeta b_1 &= 2 \left( q e_d - R_2 z_* k \right) b_0 b_1 + b_0^2 - \left( 1 + 2 R_2 m \right) b_1 + 1, \\
   & \ldots
}
Example of flows for $b_0$ and $b_1$ are shown in Figure~\ref{fig:fermionicflowb0rb1r}.

\paragraph{Merger and disappearance of fixed points}

Figure~\ref{fig:mergingFPs} shows flow diagrams in the $b_0-b_1$ plane. There are two fixed points, given by $b_0 = \frac{R_2 m \pm \nu_k}{q e_d - R_2 z_* k}$. As one approaches the oscillatory region, where $\nu_k$ becomes imaginary, the two fixed points approach each other until they merge and disappear (become imaginary). Such a merger of fixed points is reminiscent of similar discussions
in \cite{Kaplan:2009kr}.

\begin{figure}[t]
\begin{center}
\begin{picture}(500,135)
\put(10,0){\includegraphics[height=4.5cm]{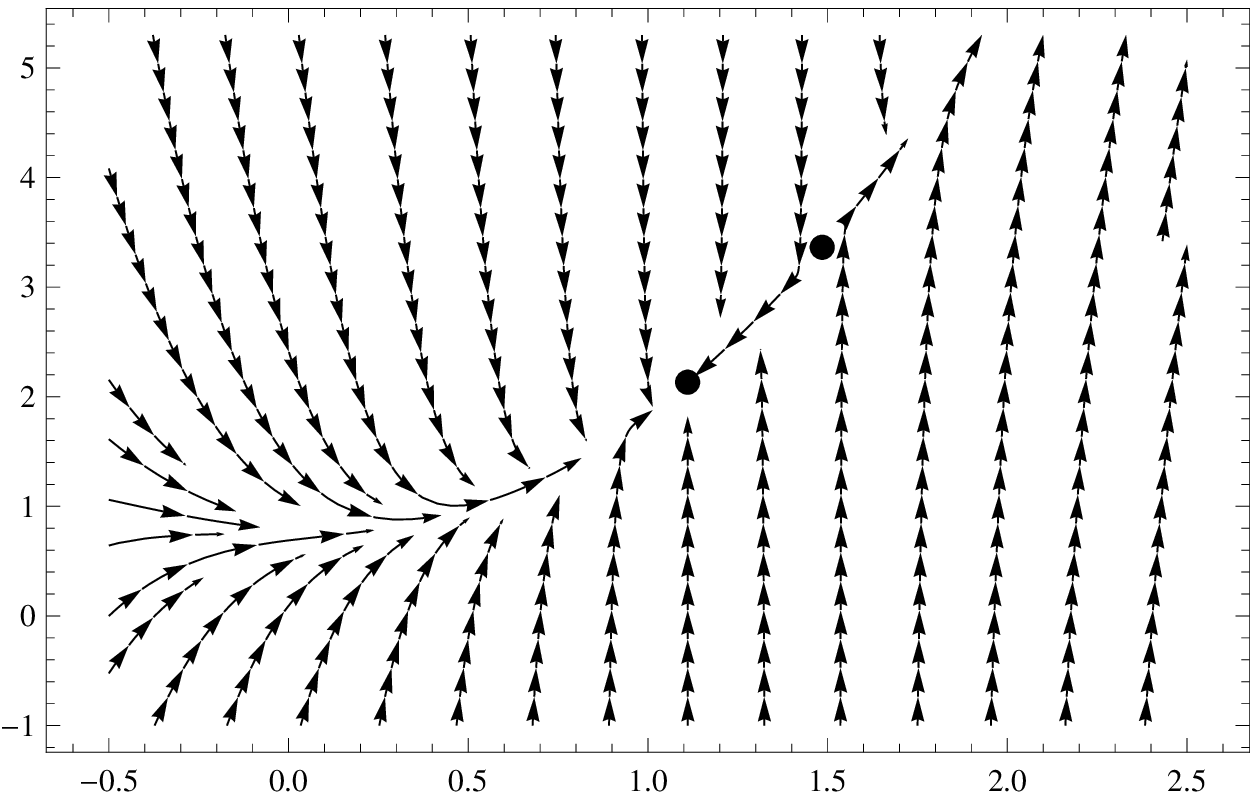}}
\put(243,0){\includegraphics[height=4.5cm]{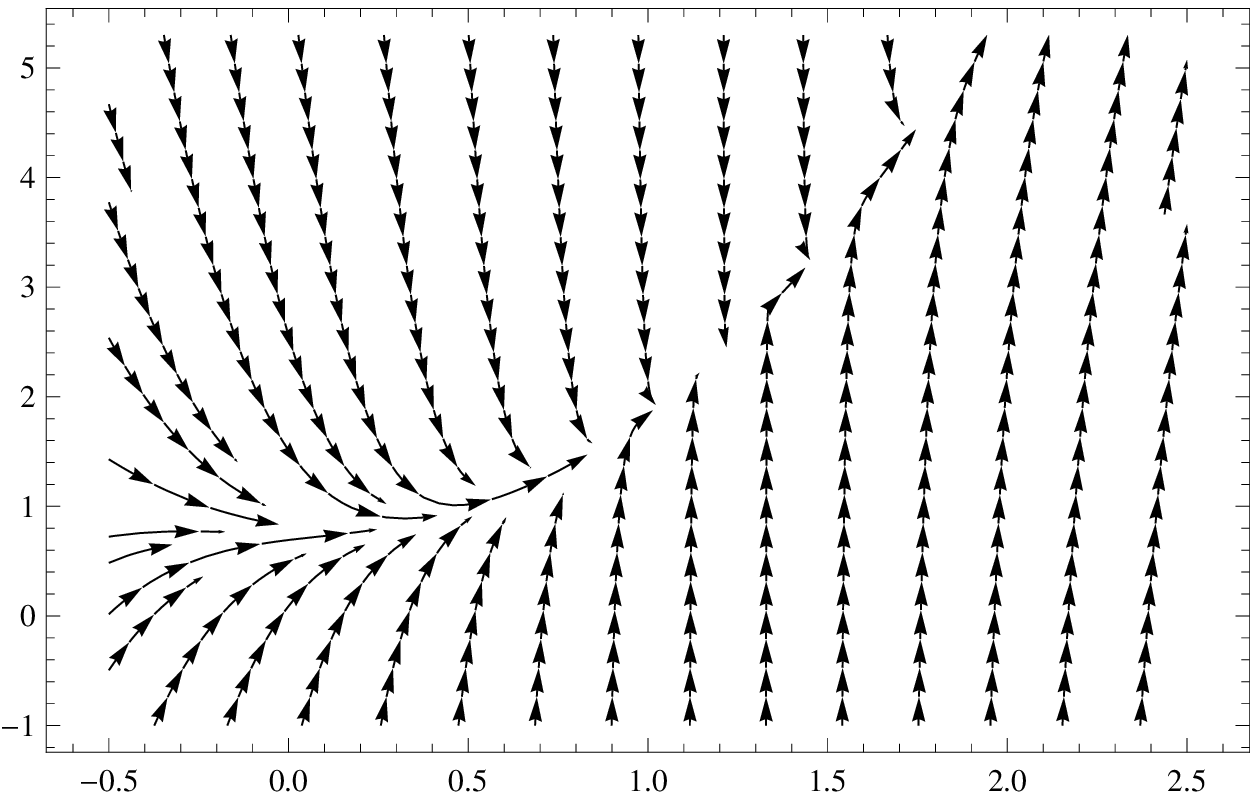}}
\put(217,6){$b_0$}
\put(450,6){$b_0$}
\put(16,135){$b_1$}
\put(248,135){$b_1$}
\end{picture}
\caption{The left panel shows the RG flow in the $b_0-b_1$-plane for the near horizon limit, and $d=3$, $m=0.4$, $k=0.1$, $q e_d = 1/6$, $z_*=1$, making $\nu_k = 0.02$. The right panel shows the same, but for $m=0.395$ (all other parameters unchanged), making $\nu_k = 0.01 i$ imaginary.}
\label{fig:mergingFPs}
\end{center}
\end{figure}

\begin{figure}[t]
\begin{center}
\begin{picture}(500,135)
\put(10,0){\includegraphics[height=4.5cm]{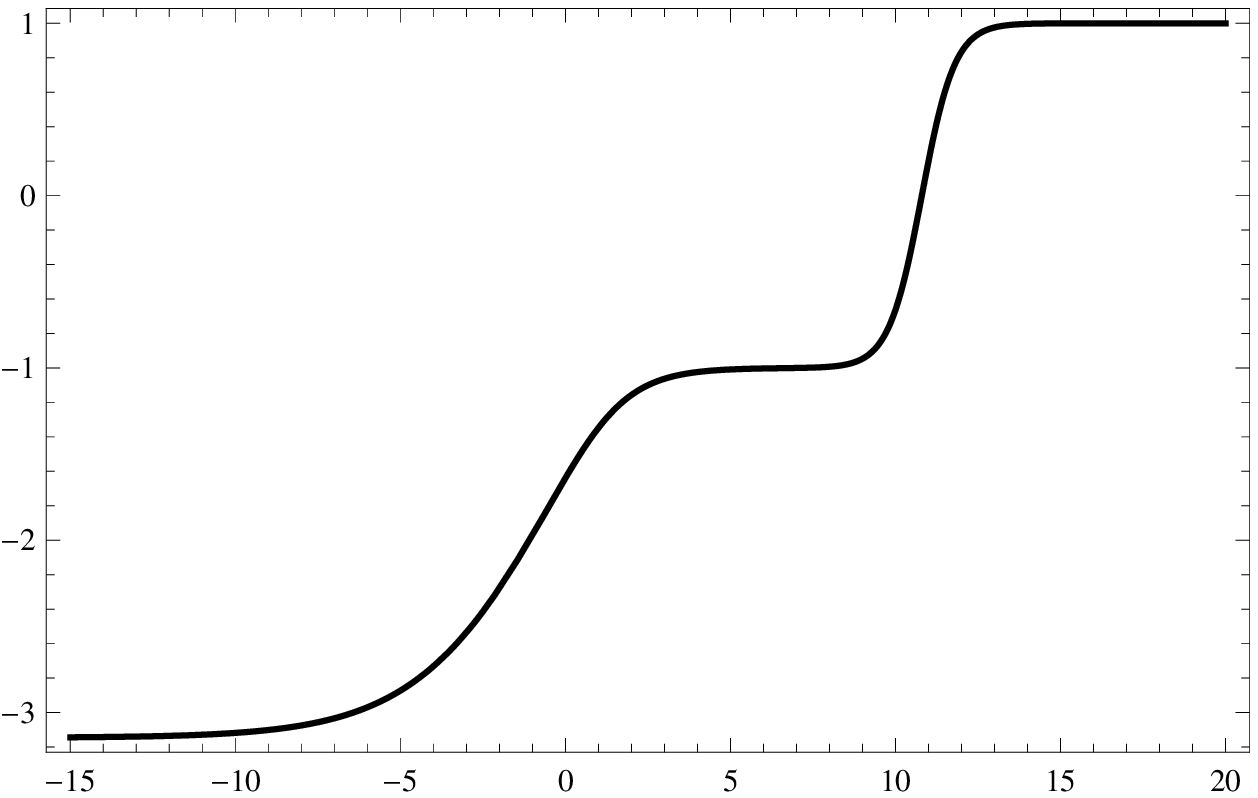}}
\put(243,0){\includegraphics[height=4.5cm]{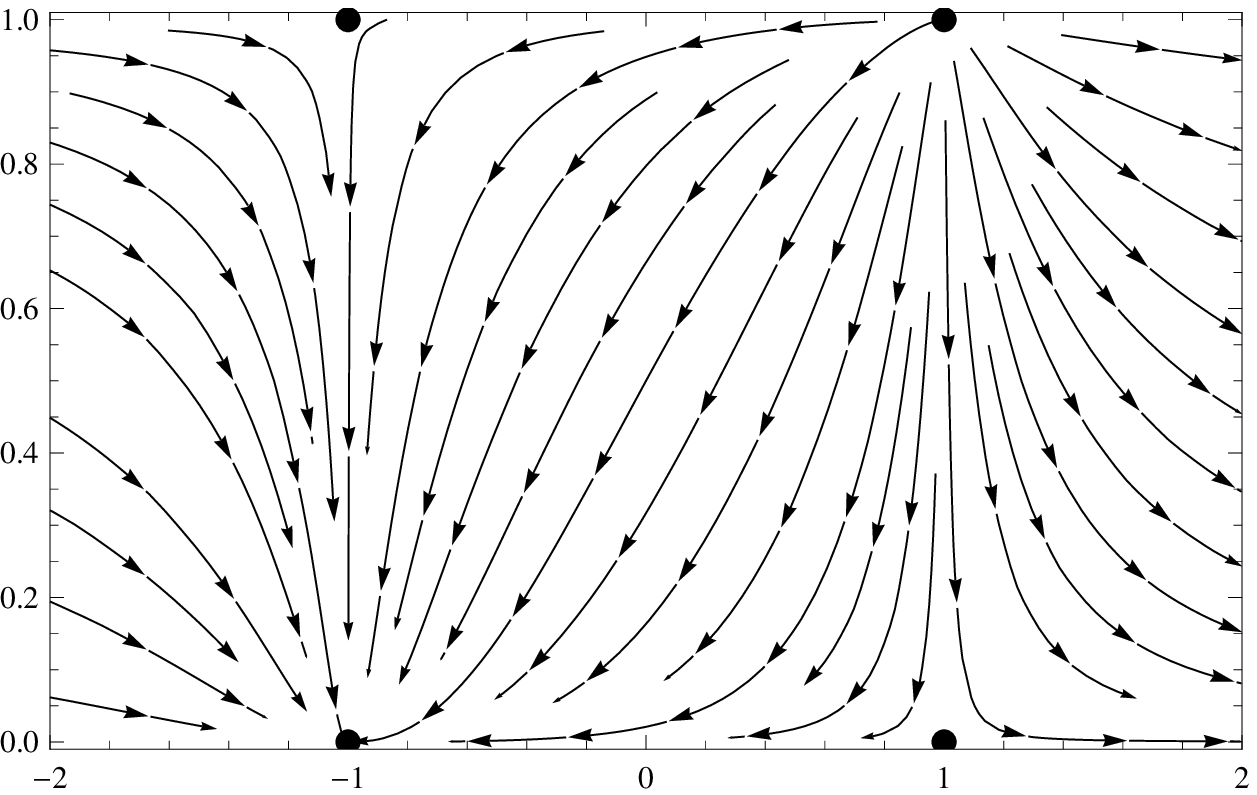}}
\put(215,6){$r$}
\put(447,6){$\tilde F$}
\put(16,134){$\tilde F$}
\put(248,134){$H$}
\end{picture}
\caption{RG flows in the ECBH background and $d=3$, $m = 1$, $k=0$, $\omega=0$. The left panel shows $\tilde F$ as a function or $r = \log \left( z^{-1} - z_*^{-1} \right)$ (so that the UV is at $r=\infty$ while the horizon is at $r=-\infty$) for $q\mu =-2$, $z_*=1$. The right panel shows the RG flow in the $\tilde F$-$H$ plane for $q=0$; as in the bosonic case (Figure \ref{fig-scalar-flow}),
there are two fixed points in the AdS$_4$ region ($H \sim 1$) and
two fixed points in the AdS$_2$ region ($H \sim 0$).}
\label{fig:fermionicECBH}
\end{center}
\end{figure}

\subsection{Flows in the full ECBH background}

Let us now study flows in the full ECBH background, i.e connecting the two asymptotic regions studied in the previous subsections. After changing variables to
(see \eq{special-basis}) 
\SP{
	\tilde F = \frac{F - 1}{F + 1},
}
the flow equation \eqref{eq:floweqECBH} becomes
\SP{
	\sqrt{H} \epsilon 
   \partial_\epsilon \tilde F = \left(m+\frac{\epsilon  \left(\omega +q A_t\right)}{\sqrt{H}}\right) \tilde F^2-2 \epsilon k \tilde F-m+\frac{\epsilon  \left(\omega +q A_t\right)}{\sqrt{H}}.
}
The left panel of Figure~\ref{fig:fermionicECBH} shows a solution to this equation, which starts by flowing from the alternative to the standard fixed point of the AdS$_4$ region, and then finally reaches the standard fixed point of the AdS$_2$ near horizon region. The the right panel of the same figure shows flows in the $\tilde F$-$H$ plane obtained for $q=0$, and $k=\omega=0$ (thereby focusing on the homogeneous mode), the fixed points of which are at $(\tilde F,H) = (\pm 1,1)$ and $(\tilde F,H) = (\pm 1,0)$. The flow of $H$  is
interpreted, as in Section~\ref{sec:scalar} (see Section~\ref{sec:FT} for
further details).

\section{\label{sec:FP}Emergent fermions and semi-holography}

\paragraph{The semi-holographic approach and goal of this section}

The main goal of this section is to understand the semi-holographic
approach to fermionic field theories \cite{Faulkner:2010tq} from the
Wilsonian perspective followed in this paper. In this approach, one
considers a weakly coupled fermionic field $\chi_a$ coupled to
fermionic fields ${\mathbf \Psi_a}$ \footnote{The label $_a$ is
a Dirac spinor index; the 
notation ${\mathbf \Psi}$ for the
field theory operator should
  not be confused with the wavefunction $\Psi$, {\em
e.g.}, 
  \eq{wvf}.} belonging to a strongly
interacting sector \footnote{$\varepsilon(k)$ in
\eq{semi-hol} represents the single particle energy of
the field $\chi_a$.}
\begin{align}
S= S_{\rm strong}  +  \int dt~\int d^d k
\left[ \chi^{\dagger}(k) \left(i\del_t - \varepsilon(k) + \mu q\right) 
\chi(k) +   g_k ~ \chi^{\dagger}(k) {\mathbf \Psi}(k) 
+ g^*_k ~ \mathbf{\Psi}^{\dagger}(k)
\chi(k)  \right] \,.
\label{semi-hol}  
\end{align}
While $\chi_a$ is described by standard field theory methods, the
strongly coupled field ${\mathbf \Psi_a}$ is described holographically
using AdS/CFT. In particular, by usual field theory arguments one can
derive \cite{Faulkner:2010tq}
\begin{align}
G_{g}(\om, k) = \frac{1}{G_0(\om, k)^{-1} - |g_k|^2  
{\cal G}_0(\om, k)} \,,
\label{dyson}
\end{align}
where $G$ and ${\cal G}$ are the Feynman propagators for the $\chi$
and the ${\mathbf \Psi}$ fields, respectively. The subscript $_0$
indicates the decoupled limit where $g_k=0$; $G_g$ denotes the $\chi$
propagator in presence of the coupling.  The two-point function ${\cal
  G}_0(k)$ from the strongly coupled sector is derived holographically
using an AdS dual.  The influence of the strong sector can clearly
have interesting consequences for Equation \eq{dyson}. An example is the
non-Fermi-liquid behaviour described in \cite{Faulkner:2009wj}.

In \cite{Faulkner:2010tq} a qualitative origin of the fields $\chi$
and ${\mathbf \Psi}$ is described in the context of the charged AdS
black hole \eq{bh-metric} discussed in this paper. In particular,
${\mathbf \Psi}$ are operators in the field theory dual to the
near-horizon AdS$_2 \times {\mathbb R}^2$ region, whereas $\chi$ is supposed to
be a `domain wall fermion' near the AdS$_2$ boundary. In what follows,
we will describe the precise origin of the field $\chi$, from the
Wilsonian RG perspective, as an emergent fermion. A hint of such an
approach appeared earlier in \cite{Faulkner:2010jy}.

\subsection{\label{sec:FP-gist}Appearance of massless fermions and derivation of semi-holography}
We start by applying \eq{slicing-f} to the charged black hole
background \eq{bh-metric}, which has an emergent near-horizon AdS$_2
\times \mathbb R^{d-1}$ region.  We put an intermediate cut-off $\ep$
at the AdS$_2$ boundary.  In other words, we split the bulk transition
amplitude from the black hole horizon to the AdS$_{d+1}$ boundary into
two parts: (a) from the horizon to the AdS$_{2}$ boundary, and (b)
from the AdS$_{2}$ boundary to the AdS$_{d+1}$ boundary. We rewrite
\eq{slicing-f} as
\begin{align}
\lan \IR|\wh{U}(\ep_{\IR}, \ep_0)|\UV\ran 
& = \int\! D^4\tilde \chi  \lan \IR|\wh{U}(\ep_{\IR}, \ep)
|\ol{\tilde\chi}_+, \tilde\chi_-\ran
e^{\int_k\ol{\tilde\chi}_+ \tilde\chi_+ + \ol{\tilde\chi}_- \tilde\chi_-} 
\Psi[\tilde\chi_+, \ol{\tilde\chi}_-] \,,
\label{div1}
\\
\Psi[\tilde\chi_+, \ol{\tilde\chi}_-] &=  
\lan \tilde\chi_+,
\ol{\tilde\chi}_-|\wh{U}(\ep,\ep_0)|\UV\ran
\label{wvf} \,,
\end{align}
where we have added a `tilde' to $\chi$ to indicate the possibility of
a different basis of coherent states at the intermediate cut-off (we
will see below in \eq{ads2-twist} that such a change of basis is
induced when we enter the AdS$_2$ region). 
Note that $\Psi[\tilde\chi_+, \ol{\tilde\chi}_-]$
does not have to be a transition amplitude; it is enough to assume
that a Gaussian wave-functional is given at the cut-off surface $\ep$.
Later, we adopt this viewpoint, regarding the AdS$_2$ boundary as the
cut-off surface.

By using the quadratic ansatz \eq{g_wf} for the wavefunction $\Psi$,
\begin{align}
 \Psi[\tilde\chi_+, \ol{\tilde\chi}_-] :=
\exp\left\{
-\frac{1}{\ka^2}\int_k
\left[
\ol{\tilde\chi}_-(k)\tilde F(k,\ep)\tilde \chi_+(k) + 
\ol{\tilde\chi}_-(k)\tilde J_-(k,\ep) + 
\ol{\tilde J}_+(k,\ep)\tilde\chi_+(k) + \tilde C(\ep)
\right]
\right\} \,, 
\label{ansatz}
\end{align}
and performing the $\int D^4\tilde\chi$ integral, we arrive, as in
\eq{double-trace-fermion}, at the following field theory at the
intermediate cut-off
\begin{align}
\Bbra \exp\int_k 
\left[-\ol O_-(k)\tilde F(k,\ep)O_+(k) +
\ol{\tilde J}_+(k, \ep) O_+(k) +\ol{O}_-(k)\tilde J_-(k,\ep)\right]\Kket_{\ep}^{\std} \,.
\label{irgen}
\end{align}
Now, suppose after solving the
flow equations for $\tilde F$ up to $z=\ep$, we find that $\tilde F(k,\ep)$ develops a single
pole at some $k_{\mu}=k_{\ast\mu}$\footnote{\label{ftnt:c-mu}
Let us see the structure of $c_{\mu}$.
Here let us adopt the representation \eqref{deven} or \eqref{dodd} of the Dirac matrices,
which is different from the representation in the next subsection.
From the equation of motion, we find
$K^{\nu}K_{\nu}\Phi_{\mp}=-i\ga^{\mu}K_{\mu}(\pm\pd_z-m\sqrt{g_{zz}})\Phi_{\pm}$.
Since we can show that $\Phi_-(\Phi_+)^{-1}$ satisfies the flow equation of $F$,
$F$ is proportional to $\ga^{\mu}K_{\mu}$ 
and thus $c_{\mu}(k-k_{\ast})_{\mu}$ can be expressed as $A_{\mu}\ga^{\wh{\mu}}$.
Now, since $A_{\mu}$ is a linear function of $k-k_{\ast}$, 
$A_{\mu}$ can be rewritten as $A_{\mu}=A_{\mu\nu}(k-k_{\ast})_{\nu}$ with some tensor $A_{\mu\nu}$.
Using the rotational symmetry in the spatial directions,
we find $A_{ti}=0$, $A_{ij}=B\de_{ij}$.
We thus find $c_t=B_{tt}\ga^{\wh{t}}$, $c_i=B\ga^{\wh{i}}$.}
\begin{align}
\tilde F(k, \ep ) &\sim \frac{1}{c_{\mu}(k-k_{\ast})_{\mu}}  \,.
\label{poleF}
\end{align}
We will see below that this indeed happens in the case
of the extremal charged black hole background, where $k_{\ast \mu}$ 
in \eq{poleF} is
given by $k_\ast^0= \om = 0$ and $\vec k_\ast =  k_F \vec k/|\vec k|$,
with $k_F$ the
Fermi momentum,  and
$\ep$ represents the AdS$_2$ boundary. Substituting the above form of
$\tilde F$ into the IR generating functional \eqref{irgen}, we find the following
non-local expression,
\begin{align}
\Bbra \exp \left( \int_k -\ol{O}_-\frac{a_1}{c_{\mu}(k-k_{\ast})_{\mu}}O_+ 
+ \cdots \right) \Kket_{\ep}^{\std} \,.
\end{align}
where $\cdots$ denotes terms regular in $(k-k_{\ast})_{\mu}$.  The
appearance of such non-local operators is a typical signal of
appearance of massless modes in the system which we have somehow
integrated out. To figure out what might have happened, recall that
\eq{irgen} was arrived at by performing the integral $\int D^4\tilde \chi$
in \eq{div1}.  Suppose, instead, we leave the integration over
$\ol{\tilde \chi}_+, \tilde\chi_-$ {\em undone}; in that case the field theory
correlator \eq{irgen} is replaced by\footnote{Here we dropped the
  quantum correction $\det \tilde F$ and terms  constant in $\ol{\tilde \chi}_+,
  \tilde \chi_-$.}
\begin{align}
\int D\ol{\tilde \chi}_+D\tilde \chi_-
\Bbra \exp\int_{k} [\ol{\tilde \chi}_+O_+ 
+ \ol{O}_-\tilde\chi_-] \Kket_{\ep}^{\std}
\exp\int_k
[\ol{\tilde \chi}_+\tilde F^{-1}\tilde\chi_- - \ol{\tilde J}_+\tilde F^{-1}\tilde\chi_- - 
\ol{\tilde \chi}_+\tilde F^{-1}\tilde J_-]  \,.
\label{interim}
\end{align}
Note that $\tilde\chi_+, \ol{\tilde\chi}_-$ in \eqref{div1}, which do
not couple to $O_+, \ol{O}_-$, have been integrated out. Writing 
$\bbra~ (...)~ \kket_{\ep} :=  $ \\ $\int\!DM|_{\Lm(\ep)} ~(...)~ e^{S[M,\ep]}$, 
where $M$
is understood to be some large $N$ matrix field with action
$S[M]$, \eq{interim} can be rewritten as \footnote{The form
  of the kinetic term of the emergent fermions in \eq{semi} is
  appropriate near the pole $k= k_\ast$ with
$k_\ast^0= \om = 0, |\vec k\ast|= k_F$, the Fermi
momentum; if one is
  interested in physics away from the pole these new fermions can be
  integrated out, leading to local operators in terms of the original
  theory.}
\begin{eqnarray}
& \displaystyle
\int\! DM~D\ol{\tilde\chi}_+D\tilde\chi_-~e^{S_{\rm total}[\ep]} \,, \nonumber \\
& \displaystyle
S_{\rm total}[\ep] := 
S[M,\ep]-\!\int_{k}a_1^{-1}\ol{\tilde\chi}_+(k)\left[
c_{\mu}(k-k_{\ast})_{\mu} + \cdots \right]
\tilde\chi_-(k)
+\int_{k}(\ol{\tilde\chi}_+\mathbb{O}_+ + 
\ol{\mathbb{O}}_-\tilde\chi_-) \,, \label{semi}
\end{eqnarray}
where we defined modified the new composite fields
\begin{align}
\mathbb{O}_+ &:= O_+ - \tilde F(\ep)^{-1}\tilde J_-(\ep) \,, \quad
\ol{\mathbb{O}}_- := \ol{O}_- - \ol{\tilde J}_+(\ep)\tilde F(\ep)^{-1} \,.
\end{align}
Thus, $S_{\rm total}$ is the new action at the AdS$_2$ boundary which includes
the emergent massless fermions $\ol{\tilde\chi}_+, \tilde\chi_-$;
the action has the form of the semi-holographic action 
\eq{semi-hol}, provided we make the identifications shown
in Table \ref{table:faulker-polchinski}.

\begin{table}[h]
\begin{center}
\begin{tabular}{|l|l| l|l|l|}
\hline
Faulkner-Polchinski &  $\chi_a$ & $\om=0, |\vec k|=
k_F$ &   ${\mathbf \Psi}_a$ & 
$S_{\rm strong}$ \\
\hline
Here & $\ol{\tilde \chi}_+, \tilde \chi_-$ & $k_{\ast,\mu}$ (see
\eq{poleF}) &
$\mathbb{O}_+, \ol{\mathbb{O}}_-$
& $S[M, \ep]$\\
\hline
\end{tabular}
\end{center}
\caption{Emergent fermions from Wilsonian RG. With the
identification in this Table, with an appropriate
rescaling of $\tilde \chi$, our equation \eq{semi} reproduces
the equation \eq{semi-hol} of  \cite{Faulkner:2010tq}, including
a determination of the couplings $g_k$. }
\label{table:faulker-polchinski}
\end{table}
\ni The identification clearly shows that the `domain wall' fermion
$\chi_a$ of \cite{Faulkner:2010tq}, appearing in Equation \eq{semi-hol},
can be identified with the emergent fermions $\ol{\tilde \chi}_+,
\tilde \chi_-$ at the AdS$_2$ boundary.

In the rest of this section, we will explain how the pole structure
\eq{poleF} arises at the AdS$_2$ boundary, with $k_\mu=k_{\ast\mu}$
identified with $\om=0, |\vec k| =  k_F$.

\subsection{Double trace coupling of the IR theory}
We begin by rewriting the ECBH background
\begin{align}
ds^2 &= \frac{R^2}{z^2}(-h(z)dt^2+dx^idx^i)+\frac{R^2}{z^2}\frac{dz^2}{h(z)} \,, \quad
A(z) = \mu\left(1-\frac{z^{d-2}}{z_{\ast}^{d-2}}\right)dt \,,
\label{ecbh}
\end{align}
where $R$ is the AdS$_{d+1}$ radius.\footnote{
In order to clarify the meaning of near horizon limit given later,
we exhibit the AdS$_{d+1}$ radius explicitly.}
Note that $z=0$ is the AdS$_{d+1}$ boundary,
$z_{\ast}$ is the horizon radius, and $\mu$ is the chemical potential in the field theory
(for details, see \cite{Faulkner:2009wj}).
As in the previous section, we consider only the $\al=1$ sector (this can be done without loss of generality 
since the $\al=1$ sector is decoupled from $\al=2$ 
and these two are related by $k_i\leftrightarrow -k_i$).\footnote{
For the definition and properties of $\al$, see Appendix \ref{app:rot-sym}.}
\paragraph{Two solution bases of Dirac equations}
We now relate the double trace coupling in the UV region (the
AdS$_{d+1}$ boundary) to the double trace coupling in the IR region
(the AdS$_2$ boundary).  For this purpose, we introduce two solution
bases \cite{Faulkner:2009wj} of the bulk Dirac equations of motion
\eqref{eq:Dirac} on the full ECBH background \eqref{ecbh},
\begin{alignat}{3}
&1.  &\qquad 
&(\phi_{\pm}, \vphi_{\pm}) ~ : ~
\phi_- \sim z^{-m} \,, \quad \phi_+ \sim z^{-m+1} \,, \quad 
\vphi_- \sim z^{m+1} \,, \quad \vphi_+ \sim z^{m}  &\qquad
&(z \sim 0) \,,
\label{sbasis1}\\
&2. &\qquad 
&(\rho_{\pm}, \psi_{\pm}) ~ : ~
\rho_{\pm} \sim \xi^{-\nu_k} \,, \quad
\psi_{\pm} \sim \xi^{\nu_k} &\qquad
&(\om\xi \sim 0) \,,
\label{sbasis2}
\end{alignat}
where 
$\nu_k$ is given in \eqref{fptwist} or \eqref{dirac-ads2},
and the new coordinate $\xi$ was introduced as follows:
\begin{align}
\xi &:= \frac{1}{d(d-1)}\frac{z_{\ast}^2}{z_{\ast}-z} \,, \quad R_2:=\frac{R}{\sqrt{d(d-1)}} \,.
\label{xi}
\end{align}
In terms of those solution bases, any solution is expressed as
\begin{align}
\Phi
&= 
A\begin{pmatrix} \phi_+ \\ \phi_- \end{pmatrix}
+ B\begin{pmatrix} \vphi_+ \\ \vphi_- \end{pmatrix}
=
C\begin{pmatrix} \rho_+ \\ \rho_- \end{pmatrix}
+ D\begin{pmatrix} \psi_+ \\ \psi_- \end{pmatrix} \label{basis} \,.
\end{align}

First, let us explain the first solution basis \eq{sbasis1}.
The asymptotic behavior \eqref{sbasis1}
can be found from the explicit solutions \eqref{dirac-sols} to the Dirac equations
on the near-boundary AdS$_{d+1}$ geometry;
explicitly, $\phi_{\pm}$ asymptote to $\sqrt{z}I_{-\nu_{\mp}}(zK)$
and $\vphi_{\pm}$ asymptote to $\sqrt{z}I_{\nu_{\mp}}(zK)$.\footnote{
  $(\phi_-,\vphi_-)$ are normalized so that the
  first coefficients of the Taylor expansions of them are 1, while
  $(\phi_+,\vphi_+)$ are defined by \eqref{eq:Dirac}
  from $(\phi_-,\vphi_-)$ and thus are not normalized 
  so that any solution be expressed as \eqref{basis}.}
The  basis  \eqref{sbasis1}  is used for identifying source fields at the 
AdS$_{d+1}$ boundary.

To understand the second solution basis \eq{sbasis2}
and to see the role played by $\xi$,
let us define the inner and outer regions (for all
details, see \cite{Gauntlett:2011wm})
\begin{align}
\mbox{the inner region } ~ 
&: ~ 0<\frac{z_{\ast}-z}{z_{\ast}}<\lm \,, \quad \om z_{\ast}<\lm \,, \quad \lm\ll\mu z_{\ast} \,, \\ 
\mbox{the outer region } ~ 
&: ~ \lm^{\pr}<\frac{z_{\ast}-z}{z_{\ast}}<\infty \,, \quad \om z_{\ast}\ll\lm^{\pr} \,.
\end{align}
The matching region is defined as the following double scaling limit
\begin{align}
\mbox{the matching region } ~ 
&: ~ \om z_{\ast} \ll \lm^{\pr}<\frac{z_{\ast}-z}{z_{\ast}}<\lm\ll\mu z_{\ast} \,.
\label{match}
\end{align}
Then, the spacetime metric and the gauge field in the inner region 
in the $\lm\to 0$ limit become ({\em cf.} \eq{near-horizon}) \footnote{
The two coordinates $\zt$ and $\xi$ are related by $\Omega\zt=\xi\om$, 
where $\Omega=\lm^{-1}\om$ is kept fixed in the limit $\lm\to 0$ (see also Appendix \ref{app-nh}).}
\begin{align}
ds^2 &= \frac{R_2^2}{\zt^2}(-d\tau^2+d\zt^2)+\frac{R^2}{z_{\ast}^2}dx^idx^i \,, \quad
A = \frac{e_d}{\zt}d\tau \,, \quad
\zt := \lm\xi \,, \quad \tau := \lm t \,, \label{nhads2}
\end{align}
where $\zt, \tau$ are kept finite in $\lm\to 0$,
and $e_d$ is proportional to the gauge coupling constant 
(for the explicit definition, see \cite{Faulkner:2009wj}).
This shows that $\xi$ (or $\zeta$) is the radial coordinate
of the AdS$_2\times\mathbb{R}^{d-1}$.

The asymptotic behavior of the second solution basis \eqref{sbasis2}
can be found from the near-AdS$_2$-boundary behavior of solutions
to the Dirac equations of motion in the inner region 
in the leading order in $\lm$, given by \eqref{dirac-ads2} with \eqref{eq:Dirac},
which are the Dirac equations in the AdS$_2\times\mathbb{R}^{d-1}$ background \eqref{nhads2}.
In fact, this inner region Dirac equation yields an analytic expression of the solution basis 
$(\rho_{\pm}, \psi_{\pm})$ in terms of the Whittaker functions
(see also \cite{Faulkner:2011tm})\footnote{
The Whittaker functions ${\rm M}_{\ka,\mu}(z)$, defined by
\begin{align}
{\rm M}_{\ka,\mu}(z) &:= 
z^{\mu+\frac{1}{2}}e^{-z/2}\sum_{n=0}^{\infty}
\frac{\Ga(2\mu+1)\Ga(\mu-\ka+n+\frac{1}{2})}{\Ga(2\mu+n+1)\Ga(\mu-\ka+\frac{1}{2})} \,,
\end{align}
are solutions to the Whittaker differential equation
\begin{align}
\left( \pd_z^2-\frac{1}{4}+\frac{\ka}{z}-\frac{\mu^2-\frac{1}{4}}{z^2} \right)
{\rm M}_{\ka,\mu}(z)=0 \,,
\end{align}
and a pair $({\rm M}_{\ka,\mu}(z),{\rm M}_{\ka,-\mu}(z))$ forms a solution basis.
Note that $(\rho_-,\psi_-)$ are normalized so that the
first coefficients of the Taylor expansions of them are 1, while
$(\rho_+,\psi_+)$ are defined by \eqref{eq:Dirac}
from $(\rho_-,\psi_-)$ so that any solution be expressed as \eqref{basis}.}
\begin{align}
\begin{pmatrix} \rho_+ \\ \rho_- \end{pmatrix}
&= (2i\om)^{\nu_k-\frac{1}{2}}\left(1-ic_k\right)^{-1}
\begin{pmatrix} 1 & -i \\ -i & 1 \end{pmatrix}
\begin{pmatrix} 
c_k{\rm M}_{-iqe_d-\frac{1}{2},-\nu_k}(2i\om\xi) \\
{\rm M}_{-iqe_d+\frac{1}{2},-\nu_k}(2i\om\xi)
\end{pmatrix} \,, \nonumber\\
\begin{pmatrix} \psi_+ \\ \psi_- \end{pmatrix}
&= (2i\om)^{-\nu_k-\frac{1}{2}}\left(1-ic_k\right)^{-1}
\begin{pmatrix} 1 & -i \\ -i & 1 \end{pmatrix}
\begin{pmatrix} 
c_k{\rm M}_{-iqe_d-\frac{1}{2},\nu_k}(2i\om\xi) \\
{\rm M}_{-iqe_d+\frac{1}{2},\nu_k}(2i\om\xi)
\end{pmatrix} \,, \quad
c_k :=\frac{-\nu_k+iqe_d}{R_2(m+kz_{\ast}R^{-1})} \,.
\end{align}
In addition, this inner region solution basis
can be extended to the outer region 
by matching the above inner region solutions  
with the solutions to the outer region Dirac equation 
(for details, see \cite{Faulkner:2009wj,Gauntlett:2011wm}).
This is the second solution \eqref{sbasis2} 
{\it defined all over the ECBH background}.

Here let us make an important remark on the matching region.
We can show that $\xi\om\ll 1$ holds
for any $\xi$ in the matching region \cite{Gauntlett:2011wm}.
This means that any point in the matching region is very close to the AdS$_2$ boundary.
Thus, any $\xi$=constant surface in the matching region can be interpreted 
as a surface which regularizes the AdS$_2$ boundary.
We will use this interpretation later when we consider 
the holography of the inner region with AdS$_2\times\mathbb{R}^{d-1}$ metric \eqref{nhads2}.

\paragraph{Renormalized double trace coupling in the continuum limit at UV}
Let us consider the renormalized double trace operators in the continuum limit
of the UV theory.
The double trace coupling flow can be expressed 
in terms of the classical solution basis \eqref{sbasis1} using \eqref{flowsol},
\begin{align}
F(z) 
= \frac{\phi_-(z) + \chi_{\UV}\vphi_-(z)}{\phi_+(z) + \chi_{\UV}\vphi_+(z)} \,,
\label{fchiuv}
\end{align}
where $\chi_{\UV}$ is a constant under the flow 
and is fixed by the boundary conditions at the AdS$_{d+1}$ boundary.
Applying the general formula \eqref{genct_f},
we can find the asymptotic behavior of the double trace coupling near the boundary $z\sim 0$,
\begin{align}
\frac{1}{F(z)^{-1}-A^{\UV}_0} \sim \chi_{\UV}^{-1}z^{-2m} \quad (z\sim 0)\,,
\end{align}
where $A^{\UV}_0$ comes from the counterterm action. 
If it is given by $A^{\UV}_0=\phi_+/\phi_-$ (see \eqref{fermion-s-ct}), 
the renormalized double trace coupling becomes $\chi_{\UV}^{-1}$.

\paragraph{Renormalized double trace coupling at IR}
Let us apply the holography in the matching region.
As mentioned above, 
since any point in the matching region is very close to the AdS$_2$ boundary,
we can expect that any cutoff theory in the matching region 
can be well approximated by the continuum theory, even
though the boundary of AdS$_2\times\mathbb{R}^{d-1}$ itself 
is, strictly speaking, not in the inner (matching) region.

Near the AdS$_2$ boundary $\om\xi \sim 0$,
the bulk Dirac equation in the background \eqref{nhads2} can
be approximated by
\begin{align}
\zt\pd_{\zt}\Phi &= U\Phi \,, \quad 
U:=
\begin{pmatrix}
mR_2 & \wt{m}R_2-qe_d \\ \wt{m}R_2+qe_d & -mR_2
\end{pmatrix} \,, \quad
\wt{m} :=\frac{kz_{\ast}}{R} \,.
\label{eq:DiracAdS2UV}
\end{align}
and, as seen from \eqref{sbasis2}, two components of each solution basis
have the same powers.
Thus, it is ambiguous to choose which component is set to be a non-normalizable mode
in formulating the GKP-Witten relations.\footnote{In other words, 
there is an ambiguity in the statement
that $|\ol{\chi}_+, \chi_-\ran$ can be an initial state for the standard quantization
(see \eqref{standard}).}
In order to avoid this ambiguity, let us diagonalize the Dirac equation,
\begin{align}
\zt\pd_{\zt}\wt{\Phi} &= 
\begin{pmatrix}
\nu_k & 0 \\ 0 & -\nu_k
\end{pmatrix}
\wt{\Phi} \,, \quad
\wt{\Phi} := V^{-1}\Phi \,, \quad 
V := (\mathbf{v}_+, \mathbf{v}_-) \,, \quad
\nu_k := \sqrt{(m^2+\wt{m}^2)R_2^2 - q^2e_d^2} \,,
\label{fptwist}
\end{align}
where $\mathbf{v}_{\pm}$ are eigenvectors of $U$ of eigenvalues $\pm\nu_k$,
and $V$ is a $2\times 2$ matrix. 
Note that we can always have $\det V=1$ by normalizing $\mathbf{v}_{\pm}$ appropriately.
Thus, the field redefinition $\Phi\to\wt{\Phi}$
is nothing but the twisting \eqref{twist} 
and hence we rewrite $\wt{\Phi}$ as $\Phi^{\sh}$. 
In this twist, the four coefficients in \eqref{twist} are given by\footnote{A convenient choice 
is given by $a=\frac{1}{\sqrt{2}} \frac{q e_d + \wt{m} R_2}{\nu_k}$, 
$b=\frac{1}{\sqrt{2}} \left( 1 - \frac{m R_2}{\nu_k} \right)$, $c=-\frac{1}{\sqrt{2}}$, 
$d=\frac{1}{\sqrt{2}} \frac{m R_2 + \nu_k}{q e_d + \wt{m} R_2} $.}
\begin{align}
\begin{pmatrix}
a & b \\ c & d 
\end{pmatrix}
= V^{-1} \,.
\end{align}
In terms of the {\it twisted fields}, 
it is possible to formulate the GKP-Witten relation of the AdS$_2 \times \mathbb R^{d-1}$ region unambiguously
in a parallel manner to Section~\ref{sec:beta} (in particular \eqref{standard}, \eqref{alternative}).

Now, let us obtain the double trace coupling at the AdS$_2$ boundary.
The expression $F(z)$ of \eqref{fchiuv}, 
which is defined {\it all over the ECBH background},
can also be expressed in terms of the second solution basis \eqref{sbasis2},
\begin{align}
F(z) 
= \frac{\phi_-(z) + \chi_{\UV}\vphi_-(z)}{\phi_+(z) + \chi_{\UV}\vphi_+(z)}
= \frac{\rho_-(\xi(z)) + \chi_{\IR}\psi_-(\xi(z))}{\rho_+(\xi(z)) + \chi_{\IR}\psi_+(\xi(z))} \,,
\quad ~
z\in\mbox{(the inner region)}\cup\mbox{(the outer region)} \,,
\label{fchiir}
\end{align}
where $\xi(z)$ is given in \eqref{xi}.
Note that $F$ is defined in terms of the original {\it untwisted} fields as \eqref{psi-ansatz}.
However, the appropriate double-trace coupling flow in the matching region
is the twisted one $F^{\sh}$.
Plugging \eqref{fchiir} into \eqref{twistfj}, we find
\begin{align}
F^{\sh}(\xi) =
\frac{\rho_-^{\sh}(\xi) + \chi_{\IR}\psi_-^{\sh}(\xi)}
{\rho_+^{\sh}(\xi) + \chi_{\IR}\psi_+^{\sh}(\xi)} \,, \quad\quad
\xi\in\mbox{(the matching region)} \,,
\label{ads2-twist}
\end{align}
where we defined the twisted solution basis  
following the twist of the Dirac fields \eqref{fptwist},
\begin{align}
\begin{pmatrix} \rho_+^{\sh}(\xi) \\ \rho_-^{\sh}(\xi) \end{pmatrix}
:= V^{-1}\begin{pmatrix} \rho_+(\xi) \\ \rho_-(\xi) \end{pmatrix} \,, \quad
\begin{pmatrix} \psi_+^{\sh}(\xi) \\ \psi_-^{\sh}(\xi) \end{pmatrix}
:= V^{-1}\begin{pmatrix} \psi_+(\xi) \\ \psi_-(\xi) \end{pmatrix} \,. \quad
\label{twisted-basis}
\end{align}
Near the AdS$_2$ boundary, i.e. using the approximation to Dirac's equation 
given in \eqref{eq:DiracAdS2UV}, the twisted solutions behave as
$\rho_-^{\sh} \sim \xi^{-\nu_k}$, $\psi_+^{\sh} \sim \xi^{\nu_k}$ and $\rho_+^{\sh} = \psi_-^{\sh} = 0$.
Therefore, the twisted double-trace coupling $F^{\sh}$ in the matching region behaves as
\begin{align}
F^{\sh}(\xi) \sim \chi_{\IR}^{-1}\xi^{-2\nu_k} \quad (\om\xi\sim 0) \,,
\label{FshIR}
\end{align}
so that the renormalized double trace coupling at the AdS$_2$ boundary 
is $\chi_{\IR}^{-1}$.\footnote{Note that since $A^{\rm IR} = \rho_+^\sh / \rho_-^\sh$ trivially vanishes, it is unnecessary to consider counterterms in this discussion.}

\paragraph{Relation between $\chi_{\UV}$ and $\chi_{\IR}$}
In order to relate $\chi_{\UV}$ and $\chi_{\IR}$, let us relate the
two solution bases \cite{Faulkner:2009wj,Faulkner:2010jy},
\begin{align}
\rho_{\pm} = a^{(+)}\phi_{\pm} + b^{(+)}\vphi_{\pm} \,, \quad
\psi_{\pm} = a^{(-)}\phi_{\pm} + b^{(-)}\vphi_{\pm} \,. \label{2bases}
\end{align}
Here $(\pm)$ implies the powers $\pm 2\nu_k$ 
in the asymptotic behavior given in \eqref{sbasis2}.\footnote{
$a^{(\pm)}, b^{(\pm)}$ are analogous to $a_{\pm}, b_{\pm}$ in \cite{Faulkner:2009wj}.
We use superscripts $(\pm)$ in order to avoid the confusion with the chirality indices,
that are written as subscripts $\pm$ in our paper.}
Thus, plugging \eqref{2bases} into \eqref{fchiir}, 
we find\footnote{One can see that this relation has the same structure as
the relation between $G_R$ and $\mathcal{G}_R$ given in \cite{Faulkner:2009wj},
which, in fact, is a special case of \eqref{chichi}
because, with the infalling boundary condition at the horizon, 
$\chi_{\IR}$ becomes $\mathcal{G}_R$ and $\chi_{\UV}$ becomes $G_R$ \cite{Iqbal:2009fd}.}
\begin{align}
\chi_{\IR}^{-1} = -\frac{a^{(-)}-\chi_{\UV}^{-1}b^{(-)}}{a^{(+)}-\chi_{\UV}^{-1}b^{(+)}} \,. \quad
\label{chichi}
\end{align}
$a^{(\pm)}, b^{(\pm)}$ can be expanded in $\om R$ (which is very small in the sense of \eqref{match}):
\begin{align}
a^{(\pm)}=a^{(\pm)}_0+(\om R)a^{(\pm)}_1+\cdots \,, \quad
b^{(\pm)}=b^{(\pm)}_0+(\om R)b^{(\pm)}_1+\cdots \,. \label{expab}
\end{align}
Using this expansion together with \eqref{chichi} and \eqref{FshIR}, 
we can obtain information about the behaviour of $F^{\sh}$ in the UV of the near horizon region.

\paragraph{Semi-holographic action}
Let us consider a holographic dual of the bulk theory in the inner region,
which has the emergent AdS$_2\times\mathbb{R}^{d-1}$ metric \eqref{nhads2}.
Let the AdS$_2\times\mathbb{R}^{d-1}$ boundary condition
be given by a wave-functional of the form \eqref{wvf}
at any point in the matching region, which can be interpreted as a cut-off 
of the AdS$_2\times\mathbb{R}^{d-1}$ boundary.
Now, we focus on the double trace coupling.
As shown above in \eqref{FshIR}, the renormalized double trace coupling is given by $\chi_{\IR}^{-1}$.
This means that the wave-functional at the AdS$_2\times\mathbb{R}^{d-1}$ boundary
in the continuum limit is given by 
\begin{align}
\Psi_{{\rm AdS}_2}[\eta_+^{\sh},\ol{\eta}_-^{\sh}]
:= \lan\eta_+^{\sh},\ol{\eta}_-^{\sh}|\Psi_{{\rm AdS}_2}\ran = \exp(-\ol{\eta}_-^{\sh} F^\sh \eta_+^{\sh} + \cdots) = \exp(-\ol{\eta}_{-R}^{\sh}\chi_{\IR}^{-1}\eta_{+R}^{\sh} + \cdots) \,.
\label{wvads2}
\end{align}
Here we have used a twisted basis $\lan \eta_+^\sh, \ol \eta_-^\sh|$
which are eigenstates of the twisted field 
$\wh{\Phi}^{\sh}$ $\equiv \wh{\tilde \Phi} := 
(\wh{\chi}^{\sh}_+, \wh{\chi}^{\sh}_-)^{\rm T}$ (see
\eq{fptwist}). $\ol{\eta}_{-R}^{\sh} \sim \xi^{-\nu_k} \ol{\eta}_-^{\sh}$ and $\eta_{+R}^{\sh} \sim \xi^{-\nu_k} \eta_+^{\sh}$ are renormalized fields.
The wavefunction 
$\Psi_{{\rm AdS}_2}$ is to be identified with
$\Psi$ of \eq{div1}, while $(\ol{\eta}_{-R}^{\sh},\eta_{+R}^{\sh})$ is to be identified with $(\ol{\tilde{\chi}}_+, \tilde{\chi}_-)$.

It can be verified numerically that for specific values of $k = k_F$, one has $b_0^{(+)} = 0$. 
In this case, writing $\vec k:=
k_F \hat k +  \Delta \vec  k_\parallel$ (where $\hat k$
is the unit vector along $\vec k$),  
$b^{(+)}$ can be expanded in  $\om$ and $ \Delta \vec k_\parallel$ as 
\begin{align}
b^{(+)} \sim c_t\om+c_i \Delta k_{\parallel i}+\cdots \,,
\label{bexpansion}
\end{align}
where $c_t, c_i$ are numerical constants (see
footnote \ref{ftnt:c-mu}). If we choose $\chi_{\rm UV} = 0$, 
then we obtain that
\SP{\label{eq:Fshpole}
	F^\sh \sim \chi_{\IR}^{-1}=b^{(-)}/b^{(+)}
}
has a single pole at the Fermi momentum $k_F$ for $\omega=0$. (For generic values of $\chi_{\rm UV}$, the pole shifts relative to $k_F$.) Therefore, \eqref{wvads2} is of the form \eqref{wvf} with $\tilde F$ replaced by $\chi_{\IR}^{-1}$, which is of the form \eqref{poleF} with $k^0_{\ast}= 0, | \vec k_\ast|= k_F$. As mentioned above, $(\ol{\eta}_{+R}^{\sh}, \eta^{\sh}_{-R})$ correspond to $(\ol{\tilde{\chi}}_+, \tilde{\chi}_-)$ in \eqref{semi}. Let us now study all this in more detail numerically.

\paragraph{Numerical study}

\begin{figure}[t]
\begin{center}
\begin{picture}(500,135)
\put(13,0){\includegraphics[height=4.5cm]{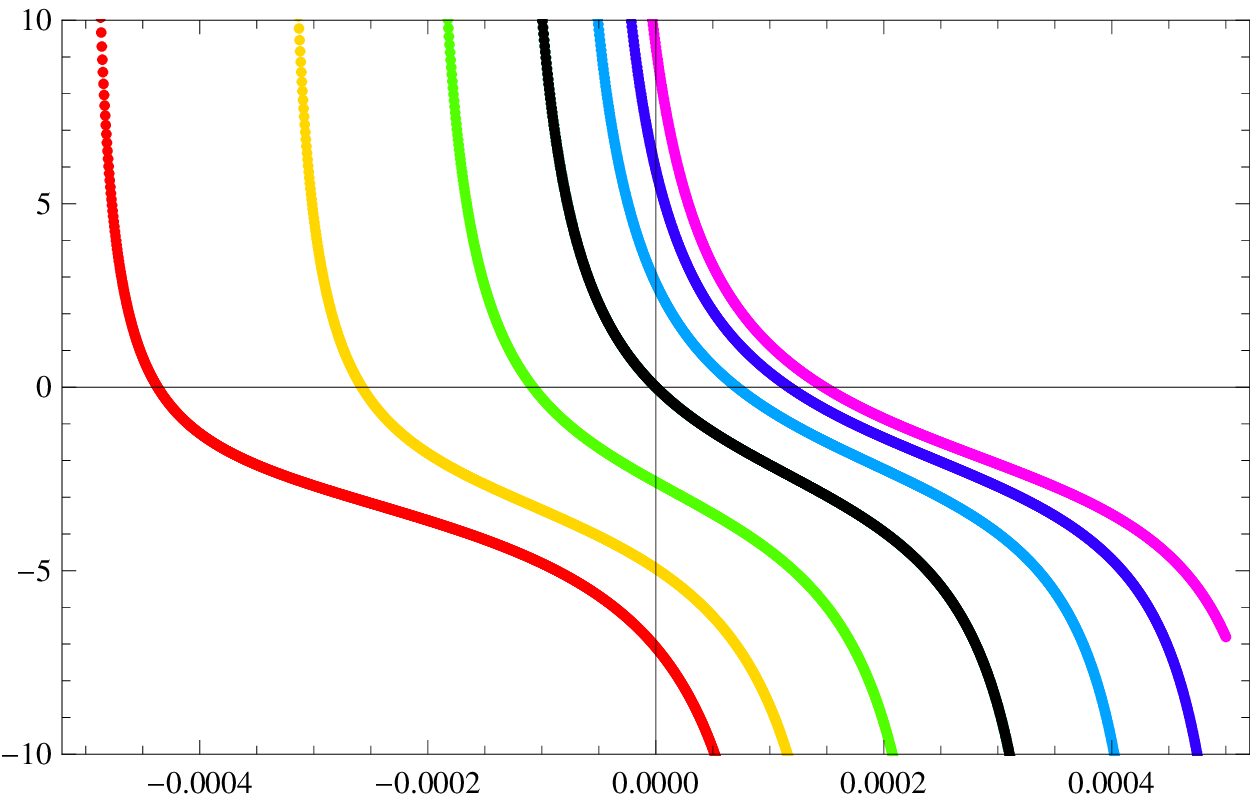}}
\put(243,0){\includegraphics[height=4.5cm]{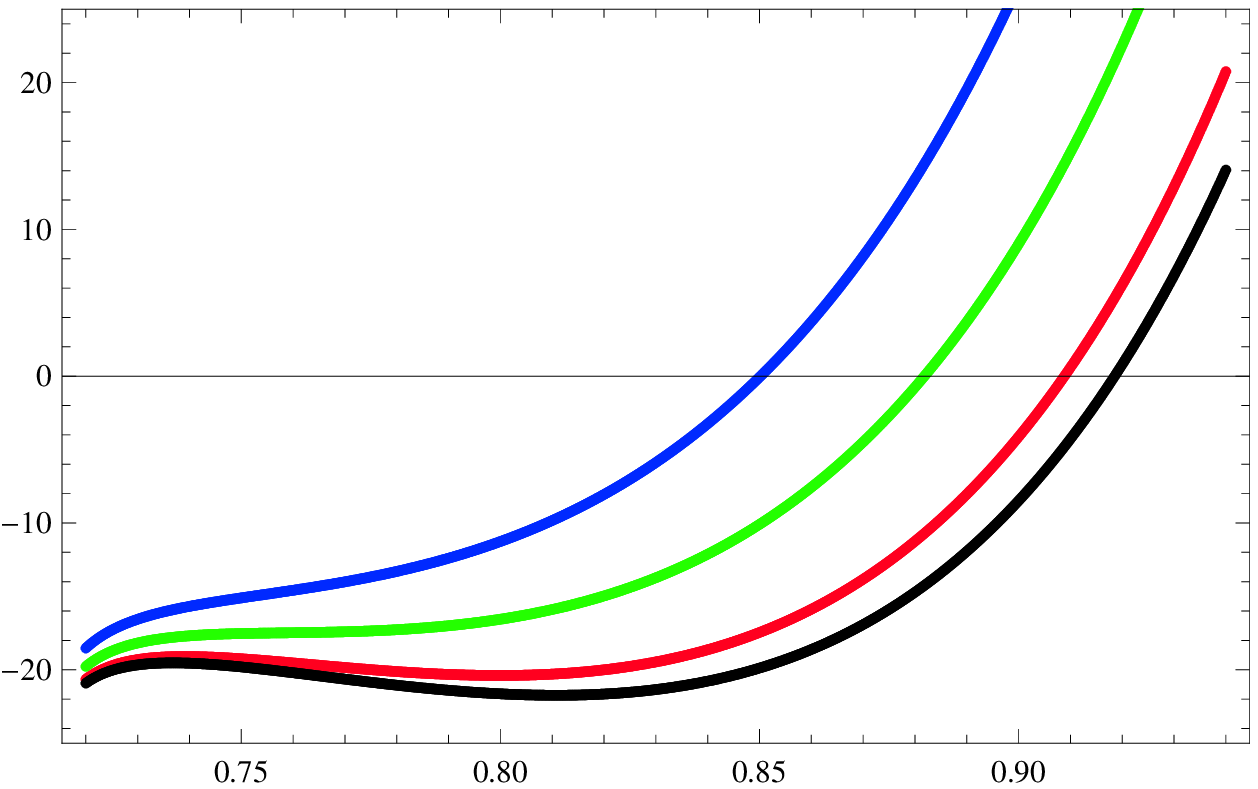}}
\put(218,6){$\omega$}
\put(453,6){$k$}
\put(20,135){${F^\sh_{\rm IR}}^{-1}$}
\put(250,135){${F^\sh_{\rm IR}}^{-1}$}
\end{picture}
\caption{Plots obtained numerically for $m=0$, $z_*=1$ $q=1$, $\mu=\sqrt{3}$. The left panel shows $F^{\sh -1}$ evaluated near the horizon ($r=\log \left( z^{-1} - z_*^{-1} \right)=-10$) as a function of $\omega$ for $\chi_{\rm UV}=0$ and $k=k_F=0.91853$ (black), $k=k_F \pm 0.005$, $k=k_F \pm 0.01$, and $k=k_F \pm 0.015$. The right panel shows $F^{\sh -1}$ evaluated near the horizon ($r=-10$) as a function of $k$ for $\omega=0$ and $\chi_{\rm UV}=0$ (black), $\chi_{\rm UV}=0.01$ (red), $\chi_{\rm UV}=0.04$ (green), and $\chi_{\rm UV}=0.08$ (blue).}
\label{fig:FPm0fig1}
\end{center}
\end{figure}

\begin{figure}[t]
\begin{center}
\begin{picture}(500,135)
\put(13,0){\includegraphics[height=4.5cm]{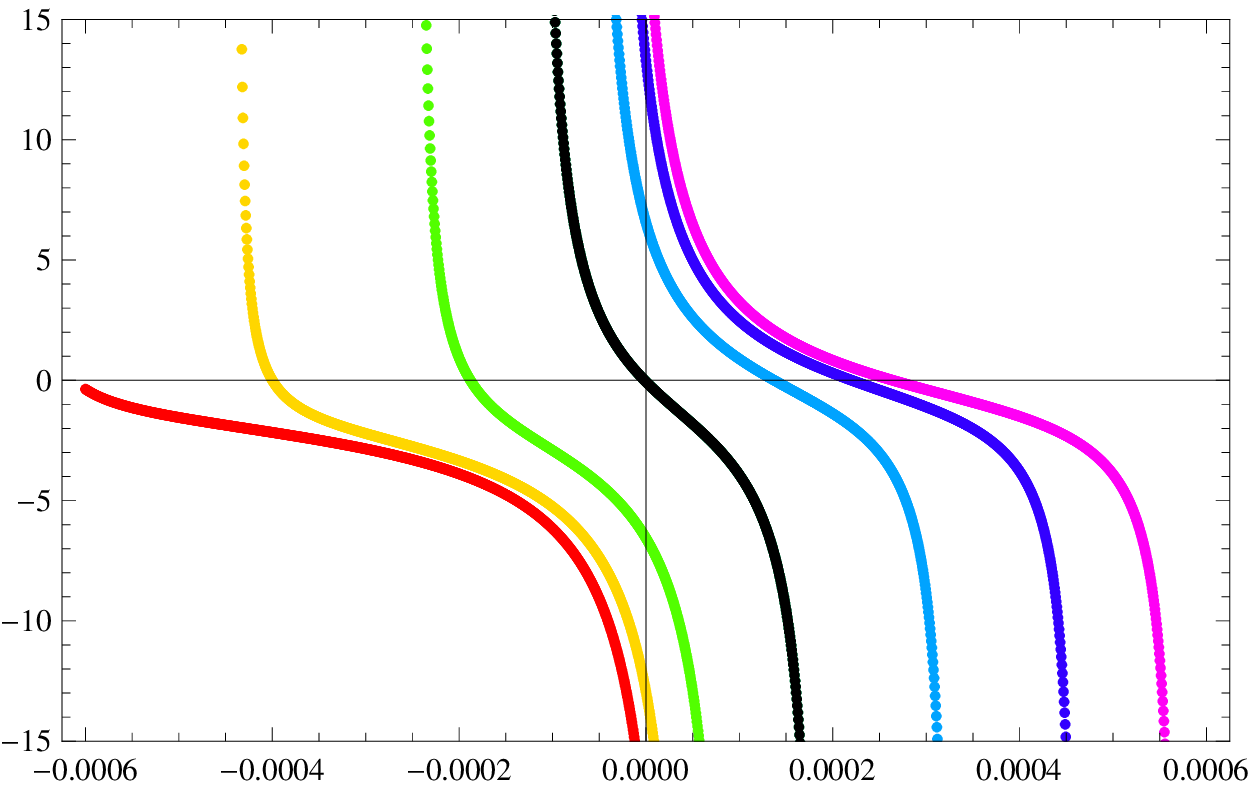}}
\put(243,0){\includegraphics[height=4.5cm]{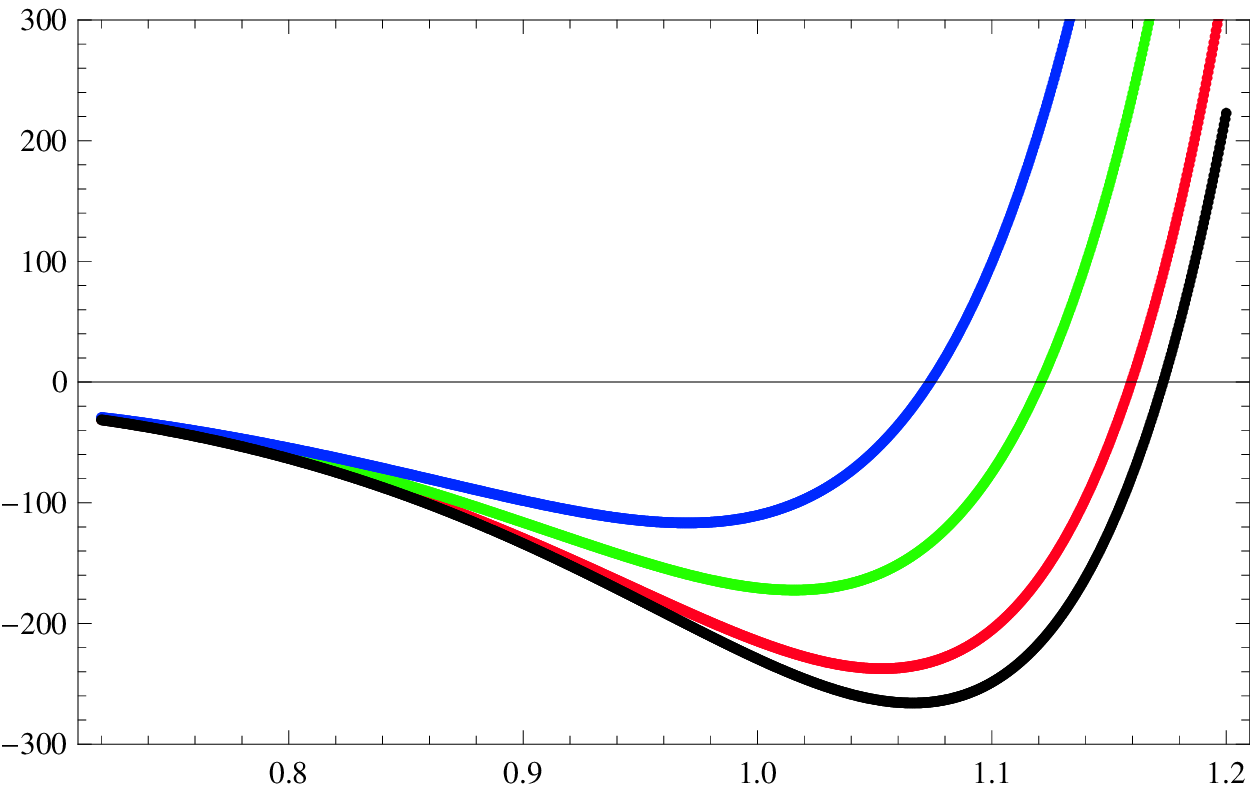}}
\put(218,6){$\omega$}
\put(453,6){$k$}
\put(20,135){${F^\sh_{\rm IR}}^{-1}$}
\put(250,135){${F^\sh_{\rm IR}}^{-1}$}
\end{picture}
\caption{Plots obtained numerically for $m=0.2$, $z_*=1$ $q=1$, $\mu=\sqrt{3}$. The left panel shows $F^{\sh -1}$ evaluated near the horizon ($r=\log \left( z^{-1} - z_*^{-1} \right)=-10$) as a function of $\omega$ for $\chi_{\rm UV}=0$ and $k=k_F=1.17309$ (black), $k=k_F \pm 0.001$, $k=k_F \pm 0.002$, and $k=k_F \pm 0.003$. The right panel shows $F^{\sh -1}$ evaluated near the horizon ($r=-10$) as a function of $k$ for $\omega=0$ and $\chi_{\rm UV}=0$ (black), $\chi_{\rm UV}=0.01$ (red), $\chi_{\rm UV}=0.04$ (green), and $\chi_{\rm UV}=0.08$ (blue).}
\label{fig:FPm02}
\end{center}
\end{figure}

\begin{figure}[t]
\begin{center}
\begin{picture}(500,135)
\put(13,0){\includegraphics[height=4.5cm]{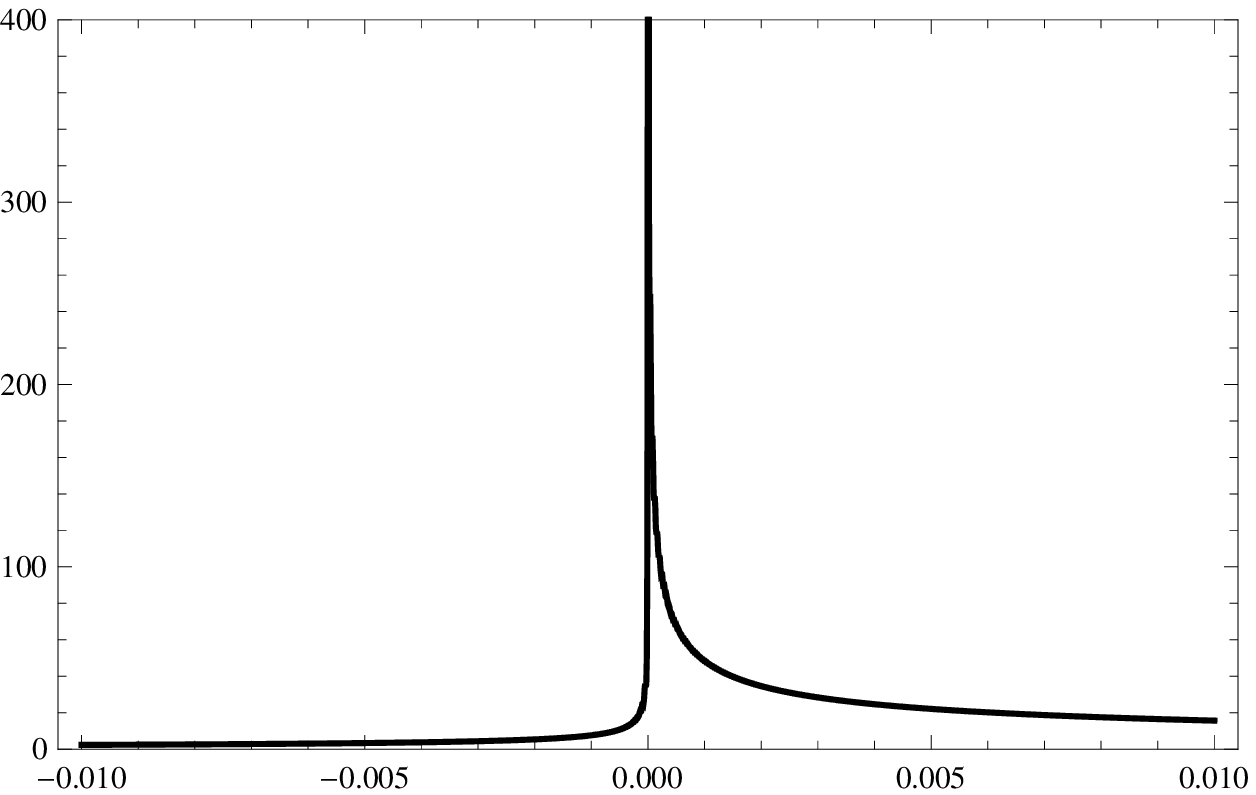}}
\put(243,0){\includegraphics[height=4.5cm]{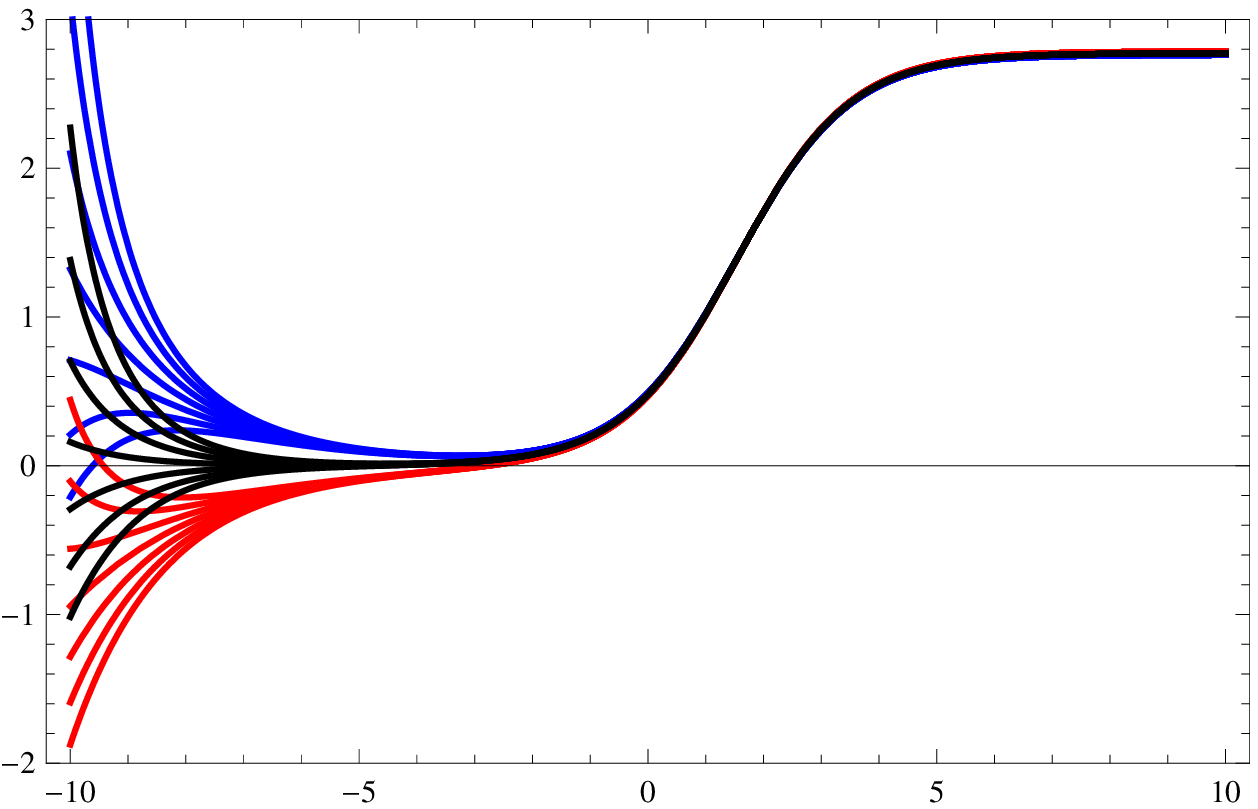}}
\put(218,6){$\omega$}
\put(445,6){$r$}
\put(18,135){${\rm Im} \, G_R$}
\put(250,135){${F^\sh}^{-1}$}
\end{picture}
\caption{Plots obtained numerically for $m=0$, $z_*=1$ $q=1$, $\mu=\sqrt{3}$. The left panel shows the imaginary part of the UV retarded Greens function $G_R$ for $k=k_F=0.91853$ as a function of $\omega$. The right panel shows $F^{\sh -1}$ as a function of $r = \log \left( z^{-1} - z_*^{-1} \right)$ for $\chi_{\rm UV}=0$ and different values of $k$ and $\omega$. The black lines are for $k=k_F$, while the red and blue lines are for $k=k_F-0.002$ and $k=k_F+0.002$, respectively. The values of $\omega$ fall between $\omega = \pm 4.5 \times 10^{-5}$.}
\label{fig:FPm0fig2}
\end{center}
\end{figure}

The left panels of Figure~\ref{fig:FPm0fig1} and Figure~\ref{fig:FPm02}, show the $\omega$-dependence of ${F^{\sh -1}}$ evaluated in the IR for $\chi_{\rm UV} = 0$ and different values of $k$. As can been seen, we obtain the characteristic linear behaviour anticipated by \eqref{eq:Fshpole}. The right panels of Figure~\ref{fig:FPm0fig1} and Figure~\ref{fig:FPm02}, show the $k$-dependence of $F^{\sh -1}$ evaluated in the IR for $\omega = 0$ and different values of $\chi_{\rm UV}$. Making $\chi_{\rm UV}$ non-zero has the effect of shifting the location of the pole of $F^\sh$. The left panel of Figure~\ref{fig:FPm0fig2} shows that the imaginary part of the AdS$_4$ retarded Greens function has a peak for $k=k_F$. This gives an independent verification of the location of $k_F$, which agrees with the plots of Figure~\ref{fig:FPm0fig1} for $\chi_{\rm UV} = 0$.

The right panel of Figure~\ref{fig:FPm0fig2} shows the flow of $F^{\sh -1}$ as a function of the radial coordinate, and $\chi_{\rm UV} = 0$. $F^{\sh -1}$ starts from the $F^{\sh -1}=b/d$ in the UV (corresponding to $F = \infty$ or alternative quantization). For $k=k_F$ and $\omega=0$, the flow approaches $F^{\sh -1} = 0$ or the alternative quantization in the near horizon region. Non-zero values of $\omega$ eventually drive the flow away from $F^{\sh -1} = 0$ in the deep IR. As $k$ is made to differ from $k_F$ by a small amount, the flow only stays near the alternative fixed point of AdS$_2 \times \mathbb R^{d-1}$ for a while, until it starts flowing towards the attractive standard fixed point at $F^{\sh -1} = \infty$. This shifts the value of $\omega$ for which $F^{\sh -1}$ hits a zero.

\vspace{5ex}

\section{\label{sec:FT}Field theory summary}

In this section, we will transcribe our findings in the previous
sections in field theoretic terms.

\subsection{Scalar operators}

We will denote the matrix field theory, implicit on the RHS of
\eq{slicing-ft}, as follows
\begin{align}
Z[\ep]= \int {\cal D}M_\ep~ \exp\left[S[M] + \int_k h(k,\ep) O(-k)
    + \int_k \frac12 g(k,\ep) O(k) O(-k) + C(\ep)\right], 
\label{scalar-operator}
\end{align}
where $M$ is a $d$-dimensional matrix field, and $O$ is a certain
single trace scalar operator. Typically, a perturbation by $O$ would
mix with other operators under Wilsonian RG; the generalization to
multiple scalar operators (dual to multiple bulk scalar fields) is
straightforward if all bulk scalars are given by quadratic actions
such as in footnote \ref{ftnt:scalar action}. However, for simplicity,
we will assume a situation in which the operator $O$ here does not mix
with others under RG flow, and under $\ep \to \ep'$, the partition
function $Z$ can be kept invariant by appropriate change of the
couplings $h, g$; we have also kept an $\ep$-dependent zero-point
energy $C(\ep)$.

The field theory couplings $g(k)$ and $h(k)$ (in the standard
quantization) are related to $f(k),
J(k)$ of \eq{s-uv} as
\begin{align}
g(k)= 1/f(k),\quad h(k)=J(k)/f(k) \,.
\label{g-f-relation}
\end{align}
Using \eq{scalar-flow} we get the beta-functions for $g, h$.

\paragraph{Example 1}

We will consider $S[M] \equiv S_1[M]$ which admits a pure AdS$_{d+1}$
dual.  The beta-functions of $h, g$ (for the $k=0$ mode) are, in this
case, given by (recalling $\ep \del/\del\ep = - \L
\del/\del \L$)
\begin{align}
- \b_g = \dot g = 1 - d g - m^2 g^2, \quad 
- \b_h= \dot h= - (d + m^2 g) h \,.
\label{g-h-beta}
\end{align}
The double trace flow has two fixed points: \\ (a) $g_* = 1/\Delta_+$;
this CFT corresponds to the standard CFT. {\em Proof}: Recall that for a CFT
deformed by a coupling $\int g {\mathbb O}$, the scaling dimension
$\g$ of $ {\mathbb O}$, also denoted as $[ {\mathbb O}]$, is related
to the $\b$-function of $g$ as follows
\begin{align}
\g = [ {\mathbb O}] = \del \b_g / \del g.
\label{beta-gamma}
\end{align}
For the double trace coupling under discussion, the fixed point value
is $g_*$, hence the deformation from the CFT is $\int_k \delta g
{\mathbb O} $, where $\delta g= g-g_*$ and ${\mathbb O}= O(k)
O(-k)$. From \eq{g-h-beta}, $\b_{\delta g}= 2\nu \delta g + m^2 \delta
g^2$, hence $\g=2\nu + m^2 \delta g$. Thus, at the fixed point
$[O(k)O(-k)] \equiv \g=2\nu$; hence $[O(x)^2]= d+2\nu= 2\Delta_+$.  By
large $N$ factorization, $[O(x)]= \sqrt{[O(x)^2]} = \Delta_+$,
implying standard quantization. \\ 
(b) $g_*= 1/\Delta_-$:
alternative quantization, in the Klebanov-Witten
window $0 \le \nu \le 1$ \cite{Klebanov:1999tb}. The proof is similar
to the above case. In this case
\begin{align}
\beta_{\delta g}= - 2\nu \delta g + m^2 \delta g^2.
\label{exact-beta}
\end{align}
Note that outside the KW window, the fixed point $g_*= 1/\Delta_-$,
although it formally exists, does not represent a valid quantization.
See Table \ref{table-anomalous} for a summary of the two fixed points.

\paragraph{Comparison with a toy model with a Wilson-Fisher fixed point}
In \cite{EIM2} we consider a $d=4-\ep$ dimensional free matrix field
theory $S[M]= \int d^dx \frac12 {\rm Tr} (\del M)^2 $, perturbed by
$\int h' O(x) + \frac12 g' O(x)^2$, with $O(x)= {\rm Tr} M(x)^2$. We
find
\begin{align}
\b_{g'} = - \ep g' + \a^2 g'^2  + ... \,,
\label{series-beta}
\end{align}
where $\a^2$ is a numerical constant \footnote{We note that in this
  field theory toy model, the other marginal operator Tr $M(x)^4$, if
  absent initially, does not get generated by the flows of the
  operators $O(x), O(x)^2$.}. The free field UV fixed point at $g'=0$
is superficially comparable with the UV CFT for alternative
quantization described by the beta-function \eq{exact-beta}.  The
point to stress here is that the beta-function \eq{exact-beta},
obtained by holographic means, is exact even for strong coupling.

\paragraph{Example 2}

The field theory $S[M] \equiv S_2[M]$ has a global $U(1)$ symmetry,
and admits a charged black hole \eq{bh-metric} dual. 
The existence of a normalizable $\delta g_{tt}$, at least for small
$z= \ep$, can be interpreted as an expectation value $\lan T_{tt}
\ran$. The running of $\delta g_{tt}$, interpreted as a flow of $\lan
T_{tt} \ran$ \footnote{Similar interpretations have been used by, {\em
    e.g.}  \cite{Iqbal:2008by, Jain:2010ip}, for two-point functions
  of currents or stress tensors in the context of cut-off dependence
  of transport coefficients.}, can be read off from the exact
dependence of $g_{tt}$ on $z$ as given by $H(z)$ in
\eq{bh-metric}. With this interpretation of the background geometry in
mind, we again perturb the theory as in \eq{scalar-operator} and
obtain the beta-functions for $g,h$ by substituting \eq{g-f-relation}
in \eq{f-flow}, \eq{scalar-flow} and \eq{r-rgf-bh}.

Along with $\lan T_{tt} \ran$ there is also a $\lan J_t \ran$ caused
by the presence of the U(1) gauge field $A_t$ in the dual
geometry. The fact that the dual charged black black hole is extremal
means that $\lan J_t \ran$ is determined by $\lan T_{tt} \ran$ and
need not be considered separately. Extremality of the dual also
implies the existence of 2 additional fixed points as $z=\ep \to
z_\ast$ which are the AdS$_2$ analogs of the 2 fixed points discussed
in Example 1: the non-zero values of $\lan T_{tt} \ran$, $\lan J_t
\ran$ are interpreted in terms of a new $d$-dimensional CFT on
$\mathbb R^1 \times \mathbb R^{d-1}$. The RG flows and the four fixed
points are described in Section~\ref{sec:scalar} and in Figure
\ref{fig-scalar-flow}.

\subsection{\label{sec:FFT}Fermionic operators}

The difference with \eq{scalar-operator} is that this time
we perturb the theory by a (single-trace) fermionic operator
$O_a$ (where $a$ is a Dirac index)
\begin{align}
Z[\ep]= \int \!\! {\cal D}M_\ep~ \exp\left[S(M,\ep) + 
\int_k \ol h_a(k,\ep) O_a(k) + \int_k \ol O_a(k)  h_a(k)
    + \int_k \frac12 \ol O_a(k)  g_{ab}(k,\ep) O_b(k)\right] \,.
\label{fermionic-operator}
\end{align}
Comparing with \eq{double-trace-fermion} we find that in the spinorial
basis we use
\[
O_a(k) g_{ab}(k,\ep) O_b(k) = \ol{O}_-F(\ep)O_+ \,, \quad
\ol h_a(k,\ep) O_a(k) = \ol{O}_-J_-(\ep) \,, \quad
\ol O_a(k)  h_a(k)= \ol{J}_+(\ep)O_+ \,. 
\]
In other words, double trace and single trace couplings are given
directly in terms of $F, J$, whose running is given by
\eq{fermionic-flow}.

\paragraph{Example 1}

As in the bosonic case, the field theory $S_1[M]$ is assumed to admit
a pure AdS$_{d+1}$ dual. The beta-function for the double trace
coupling is given by \eq{eq:flowFAdS4}, or alternatively by
\eq{eq:flowFAdS42}. In terms of the latter, the zero-momentum coupling
is $f_0$ which, according to the flow equations \eq{eq:betaf},
\eq{eq:betag}, flows to the IR fixed point $f_{0,*}=0$ from the UV
fixed point $f_{0,*}=\infty$. 

The nature of the fixed points can be found as in the bosonic case
above. Thus, near $f_{0,*}=0$, the double trace coupling is $f_0 \ol
O_- O_+$, hence by \eq{beta-gamma}, we have $\left[\kern2pt \ol O_-(k) O_+(k)
  \right] = \del \b_{f_0}/\del f_0= 2m$ (recall $\ep \del f_0/\del\ep
= - \L \del f_0/\del \L \equiv \beta_{f_0}$).  In position space we
obtain $\left[\kern2pt \ol O_-(x) O_+ (x)\right] = d + 2m$. Using large $N$
factorization, we obtain $\g_{\ol O_-}= \g_{O_+}= d/2+ m \equiv
\Delta_+$, which says that $f_{0,*}=0$ corresponds to standard
quantization. The fact that $\left[\kern2pt \ol O_- O_+\right] > d$ 
implies that it is an
irrelevant operator, hence the standard fixed point is an IR fixed
point; this agrees with the attractive nature of the fixed point, as
found in \eq{eq:betaf}.

The nature of the fixed point $f_{0,*}=\infty$, or equivalently
$g_{0,*}=0$ can be similarly found. The coupling now is $g_0 \ol O_+
O_-$ which, using \eq{beta-gamma}, gives $\left[ \ol O_+ O_- \right]
=-2m$, or $\left[ \ol O_+(x) \right]$ $=\left[ O_- (x)\right] = d/2 -
  m$, which signifies alternative quantization. The fact that $\ol O_+
  O_-$ turns out to be a relevant operator accords with the fact that
  the alternative fixed point, by \eq{eq:betag}, is repulsive, and is
  hence a UV fixed point.

\paragraph{Example 2}

As in the bosonic case, the field theory $S_2[M]$ is assumed to admit
an extremal charged black hole dual, with scale-dependent $\lan T_{tt}
\ran$, $\lan J_t \ran$.  The beta-function for the double trace
coupling is given by \eq{eq:floweqECBH}. As in the bosonic case, the
double trace flow, together with the flow of $\lan T_{tt} \ran$,
connect the two fixed points at very high energies (the boundary
AdS$_{d+1}$ region) to two fixed points in the IR (the near-horizon
AdS$_2$ region), see Figure \ref{fig:fermionicECBH}.

\subsection{\label{sec:random-source}Emergence of new 
(bosonic/fermionic) degrees of freedom}

As discussed in Section~\ref{sec:FP}, the double trace coupling $g(k)$
in Example 2 above can develop a pole singularity at the
Fermi surface\footnote{ For simple perturbations around standard free field
  theories, operators with zero momentum are more relevant than those
  with finite momenta (higher derivatives); the fact that the double
  trace coupling at a finite momentum blows up is due to the black
  hole background which represents strong coupling physics.}  This
implies emergence of new light degrees of freedom in the field theory.
We can trace their origin as follows.

\paragraph{Low energy theory $\to$ random (dynamical) source}
Equations \eq{slicing-ft} and \eq{ep0-ep-f} can be interpreted as follows:
holographic version of Wilsonian RG can be regarded as appearance of
new dynamical degrees of freedom, in the form of random
(fluctuating) sources \footnote{The field theory variables
are generically denoted as a matrix field $M(x)$, as below
\eq{interim}. $O[M]$ is a single trace operator; the
Equation \eq{random-source} is valid if this operator does
not mix with others under RG. More generally, one should
include all single trace operators $O_n[M]$ and the
corresponding sources $J_n$.}
\begin{align}
Z[\ep_0]=\int DM_{\ep_0} \exp\left( S_0[M] + \int J_0 O[M] \right) = Z[\ep] =
\int DM_\ep~DJ \exp\left({\cal S}[J]\right) \exp\left( S[M] + \int J O[M]
\right).
\label{random-source}
\end{align}
Generically we can integrate out the random sources, yielding a low
energy effective action in terms of the original field
variables. However, in special situations (e.g.  on the Fermi surface,
as discussed above in Section~\ref{sec:FP-gist}), $J(k)$ becomes
`massless' and can be identified as an emergent degree of freedom
(bosonic or fermionic, depending on whether $O$ is bosonic of
fermionic), which must be kept in the dynamics to keep a local
description of the theory; see \eq{semi} as an example.

\paragraph{Comparison with Gross-Neveu model and QCD}

The essence of holographic Wilsonian RG, as captured by
\eq{random-source}, can be compared with the Nambu$-$Jona-Lasinio
(NJL) approach to QCD (see, e.g. \cite{Dhar:1983fr,Dhar:1985gh}, or in
a simpler setting, with the Gross-Neveu model \cite{Gross:1974jv}. By
introducing a dynamical source term the partition function of the
latter model can be written as
\begin{align}
Z=  \int D\ol\psi~D \psi~ \exp  \left[\int d^2 x\left( 
\ol\psi \del \!\!\!/
\psi + \frac{g^2}{2} \left( \ol\psi \psi \right)^2\right) \right]
=  \int DJ~ D\ol\psi~D \psi~ \exp\left[- \int \frac{J^2}{2g^2}\right]   
\, \exp\left[\int~ \left(\ol\psi 
\del \!\!\!/ \psi + J~ \ol\psi \psi \right) \right].
\label{g-n}
\end{align}
The fermions transform in the fundamental representation of
$U(N_f)$. In the second expression for the partition function,
integrating out the fermions gives an effective action $S_{\rm
  eff}[J(k)]$.  The source field $J(x) \sim \ol \psi(x) \psi(x)$ is a
rough analog of mesons in the QCD problem
\cite{Dhar:1983fr,Dhar:1985gh}. The last expression of \eq{g-n} is
analogous to the last expression of
\eq{random-source}; the fluctuating source of
\eq{random-source} (for bosonic $J(k)$) can be identified with the
`meson' fields.  The effective action $S_{\rm eff}$ can be computed
exactly at large $N_f$, and exhibits a symmetry breaking potential.

A massless $J(k)$ field, in such contexts, can emerge as follows:
suppose that 
the fermions in \eq{g-n} also transform under a second flavour group
$U(N_f')$, under which the mesons transform nontrivially.  In this
case $S_{\rm eff}[J(k)]$ is invariant under $U(N_f'$); in the
symmetry broken phase, some of the source fields $J(k)$ can be
identified as Goldstone bosons (pions). See \cite{Nickel:2010pr}
for discussions along these lines. 
It would be interesting to find an analog of the above picture
for emergent massless fermions; see a related discussion 
in \cite{Sachdev:2010uz} in the context of the `FL$*$' phase .

\section{\label{sec:future}Concluding remarks}

In this work, we have described a framework of how to carry out the
program of holographic Wilsonian RG (hWRG) with fermions. More details
will appear in \cite{EIM2}. As mentioned in the Introduction, we found
a precise meaning of the existence of a Fermi surface in terms of
singular effective double trace coupling in the AdS$_2$ region; we
also found a dynamical origin of the semi-holographic origin of
Faulkner and Polchinski.

We end with a few open questions and generalizations:

\begin{itemize}

\item
It is of obvious interest to identify specific strongly coupled field
theories which have holographic duals of the kind described in this
paper. Work in this direction is in progress \cite{EIM2} (see the
discussion around \eq{series-beta} for an example).

\item
The random source interpretation of hWRG \eq{random-source} gives an
insight into Wilsonian RG of strongly coupled field theories
\cite{Lee:2009ij,Lee:2010ub}.  Random source models have an extensive
literature in statistical mechanics; an example is the so-called
compressible Ising model with random couplings on the links (see
\cite{Baker-Essam}, e.g.). It would be interesting to see if these
models too can have emergent degrees of freedom as we found here.

\item
The formalism developed in this paper is applicable to any
asymptotically AdS background, especially those which are
translationally invariant in the boundary directions.  The discussions
of RG flows can be easily generalized to the case of non-extremal
black hole where the theory at IR is massive; for nearly extremal
black holes, our discussion on semi-holography generalizes rather
easily, including the appearance of an emergent fermion at the AdS$_2$
boundary.  Our formalism can also be applied to Lifshitz geometries \cite{Goldstein:2010aw, Hartnoll:2010gu}
with different dynamical exponents in the UV and IR \cite{EIM2}. It
would also be interesting to see if we can apply hWRG in reverse \cite{Heemskerk:2010hk}:
starting out from a IR conformal field theory and flowing towards the UV.  The
semi-holographic viewpoint does not require the existence of a UV
fixed point; in particular the UV theory is allowed to be a lattice
theory. It would be obviously important to complete such a picture of
hWRG.

\item
In addition to the flow of couplings, it is important to compute the
flow of correlation functions; in particular, it would be interesting to
understand RG flows of transport coefficients using hWRG (see, {\em
  e.g.} \cite{Iqbal:2008by, Jain:2010ip, Eling:2011ct}).

\end{itemize}

\section*{Acknowledgement} 

We would like to thank Kedar Damle, Avinash Dhar, Deepak Dhar, Sachin Jain, Yutaka Matsuo,
Shiraz Minwalla, Maurizio Piai, Mukund Rangamani, Subir Sachdev, Dam Son, 
Tadashi Takayanagi, and Sandip Trivedi for discussions. 
We would like to thank Avinash Dhar, Shiraz Minwalla and
Mukund Rangamani for helpful comments on the manuscript.
We would also like to thank the
organizers of the meetings \cite{confs}, in TIFR, Esfahan and Milos, for
invitation to present this work and and the participants for
stimulating discussions on various aspects of the subject.

\appendix

\section{Notations} \label{notation}
\subsection{Miscellaneous notations}
\paragraph{coordinates and momenta}
\begin{align}
x^i &:= ( x^1, x^2, \cdots, x^{d-1} ) \,, \quad
x^{\mu} := ( t, x^i ) \,, \quad
x^M := ( z, x^{\mu} ) \,, \nonumber\\
k_i &:= ( k_1, k_2, \cdots, k_{d-1} ) \,, \quad
k_{\mu} := ( \vep\om, k_i ) \,, \quad
k^2 := g^{\mu\nu}k_{\mu}k_{\nu} \,, \quad
\bsk^2 := \eta^{\wh{\mu}\wh{\nu}}k_{\mu}k_{\nu} = \vep\om^2 + k_ik_i \,, \nonumber
\end{align}
where $\vep=1(-1)$ for Euclidean (Lorentzian) metric. 

\paragraph{metric, covariant derivative and spin connections}
\begin{align}
ds^2 
&= g_{MN}dx^Mdx^N 
= \eta_{\wh{a}\wh{b}}\te^{\wh{a}}\te^{\wh{b}} \,, \quad
\eta_{\wh{a}\wh{b}} := {\rm diag}\{ 1, \ep, 1, \cdots, 1 \} \,, \quad
g := \det{(g_{MN})} \,, \nonumber\\
D_M\psi &:= \pd_M\psi + \frac{1}{4}\om_{M\wh{a}\wh{b}}\Ga^{\wh{a}\wh{b}}\psi - iqA_M\psi \,, \quad
\ol{D_M\psi} := 
\pd_M\ol{\psi} - \frac{1}{4}\om_{M\wh{a}\wh{b}}\ol{\psi}\Ga^{\wh{a}\wh{b}} + iqA_M\ol{\psi} \,, \nonumber\\
\om^{\wh{a}\wh{b}} &= \om_M{}^{\wh{a}\wh{b}}dx^M \,, \quad
d\te^{\wh{a}} + \om^{\wh{a}}{}_{\wh{b}} \wg \te^{\wh{b}} = 0 \,.
\end{align}

\paragraph{Fourier transformations}
\begin{align}
\psi(z,x) &:= \int\!\frac{d^dk}{(2\pi)^d}e^{ikx}\,\psi(z,k) \,, \quad
\ol{\psi}(z,x) := \int\!\frac{d^dk}{(2\pi)^d}e^{-ikx}\,\ol{\psi}(z,k) \,, \quad
kx := k_{\mu}x^{\mu} = \vep\om t + k_ix^i \,. \nonumber
\end{align}

\subsection{Dirac action} \label{gen_act}
We give a general notation which applies to both Euclidean and Lorentz signatures.
The Dirac action is
\begin{align}
S := \cN\!\int\!dzd^dx\sqrt{\vep g}\,(\ol{\psi}\Ga^MD_M\psi - m\ol{\psi}\psi) \,. \nonumber
\end{align}
where $g:=\det(g_{MN})$.\footnote{
$g_{tt}>0$ for Euclidean case and $g_{tt}<0$ for Lorentzian case}
A numerical factor $\cN$ is chosen so that the path integral representation is $\int[d\psi]e^{-S}$
both for Euclidean and Lorentzian signature.
If we compare with the Dirac actions given in \cite{Iqbal:2009fd},
our $\cN$ is related to $\cN$ in \cite{Iqbal:2009fd} (which we write as $\cN_{\rm IL}$), 
by $\cN = \cN_{\rm IL}$ for the Lorentzian metric, 
and $\cN = -\cN_{\rm IL}$ for the Euclidean metric.
In this paper, we set $\cN=1$ just for notational simplicity\footnote{
In fact, $\cN_{\rm IL}$ cannot be arbitrary; 
it is constrained by the unitarity of the bulk theory.} 
except the next section.
In order to revive $\cN$, we just have to replace $\ka^{-2}$ by $\ka^{-2}\cN$.

\section{Generalized coherent states} \label{g_coh}
In this section, we will investigate the state space of a quantized Dirac fermion in the general background.
For simplicity, we will omit spinor indices and momenta variables 
and momentum integral symbols.
This is equivalent to picking up a Hilbert space 
with one specific momentum and one specific spinor component.
This is consistent because the anti-commutation relations we will use 
have no mixing between different momenta and spinor components.
The starting point is the anti-commutation relation
\begin{align}
\{ \hchi_{\pm}, \, \pm\cN_d{\hochi}_{\pm} \} = 1 \,, \quad\quad
\cN_d := \ka^{-2}(2\pi)^{-d}\cN \,.
\end{align}
First, we construct a Fock space with regarding $\hchi_{\pm}$ as annihilation operators
and $\pm\cN_d{\hochi}_{\pm}$ as creation operators starting from a ground state $|0,0\ran\ran$,
which is annihilated by $\hchi_{\pm}$.
The basis of the Fock space is 
$\{ |0,0\ran\ran, |1,0\ran\ran, |0,1\ran\ran, |1,1\ran\ran \}$ with
\begin{align}
|1,0\ran\ran := \cN_d{\hochi}_+|0,0\ran\ran \,, \quad
|0,1\ran\ran := -\cN_d{\hochi}_-|0,0\ran\ran \,, \quad
|1,1\ran\ran := \cN_d{\hochi}_+(-\cN_d{\hochi}_-)|0,0\ran\ran \,.
\end{align}
Next, the dual basis is defined so that 
\begin{align}
\lan\lan m,n|p,q \ran\ran = \de_{mp}\de_{nq} \,.
\end{align}
The dual basis is constructed from the dual ground state with $\lan\lan 0,0|0,0 \ran\ran = 1$,
\begin{align}
\lan\lan 1,0| := \lan\lan 0,0|\hchi_+ \,, \quad
\lan\lan 0,1| := \lan\lan 0,0|\hchi_- \,, \quad
\lan\lan 1,1| := \lan\lan 0,0|\hchi_-\hchi_+ \,.
\end{align}
We now define ``generalized coherent states'', which satisfy
\begin{alignat}{4}
{\hochi}_+|\ol{u}, v\ran &= \ol{u}|\ol{u}, v\ran \,, &\qquad
\hchi_-|\ol{u}, v\ran &= v|\ol{u}, v\ran \,, &\qquad
\hchi_+|u, \ol{v}\ran &= u|u, \ol{v}\ran \,, &\qquad
{\hochi}_-|u, \ol{v}\ran &= \ol{v}|u, \ol{v}\ran \,, \nonumber\\
\lan u, \ol{v}|\hchi_+ &= \lan u, \ol{v}|u \,, &\qquad
\lan u, \ol{v}|{\hochi}_- &= \lan u, \ol{v}|\ol{v} \,, &\qquad
\lan \ol{u}, v|{\hochi}_+ &= \lan \ol{u}, v|\ol{u} \,, &\qquad
\lan \ol{u}, v|\hchi_- &= \lan \ol{u}, v|v \,. 
\end{alignat}
They are explicitly constructed as follows\footnote{
We often see a form like $\cN_d(\ol{a}b+\ol{c}d)$ in the exponent.
In the field theory, spinor indices and momentum dependence are recovered as follows
\begin{align}
\cN_d(\ol{a}b+\ol{c}d) \longrightarrow 
\cN_d\int\!\!d^dk(\ol{a}(k)b(k)+\ol{c}(k)d(k))
=\cN\int_k (\ol{a}(k)b(k)+\ol{c}(k)d(k)) \,.
\end{align}}
\begin{align}
|\ol{u}, v\ran &:= e^{\cN_d(-\ol{u}\hchi_+-{\hochi}_-v)}|1,0\ran\ran \,, \quad
|u, \ol{v}\ran := e^{\cN_d({\hochi}_+u+\ol{v}\hchi_-)}|0,1\ran\ran \,, \nonumber\\
\lan u,\ol{v}| &:= \lan\lan 1,0|e^{\cN_d(-{\hochi}_+u -\ol{v}\hchi_-)} \,, \quad
\lan \ol{u},v| := \lan\lan 0,1|e^{\cN_d(\ol{u}\hchi_++{\hochi}_-v)} \,.
\end{align}
Inner products of generalized coherent states are
\begin{alignat}{2}
\lan p,\ol{q}|\ol{u},v \ran &= e^{\cN_d(-\ol{u}p-\ol{q}v)} \,, &\qquad 
\lan \ol{p}, q|u,\ol{v} \ran &= e^{\cN_d(\ol{p}u+\ol{v}q)} \,, \label{inpro} \\
\lan p,\ol{q}|u,\ol{v} \ran &= \cN_d\de(p-u)\de(\ol{q}-\ol{v}) \,, &\qquad
\lan \ol{p},q|\ol{u},v \ran &= \cN_d\de(\ol{p}-\ol{u})\de(q-v) \,. \nonumber
\end{alignat}
Generalized coherent states satisfy the completeness relations,
\begin{align}
\wh{1} 
&= \frac{1}{\cN_d^2}\int\! d\ol{u}dvdud\ol{v} \, |\ol{u},v \ran\lan u,\ol{v}|e^{\cN_d(\ol{u}u+\ol{v}v)}
= \frac{1}{\cN_d^2}\int\! dud\ol{v}d\ol{u}dv \, |u,\ol{v} \ran\lan \ol{u},v|e^{\cN_d(-\ol{u}u-\ol{v}v)}
\,, \label{comp1} \\
\wh{1}
&= \frac{1}{\cN_d}\int\! dvd\ol{u} \, |\ol{u},v \ran\lan \ol{u},v|
= \frac{1}{\cN_d}\int\! d\ol{v}du \, |u,\ol{v} \ran\lan u,\ol{v}| \,. \label{comp2}
\end{align}
\paragraph{Proof of the replacement rule of the Schr\"{o}dinger equation}
Here we give a derivation of the replacement rules \eqref{replace} of the Schr\"{o}dinger equation.
From the inner product above, we find
\begin{align}
&{} \quad
\lan \eta_+,\ol{\eta}_- |\wh{H}| \ol{\chi}_+,\chi_- \ran \nonumber\\
&= 
\wh{H}(
\wh{\ol{\chi}}_+ \rightarrow \ol{\chi}_+, \,
\wh{\chi}_- \rightarrow \chi_-, \,
\wh{\chi}_+ \rightarrow \eta_+, \,
\wh{\ol{\chi}}_- \rightarrow \ol{\eta}_-)
e^{\cN_d(-\ol{\chi}_+\eta_+ - \ol{\eta}_-\chi_-)} \nonumber\\
&=
\wh{H}\left(
\wh{\ol{\chi}}_+ \rightarrow \frac{1}{\cN_d}\frac{\de}{\de\eta_+}, \,
\wh{\chi}_- \rightarrow -\frac{1}{\cN_d}\frac{\de}{\de\ol{\eta}_-}, \,
\wh{\chi}_+ \rightarrow \eta_+, \,
\wh{\ol{\chi}}_- \rightarrow \ol{\eta}_- \right)
e^{\cN_d(-\ol{\chi}_+\eta_+ - \ol{\eta}_-\chi_-)} \,. \nonumber
\end{align}
One can see that the derivation is independent of the details of the Hamiltonian
and the replacement rule applies to Hamiltonians with interactions. Note that one also needs to be careful about the ordering of operators in the Hamiltonian, as can be seen from, for instance, the mass terms of \eqref{rad_H_ads}.

\section{Holographic Wilsonian RG for Dirac fermions in a rotationally symmetric background}
\label{app:rot-sym}

We consider Dirac equation on a rotationally symmetric background with
$A_i=0$ and using a specific representation of Dirac matrices
following \cite{Faulkner:2009wj}. We find that the flow equations \eq{fermionic-flow}
simplify considerably and construct their solutions in terms of
classical solutions of the Dirac equations. In this section, we will
use the Lorentzian signature.

\subsection{Dirac equation and classical solutions}
The action \eq{action-phi} in the momentum space becomes
\begin{align}
S 
= \int dz \int_k
\left(
\ol{\Phi}\Ga^{\wh{z}}\pd_z\Phi
+ \sqrt{g_{zz}}
\left[
\ol{\Phi}(-i\Ga^tu_t + i\Ga^ik_i)\Phi
- m\ol{\Phi}\Phi
\right]
\right) \,.
\end{align}
with $u_t := \om + qA_t$.
The equations of motion can be written as
\SP{
\bsD\Phi &= 0 \,, \quad
\bsD := 
\Ga^{\wh{z}}\pd_z - m\sqrt{g_{zz}} 
- i\sqrt{-g_{zz}g^{tt}}\Ga^{\wh{t}}u_t
+ i\sqrt{g_{zz}g^{ii}}\Ga^{\wh{i}}k_i \,, \\
\ol{\Phi}\ol{\bsD} &= 0 \,, \quad
\ol{\bsD} := 
\Ga^{\wh{z}}\pd_z + m\sqrt{g_{zz}} 
+ i\sqrt{-g_{zz}g^{tt}}\Ga^{\wh{t}}u_t
- i\sqrt{g_{zz}g^{ii}}\Ga^{\wh{i}}k_i \,.
}
Without loss of generality, we can choose a special $k$ 
due to the rotational symmetry in the spatial directions
\begin{align}
k &:= k_1 \,, \quad
\wh{k}_1 := 1 \,, \quad
\wh{k}_2 \,, \cdots \,, \wh{k}_{d-1} := 0 \,.
\end{align}
For simplicity, we consider the case $d=3$ (the extension
to higher dimensions is given in Appendix \ref{app:higher-dim}).
The Dirac operator commutes with following two projectors
\begin{align}
[\bsD, \Pi_{\al}] &= [\ol{\bsD}, \Pi_{\al}] = 0 \,, \quad
\Pi_{\al}\Pi_{\bt} = \de_{\al\bt}\Pi_{\bt} \,, \quad
\Pi_{\al} := \frac{1-(-)^{\al}\wh{k}_i\Ga^{\wh{i}}\Ga^{\wh{z}}\Ga^{\wh{t}}}{2} \,, \quad
\wh{k}_i\wh{k}_i = 1 \,, \quad \al=1,2 \,. 
\label{proj}
\end{align}
We adopt such a representation of the Dirac matrices that 
the Dirac operator is diagonal in terms of the projections
\SP{
&{}
\Ga^{\wh{z}}
=
\begin{pmatrix}
\sig_3 & 0 \\ 0 & \sig_3
\end{pmatrix} \,, \quad
\Ga^{\wh{t}}
=
\begin{pmatrix}
i\sig_1 & 0 \\ 0 & i\sig_1
\end{pmatrix} \,, \quad
\Ga^{\wh{1}}
=
\begin{pmatrix}
-\sig_2 & 0 \\ 0 & \sig_2
\end{pmatrix} \,, \quad
\Ga^{\wh{2}}
=
\begin{pmatrix}
0 & \sig_2 \\ \sig_2 & 0
\end{pmatrix} \,, \\
&{}
\Pi_1 = 
\begin{pmatrix}
1 & 0 \\ 0 & 0
\end{pmatrix} \,, \quad
\Pi_2 = 
\begin{pmatrix}
0 & 0 \\ 0 & 1
\end{pmatrix} \,, \quad
\Phi = 
\begin{pmatrix}
\Phi_1 \\ \Phi_2
\end{pmatrix} \,, \quad
\ol{\Phi} =
(\ol{\Phi}_1, \ol{\Phi}_2) \,.
\label{rep}
}
Then, the action and the equations of motion become
\begin{align}
&{}
S =
\sum_{\al}\int dz\int_k 
\left(
\ol{\chi}_{\al}^+\pd_z\chi_{\al}^+ - \ol{\chi}_{\al}^-\pd_z\chi_{\al}^-
+ 
\sqrt{g_{zz}} 
\left[
- \ol{\chi}_{\al}^+T_{\al}^-\chi_{\al}^-
+ \ol{\chi}_{\al}^-T_{\al}^+\chi_{\al}^+
- m\ol{\chi}_{\al}^+\chi_{\al}^+ 
- m\ol{\chi}_{\al}^-\chi_{\al}^-
\right]
\right) \,, \nonumber\\
&{}
(\sqrt{g^{zz}}\pd_z \mp m)\chi_{\al}^{\pm} = T_{\al}^{\mp}\chi_{\al}^{\mp} \,, \quad
(\sqrt{g^{zz}}\pd_z \pm m)\ol{\chi}_{\al}^{\pm} = T_{\al}^{\pm}\ol{\chi}_{\al}^{\mp} \,, \nonumber\\
&{}
\Phi_{\al} =
\begin{pmatrix}
\chi_{\al}^+ \\ \chi_{\al}^-
\end{pmatrix} \,, \quad
\ol{\Phi}_{\al} =
(\ol{\chi}_{\al}^+ , \ol{\chi}_{\al}^-) \,, \quad
T_{\al}^{\pm} := \pm\sqrt{-g^{tt}}u_t - (-)^{\al}\sqrt{g^{11}}k \,. \nonumber
\end{align}
The indices $\pm$ imply the eigenvalues $\pm 1$ of $\sig_3$,
which are actually equal to the $\Ga^{\wh{z}}$-chirality. 
The radial Hamiltonian and the anticommutation relations are defined by
\begin{align}
&{}
\wh{H} 
:=
\int\!\frac{d^dk}{(2\pi)^d} \sqrt{g_{zz}}
\left[
- \ol{\chi}_{\al}^+T_{\al}^-\chi_{\al}^-
+ \ol{\chi}_{\al}^-T_{\al}^+\chi_{\al}^+
+ m\chi_{\al}^+\ol{\chi}_{\al}^+
- m\ol{\chi}_{\al}^-\chi_{\al}^-
\right], \nonumber\\
&{}
\{\chi_{\al}^{\pm}, \ol{\chi}_{\bt}^{\pm}\} 
= \pm\ka^2(2\pi)^d \, \de_{\al\bt} \,, \quad
\label{apptwist}
\end{align}
where the summation symbol over $\al$ is omitted in the radial Hamiltonian.
We can prove that the radial Hamiltonian and the anticommutation relations
reproduce the classical equations of motion.

The component expressions of the equations of motion are given by
\begin{align}
\label{eq:Dirac}
(\sqrt{g^{zz}} \pd_z \mp m)\Phi_{\al}^{\pm} = T_{\al}^{\mp}\Phi_{\al}^{\mp} \,, \quad
(\sqrt{g^{zz}} \pd_z \pm m)\ol{\Phi}_{\al}^{\pm} = T_{\al}^{\pm}\ol{\Phi}_{\al}^{\mp} \,.
\end{align}
One can see from the equations
of motion that $\Phi_{\al}^{\pm}$ and $\ol{\Phi}_{\al}^{\mp}$ satisfy
the same equations.  From the coupled first order equations, one can
get a decoupled second order equation for each component
\SP{
\label{eq2ord} 
\left( \sqrt{g^{zz}} \partial_z \pm m -
\sqrt{g^{zz}} \partial_z \log T^\mp_\alpha \right) \left(
\sqrt{g^{zz}} \partial_z \mp m \right) \Phi^\pm_\alpha - \left(
g^{ii} k^2 + g^{tt} u_t^2 \right) \Phi^\pm_\alpha &= 0 \,, \\ \left(
\sqrt{g^{zz}} \partial_z \mp m - \sqrt{g^{zz}} \partial_z \log
T^\pm_\alpha \right) \left( \sqrt{g^{zz}} \partial_z \pm m \right)
\ol{\Phi}^\pm_\alpha - \left( g^{ii} k^2 + g^{tt} u_t^2 \right)
\ol{\Phi}^\pm_\alpha &= 0 \,.  }
Let us introduce a basis of solutions to \eqref{eq:Dirac} 
and express any solution to \eqref{eq:Dirac} in terms of the basis,
\begin{align}
\Phi_{\al}^\pm &= A_{\al}\phi_{\al}^{\pm} + B_{\al}\vphi_{\al}^{\pm} \,, \quad
\ol{\Phi}_{\al}^\pm = \ol{A}_{\al}\ol{\phi}_{\al}^{\pm} + \ol{B}_{\al}\ol{\vphi}_{\al}^{\pm} \,,
\label{solbasis}
\end{align}
where $(+)$-component and $(-)$-component are related by \eqref{eq:Dirac}.
The numerical coefficients $(A_{\al}, B_{\al}, \ol{A}_{\al}, \ol{B}_{\al})$
are fixed by boundary conditions.
Examples of the solution bases are given in Section~\ref{sec:FP}.

\subsubsection{AdS$_{d+1}$}

The solutions to \eq{eq2ord} are given by ($\nu_\pm = m \pm
\frac{1}{2}$)  
\SP{ 
\Phi^\pm &= \sqrt{z} \left( a_\pm(k_\mu)
  I_{-\nu_\mp}(z K) + b_\pm(k_\mu) I_{\nu_\mp}(z K) \right)
\,,  \\ 
\ol{\Phi}^\pm &= \sqrt{z} \left( \ol a_\pm(k_\mu)
  I_{-\nu_\pm}(z K) + \ol b_\pm(k_\mu) I_{\nu_\pm}(z K) \right) \,.  
\label{dirac-sols}}
where $K^2 \equiv k^2 - (\omega + \mu_q)^2 \neq 0, \om \neq 0$, 
and we have suppressed the $\a$-index.  
The requirement that these solutions \eq{dirac-sols} satisfy the Dirac
equation Eq.~\eqref{eq:Dirac}, relates the integration constants as
\SP{ a_-(k_\mu) &= \frac{k + \omega + \mu_q}{K} \ a_+(k_\mu) \,, \quad
  \ \ b_-(k_\mu) = \frac{k + \omega + \mu_q}{K} \ b_+(k_\mu) \,, \\ \ol
  a_+(k_\mu) &= \frac{k + \omega + \mu_q}{K} \ \ol a_-(k_\mu) \,, \quad \ \ol
  b_+(k_\mu) = \frac{k + \omega + \mu_q}{K} \ \ol b_-(k_\mu) \,.  }
For reference, we give the asymptotic behavior $(z\sim 0)$ of the Bessel functions,
\begin{align}
\sqrt{z}I_{-\nu_+}(zK) &\sim z^{-m} \,, \quad
\sqrt{z}I_{-\nu_-}(zK) \sim z^{-m+1} \,, \quad
\sqrt{z}I_{\nu_+}(zK) \sim z^{m+1} \,, \quad
\sqrt{z}I_{\nu_-}(zK) \sim z^{m} \,.
\label{bessel-small-z}
\end{align}

\subsubsection{Near horizon limit} \label{app-nh}

In the near horizon limit, i.e. small $z - z_*$, we define new coordinates $\zeta$ and $\tau$ through
\SP{
	z - z_* = - \frac{R_2^2 \lambda z_*^2}{\zeta}, \ \ t = \lambda^{-1} \tau, \ \ \lambda \ll 1,
}
where $R_2 = 1/\sqrt{d(d-1)}$. Furthermore, we study low frequencies $\omega$ such that $\Omega := \lambda^{-1} \omega$ is held fixed in the $\lambda \rightarrow 0$ limit. Expanding in small $\lambda$, we have that
\SP{
	\sqrt{g^{zz}} \partial_z \log T^\pm_\alpha = \frac{R_2^{-1} \zeta \Omega}{q e_d + \zeta \Omega \mp (-)^\alpha R_2 z_* k} + \mathcal O(\lambda),
}
so that Eq.~\eqref{eq2ord} becomes
\SP{
	\left( \zeta \partial_\zeta \pm R_2 m - \frac{\zeta \Omega}{q e_d + \Omega \pm (-)^\alpha R_2 z_* k} \right) \left( \zeta \partial_\zeta \mp R_2 m \right) \Phi^\pm_\alpha - \left( R_2^2 z_*^2 k^2 - (q e_d + \zeta \Omega)^2 \right) \Phi^\pm_\alpha &= 0, \\
	\left( \zeta \partial_\zeta \mp R_2 m - \frac{\zeta \Omega}{q e_d + \zeta \Omega \mp (-)^\alpha R_2 z_* k} \right) \left( \zeta \partial_\zeta \pm R_2 m \right) \ol{\Phi}^\pm_\alpha - \left( R_2^2 z_*^2 k^2 - (q e_d + \zeta \Omega)^2 \right) \ol{\Phi}^\pm_\alpha &= 0. \label{dirac-ads2}
}
For the special case $\Omega = 0$, alternatively in the $\zeta \rightarrow 0$ limit, one has the solutions
\SP{
	\Phi^\pm_\alpha &= a^\pm_\alpha \zeta^{-\nu_k} + b^\pm_\alpha \zeta^{\nu_k} \,, \quad
	\nu_k := \sqrt{ R_2^2 (m^2 + z_*^2 k^2) - q^2 e_d^2}, \label{diracsol-ads2}
}
for some (not independent) integration constants $a^\pm_\alpha$ and $b^\pm_\alpha$. 
Hence, in the AdS$_2$ region, $\nu_k$ plays the role of an effective mass.
Note that in fact the Dirac equation \eqref{dirac-ads2} yields analytic solutions
in terms of the Whittaker functions \cite{Faulkner:2011tm, Gauntlett:2011wm}.

\subsection{Schr\"{o}dinger equation and flow equations}
\label{app:nice-flow}
The Schr\"{o}dinger equation consistent with the Heisenberg equation is
\begin{align}
&{}
-\ka^2\pd_z\Psi[\eta^+, \ol{\eta}^-] = H\Psi[\eta^+, \ol{\eta}^-] \,, \nonumber\\
&{}
H := \wh{H}\left(
\Phi_{\al}^+ \rightarrow \eta_{\al}^+ \,,
\ol{\Phi}_{\al}^+ \rightarrow \frac{\de}{\de\eta_{\al}^+} \,,
\Phi_{\al}^- \rightarrow -\frac{\de}{\de\ol{\eta}_{\al}^-} \,,
\ol{\Phi}_{\al}^- \rightarrow \ol{\eta}_{\al}^-
\right) \,, \nonumber\\
&{}
\frac{H}{\ka^2}
=
\int_k \sqrt{g_{zz}}
\left[
(\ka_d^2)^2\frac{\de}{\de\eta_{\al}^+} T_{\al}^- \frac{\de}{\de\ol{\eta}_{\al}^-}
+ \ol{\eta}_{\al}^- T_{\al}^+ \eta_{\al}^+
+ \ka_d^2 m\eta_{\al}^+\frac{\de}{\de\eta_{\al}^+}
+ \ka_d^2 m\ol{\eta}_{\al}^-\frac{\de}{\de\ol{\eta}_{\al}^-}
\right] \,.
\label{app:schro}
\end{align}
Let us consider an ansatz for the wave-functional of the form\footnote{More generally, one could consider a term of the form $\ol{\eta}_{\al}^-F_{\al\beta}\eta_{\beta}^+$. However, since the Hamiltonian is diagonal, it is consistent, for simplicity, to concentrate on flows for which the off-diagonal components are put to zero.}
\begin{align}
\Psi[\eta^+, \ol{\eta}^-] 
&:= 
\exp\left[
-\int_k
\left(
\ol{\eta}_{\al}^-F_{\al}\eta_{\al}^+ 
+ \ol{J}_{\al}^+\eta_{\al}^+ + \ol{\eta}_{\al}^-J_{\al}^- + C_{\al}
\right)
\right] \,,
\label{psi-ansatz}
\end{align}
where coefficients $F_{\al}, J_{\al}^-, \ol{J}_{\al}^+$ depend on $z$.
Then, the Schr\"{o}dinger equation yields the equations
\eq{eq:floweqsnicegammas} for the coefficients 
with solutions given by \eq{flowsol}.

\subsection{\label{app:higher-dim}Extension to higher dimensions}
Here we give a comment on the extension to other dimensions.
The choice of the representation of the $(1+3)$-dimensional Dirac matrices 
was made to make the Dirac operator diagonal.
In order to keep our analysis above even in higher dimensions, 
we have to keep the diagonal structure of the projectors $\Pi_{\al}$ 
in the extension to higher dimensions.
For explicit constructions, it is convenient to introduce the tensor notation
\begin{align}
A:= 
\begin{pmatrix}
a & b \\ c & d
\end{pmatrix}
\longrightarrow
A \otimes B =
\begin{pmatrix}
aB & bB \\ cB & dB
\end{pmatrix} \,.
\end{align}
Then, the $(1+3)$-dimensional case is
\begin{align}
\Ga^{\wh{z}} &= 1 \otimes \sig_3 \,, \quad
\Ga^{\wh{t}} = 1 \otimes i\sig_1 \,, \quad
\Ga^{\wh{1}} = -\sig_3 \otimes \sig_2 \,, \quad
\Ga^{\wh{2}} = \sig_1 \otimes \sig_2 \,. \quad
\end{align}
In $(1+4)$-dimensions, $\Ga^{\wh{3}}$ can be chosen to be proportional to 
a product of all Dirac matrices in $(1+3)$-dimension.
\begin{align}
\Ga^{\wh{z}} &= 1 \otimes \sig_3 \,, \quad
\Ga^{\wh{t}} = 1 \otimes i\sig_1 \,, \quad
\Ga^{\wh{1}} = -\sig_3 \otimes \sig_2 \,, \quad
\Ga^{\wh{2}} = \sig_1 \otimes \sig_2 \,, \quad
\Ga^{\wh{3}} = \sig_2 \otimes \sig_2 \,.
\end{align}
In $(1+5)$-dimension, in order to keep the diagonal structure of the projectors $\Pi_{\al}$,
the first three matrices $\Ga^{\wh{z}}, \Ga^{\wh{t}}, \Ga^{\wh{1}}$ 
should have the same structure as the lower dimensional cases. 
This is realized by multiplying the three by the unit matrix of size 2.
\begin{align}
\Ga^{\wh{z}} &= 1 \otimes \sig_3 \otimes 1 \,, \quad
\Ga^{\wh{t}} = 1 \otimes i\sig_1 \otimes 1\,, \quad
\Ga^{\wh{1}} = -\sig_3 \otimes \sig_2 \otimes 1\,, \quad \nonumber\\
\Ga^{\wh{2}} &= \sig_1 \otimes \sig_2 \otimes 1 \,, \quad
\Ga^{\wh{3}} = \sig_2 \otimes \sig_2 \otimes \sig_2 \,, \quad
\Ga^{\wh{4}} = \sig_2 \otimes \sig_2 \otimes \sig_3 \,.
\end{align}
In this way, we can continue explicit constructions of the Dirac matrices.
Thus, all the analyses above hold for higher dimensions.

\section{Derivation of path integral from the operator formalism} \label{app_bdy_act}
In this section, we derive a path integral representation of a transition amplitude
from the anti-commutation relations and the Hamiltonian.
Especially, the boundary actions, 
which are usually required classically to ensure a well-defined variational problem of a classical action,
will be derived quantum mechanically.

Let us derive the path integral representation of the following transition amplitude
\begin{align}
\lan \chi_+, \ol{\chi}_-|{\rm P}e^{-\int_0^ldz\wh{H}}|\ol{\chi}_+, \chi_- \ran \,,
\end{align}
with a radial Hamiltonian in a quadratic form.\footnote{
This is just for simplicity.
Actually, we do not have to limit ourselves to a quadratic form
as seen from the following derivation.}
\begin{align}
\wh{H} &:= \int_k \sqrt{g_{zz}}
\left[
\ol{\chi}_+h_{++}\chi_+ + \ol{\chi}_+h_{+-}\chi_-
+ \ol{\chi}_-h_{-+}\chi_+ + \ol{\chi}_-h_{--}\chi_-
\right] \,.
\end{align}
A path integral representation is obtained by dividing a transition amplitude 
into a product of infinitesimal transition amplitudes
with inserting the completeness relation
\begin{align}
\prod_{k=1}^{N-1}\left[\int\!d^4\chi_k\right]T(N,N-1)C(N-1)T(N-1,N-2)\cdots T(2,1)C(1)T(1,0) \,,
\label{prod_pi}
\end{align}
where $\ep := l/N, ~ d^4\chi_k := d\chi^{(k)}_+d\ol{\chi}^{(k)}_+d\chi^{(k)}_-d\ol{\chi}^{(k)}_-$, 
and $(0)$ and $(N)$ correspond to the initial time and the final time, respectively.
$T(k+1,k)$ and $C(k)$ are the transition amplitude from 
$(k)$ to $(k+1)$ and an exponential of the completeness relation, respectively:
\begin{align}
T(k+1,k) 
&:= 
\lan \chi^{(k+1)}_+, \ol{\chi}^{(k+1)}_-|{\rm T}e^{-\ep\wh{H}}|\ol{\chi}^{(k)}_+, \chi^{(k)}_- \ran \,, 
\nonumber\\
C(k) &:= e^{\int_k(\ol{\chi}^{(k)}_+\chi^{(k)}_++\ol{\chi}^{(k)}_-\chi^{(k)}_-)} \,.
\end{align}
The infinitesimal transition amplitude $T(k,k-1)$ becomes
\begin{align}
T(k,k-1)
&= e^{-\ep H(k;k-1)+\int_k(-\ol{\chi}^{(k-1)}_+\chi^{(k)}_+-\ol{\chi}^{(k)}_-\chi^{(k-1)}_-)} \,, 
\nonumber\\
H(k,k-1) &:= \int_k\sqrt{g_{zz}}\,
\left[
\ol{\chi}^{(k-1)}_+h_{+-}\chi^{(k-1)}_-
+ \ol{\chi}^{(k)}_-h_{-+}\chi^{(k)}_+
+ \ol{\chi}^{(k)}_+h_{++}\chi^{(k-1)}_+
- \ol{\chi}^{(k)}_-h_{--}\chi^{(k-1)}_- 
\right] \,. \nonumber
\end{align}
Let us see the exponents of $T(k+1,k)C(k)T(k,k-1)C(k-1)$ in \eqref{prod_pi},
\begin{align}
&{}
-\ep H(k+1,k) -\ep H(k,k-1) \nonumber\\
&{} 
+\int_k\left[ ~~
\al({-\ol{\chi}^{(k)}_+\chi^{(k+1)}_+}-\ol{\chi}^{(k+1)}_-\chi^{(k)}_-)
+\bt(-\ol{\chi}^{(k)}_+\chi^{(k+1)}_+{-\ol{\chi}^{(k+1)}_-\chi^{(k)}_-})
\right. \nonumber\\
&{} \quad\quad\quad\quad
+\al({\ol{\chi}^{(k)}_+\chi^{(k)}_+}{+\ol{\chi}^{(k)}_-\chi^{(k)}_-})
+\bt({\ol{\chi}^{(k)}_+\chi^{(k)}_+}{+\ol{\chi}^{(k)}_-\chi^{(k)}_-}) \nonumber\\
&{} \quad\quad\quad\quad
+\al(-\ol{\chi}^{(k-1)}_+\chi^{(k)}_+{-\ol{\chi}^{(k)}_-\chi^{(k-1)}_-})
+\bt({-\ol{\chi}^{(k-1)}_+\chi^{(k)}_+}-\ol{\chi}^{(k)}_-\chi^{(k-1)}_-) \nonumber\\
&{} \left. \quad\quad\quad\quad
+\al(\ol{\chi}^{(k-1)}_+\chi^{(k-1)}_++\ol{\chi}^{(k-1)}_-\chi^{(k-1)}_-)
+\bt(\ol{\chi}^{(k-1)}_+\chi^{(k-1)}_++\ol{\chi}^{(k-1)}_-\chi^{(k-1)}_-)
\right] \,, \label{tctc}
\end{align}
where each exponent of $T,C$ is divided into two parts through $\al+\bt=1$.
First, let us extract four terms from the terms with $\al$-coefficients,
which are the first term in the second line,
the first and second terms in the third line and the second term in the fourth line,
of the last expression \eqref{tctc}.
The sum of the four terms becomes
\begin{align}
-\al\sum_{k=1}^{N-1} \ol{\chi}^{(k)}_+(\chi^{(k+1)}_+-\chi^{(k)}_+)
+\al\sum_{k=0}^{N-2} \ol{\chi}^{(k+1)}_-(\chi^{(k+1)}_--\chi^{(k)}_-) \,. \label{1stcomb}
\end{align} 
One can see that this becomes a kinetic term after the $\ep\to 0$ limit.
Similarly, let us extract four terms from the terms with $\bt$-coefficients,
which are the fourth term in the second line,
the third and fourth terms in the third line and the third term in the fourth line,
of \eqref{tctc}.
The sum of the four terms becomes
\begin{align}
\bt\sum_{k=0}^{N-2} (\ol{\chi}^{(k+1)}_+-\ol{\chi}^{(k)}_+)\chi^{(k+1)}_+
-\bt\sum_{k=1}^{N-1} (\ol{\chi}^{(k+1)}_--\ol{\chi}^{(k)}_-)\chi^{(k)}_+ \,. \label{2ndcomb}
\end{align}
One can see that this becomes a kinetic term after the $\ep\to 0$ limit.
Then, gathering all factors and extracting the exponent 
taking these two types of combinations into account, we find
\begin{align}
&{}
-\ep\sum_{k=0}^{N-1}H(k+1,k) \nonumber\\
&{}
+\int_k\ep\left[
-\al\sum_{k=1}^{N-1} \ol{\chi}^{(k)}_+(\chi^{(k+1)}_+-\chi^{(k)}_+)\ep^{-1}
+\al\sum_{k=0}^{N-2} \ol{\chi}^{(k+1)}_-(\chi^{(k+1)}_--\chi^{(k)}_-)\ep^{-1}
\right. \nonumber\\
&{} \quad\quad\quad\quad ~
\left.
+\bt\sum_{k=0}^{N-2} (\ol{\chi}^{(k+1)}_+-\ol{\chi}^{(k)}_+)\ep^{-1}\chi^{(k+1)}_+
-\bt\sum_{k=1}^{N-1} (\ol{\chi}^{(k+1)}_--\ol{\chi}^{(k)}_-)\ep^{-1}\chi^{(k)}_-
\right] \nonumber\\
&{}
+\int_k
\left[
-\al\ol{\chi}^{(N)}_-\chi^{(N-1)}_- - \bt\ol{\chi}^{(N-1)}_+\chi^{(N)}_+
-\bt\ol{\chi}^{(1)}_-\chi^{(0)}_- - \al\ol{\chi}^{(0)}_+\chi^{(1)}_+
\right] \,.
\end{align}
The second line results from the first combination \eqref{1stcomb}
and the third line results from the second combination \eqref{2ndcomb}.
After taking $N\rightarrow\infty$, we find
\begin{align}
&{} \quad \int\!dz \left[
\int_k
\left(
-\al\ol{\chi}_+\pd_z\chi_+ + \al\ol{\chi}_-\pd_z\chi_-
+\bt\pd_z\ol{\chi}_+\chi_+ - \bt\pd_z\ol{\chi}_-\chi_-
\right)
-H
\right] \nonumber\\
&{}
+\int_k\left[
-\al\ol{\chi}_-\chi_- - \bt\ol{\chi}_+\chi_+ 
\right]_{z=l}
+\int_k\left[
-\al\ol{\chi}_+\chi_+ - \bt\ol{\chi}_-\chi_-
\right]_{z=\ep} \,,
\end{align}
which consists of $-S$ plus boundary actions at two ends.
The boundary actions coincide with those obtained \cite{Henneaux:1998ch}.
One can see that partial integrations of the kinetic terms of the Dirac action
corresponds to changes of pairings in taking all products of the infinitesimal transition amplitudes.

\section{\label{app:counterterm}Counterterm action}

In AdS/CFT, if we start from the classical action with no counterterm action,
it is difficult to take a well-defined continuum limit $\ep_0\to 0$
and get a correct answer.
For example, a two-point function generally contains 
contact terms with integer powers of momentum $k_{\mu}$, 
which dominate over non-local terms with non-integer powers of $k_{\mu}$
in the continuum limit.
Since the non-local term with the lowest power
gives the correct scaling behavior of two-point functions,
we need to eliminate the dominating contact terms.
The counterterm action is added to fulfil this purpose.\footnote{
Of course, in addition, the rescaling of boundary fields is also needed
to make correlation functions finite in the $\ep_0\to 0$ limit.}
We adopt the Poincare coordinate $ds^2\sim z^{-2}(dz^2+\cdots)$
near the AdS boundary. For details of holographic renormalization and the counterterm action, see, {\em e.g.}, \cite{Skenderis:2002wp, Papadimitriou:2004ap}.

\subsection{Scalar case}
The GKP-Witten relation with the counterterm action in our language is
\begin{align}
\Bbra \exp\int_k\phi_0O\Kket_{{\rm ct}, \ep_0}^{\std} 
&= \lan \IR|\wh{U}(\ep_{\IR}, \ep_0)e^{\wh{S}_{\ct}^0}|\phi_0 \ran \,.
\label{scalar-ct}
\end{align}
Here $\ep_0$ is a UV cutoff and $S_{\ct}$ is the counterterm action.
One choice of the counterterm action is\footnote{
In general, $A_0$ can take different forms: 
for example, a minimal choice is to take terms which have smaller (integer) powers 
than the smallest non-integer power $2\nu$
in the expansion of $\ep_0^{d}A_0$ in $k\ep_0$.
The remaining integer powers become subleading in $\ep_0\to 0$ limit since
they are larger than $2\nu$.
This remark holds in the fermion case, as well.}
\begin{align}
&{} 
\wh{S}_{\ct}^0
:= \frac{1}{2}\int_kA_0\wh{\phi}^2 \,, \quad
A_0 := \frac{\pi_-(\ep_0)}{\phi_-(\ep_0)}
= -\ep_0^{-d}\left(\De_- + O((k\ep_0)^2) +\cdots\right) \,,
\label{scalar-s-ct}
\end{align}
where we introduced solutions $\phi_{\pm}$ to the classical equation of motion
and their conjugate momenta $\pi_{\pm}$,
\begin{align}
\phi_{\pm}(z) := c_{\pm}z^{d/2}I_{\pm\nu}(kz) = z^{\De_{\pm}}+\cdots \,, \quad
\pi_{\pm}(z) := -\sqrt{g}g^{zz}\pd_z\phi_{\pm}(z) \,.
\end{align}
Let us consider the following question: 
if a solution to the radial Schr\"{o}dinger equation is given,
what couplings does the UV generating functional contain?
A general solution to the radial Schr\"{o}dinger equation can be written as
\begin{align}
\lan \phi|\wh{U}(\ep,\ep_0)e^{\wh{S}_{\ct}^0}|\Psi_0\ran
=\exp\left[\int_k\left(-\frac{1}{2}f(\ep)\phi^2+J(\ep)\phi\right)+C(\ep)\right] \,.
\end{align}
Then, the above question is to find a UV generating function corresponding to $|\Psi_0\ran$,
which are derived as follows:
\begin{align}
\lan \IR|\wh{U}(\ep_{\IR},\ep_0)e^{\wh{S}_{\ct}^0}|\Psi_0\ran 
&= 
\int\!D\phi_0\lan \IR|\wh{U}(\ep_{\IR},\ep_0)e^{\wh{S}_{\ct}^0}|\phi_0\ran 
\lan\phi_0|\Psi_0\ran \nonumber\\
&=
\Bbra \exp\left(\frac{1}{2}\int_k\frac{O^2}{f(\ep_0)+A_0}
+\int_k\frac{J(\ep_0)O}{f(\ep_0)+A_0}+\cdots\right) \Kket_{{\rm ct}, \ep_0}^{\std} \,.
\label{scalar-double-trace-ct}
\end{align}

\subsection{Fermion case} \label{app:fermion-ct}
The discussion is parallel to the scalar case.
The GKP-Witten relation with the counterterm action is
\begin{align}
\Bbra \exp\left(\int_k [\ol{\chi}^0_+O_+ + \ol{O}_-\chi^0_-] \right) \Kket_{{\rm ct}, \ep_0}^{\std}
&= \lan \IR|\wh{U}(\ep_{\IR}, \ep_0)e^{\wh{S}_{\ct}^0}|\ol{\chi}_+^0, \chi_-^0\ran \,.
\label{fermion-ct}
\end{align}
One choice of the counterterm action is 
\begin{align}
\wh{S}_{\ct}^0 &:= \int_k\wh{\ol{\chi}}_+A_0\wh{\chi}_- \,, \quad
A_0 := \frac{\phi_+(\ep_0)}{\phi_-(\ep_0)} \,,
\label{fermion-s-ct}
\end{align}
where we used the same notation as \eqref{solbasis}.
Note that $\phi_-\sim z^{-m}$ if a fermion is of mass $m$.
Now, let us assume that we have a solution to the radial Schr\"{o}dinger equation given by
\begin{align}
\lan \chi_+,\ol{\chi}_-|\wh{U}(\ep,\ep_0)e^{\wh{S}_{\ct}^0}|\Psi_0 \ran
&= \exp\left(-\int_k
\left[\ol{\chi}_-F(\ep)\chi_+ + \ol{J}_+(\ep)\chi_+ + \ol{\chi}_-J_-(\ep)\right]\right) \,.
\end{align}
Then, the corresponding UV generating functional is
\begin{align}
&{} \quad
\lan\IR|\wh{U}(\ep_{\IR}, \ep_0)e^{\wh{S}_{\ct}}|\Psi_0\ran
\nonumber\\
&= 
\Bbra \exp
\int_k\left[
- \ol{O}_-\frac{1}{F(\ep_0)^{-1}-A_0}O_+ 
+ \ol{J}_+(\ep_0)\frac{F(\ep_0)^{-1}}{F(\ep_0)^{-1}-A_0}O_+
+ \ol{O}_-\frac{F(\ep_0)^{-1}}{F(\ep_0)^{-1}-A}J_-(\ep_0)
\right] \Kket_{{\rm ct}, \ep_0}^{\std} \,.
\label{genct_f}
\end{align}

\end{document}